\documentclass[11pt]{article}
\usepackage[utf8]{inputenc}
\usepackage{tensor}
\usepackage{amssymb}
\usepackage{authblk}
\usepackage{xcolor}
\usepackage{graphicx}
\usepackage{jheppub}
\usepackage{bm}
\usepackage{tikz}
\newcommand{\p}{\partial}
\usepackage[multiple]{footmisc}
\usepackage{subcaption}
\usepackage{longtable}
\usepackage{booktabs}
\usepackage[multiple]{footmisc}
\usepackage{subcaption}
\usepackage{multicol}
\usepackage{multirow}
\usepackage{rotating}
\usepackage{array}

\makeatletter
\newcommand*\bigcdot{\mathpalette\bigcdot@{.65}}
\newcommand*\bigcdot@[2]{\mathbin{\vcenter{\hbox{\scalebox{#2}{$\m@th#1\bullet$}}}}}
\makeatother

\title{Generalized ADHM equations from marginal deformations in open superstring field theory}
\author[a,b]{Jakub Vo\v{s}mera
} 
\affiliation[a]{CEICO, Institute of Physics of the AS CR,\\ Na Slovance 2, 182 21 Prague 8, Czech~Republic} 
\affiliation[b]{Institute of Particle and Nuclear Physics, Charles University,\\ V Hole\v{s}ovi\v{c}k\'{a}ch 2, 180 00 Prague 8, Czech~Republic} 
\emailAdd{vosmera@gmail.com} 
\abstract{Working within the framework of both the $A_\infty$ and the Berkovits open superstring field theory, we derive a necessary and sufficient condition for a Neveu-Schwarz marginal deformation to be exact up to third order in the deformation parameter. For a specific class of backgrounds, we find that this condition localizes on the boundary of the worldsheet moduli space, thus providing a very simple computational prescription for recovering algebraic constraints (generalized ADHM equations) which need to be satisfied by the moduli.
Applying our results to the $\mathrm{D(-1)/D3}$ system, we confirm up to third order that blowing up the size of the $\mathrm{D}$-instanton inside the D3 brane worldvolume is an exact modulus of the full string theory. We also discuss examples of more complicated backgrounds, such as instantons on unresolved ALE spaces, as well as the spiked instantons.
}
\arxivnumber{1910.00538}

\begin{document}
  \maketitle
  

\section{Introduction and summary}

It is a widely accepted conjecture that classical solutions of open superstring field theory\footnote{Over the past three decades, several constructions of open superstring field theory in both NS and R sector appeared in the literature (see \cite{Witten:1985cc,Witten:1986qs,Preitschopf:1989fc,Arefeva:1989cp,Berkovits:1995ab,Erler:2013xta}, as well as \cite{Kroyter:2012ni,Iimori:2013kha,Erler:2015lya,Kunitomo:2015usa,Erler:2016ybs,Ohmori:2017wtx,Erler:2017onq} for some recent developments). In this paper we will focus on the NS sector of the $A_\infty$ open superstring field theory \cite{Erler:2013xta} and the Berkovits (WZW-like) open superstring field theory \cite{Berkovits:1995ab}. These were recently shown to be related by a field redefinition \cite{Erler:2015rra,Erler:2015uba,Erler:2015uoa}.} (OSFT) formulated on a given consistent open superstring background (boundary SCFT), should correspond to new
consistent open superstring backgrounds. Depending on the background, there are a variety of types of classical solutions to consider. This paper offers a detailed look at a special class of classical solutions, called \emph{marginal deformations} which exist for continuous ranges of parameters and generally appear whenever the GSO-projected boundary spectrum of the background at hand contains operators with conformal dimension 1/2 in the matter sector. Restrictions, imposed on the marginal couplings by the requirement that the deformation corresponds to a consistent solution of the classical equations of motion of the OSFT, will prove to offer a direct probe into the structure of the moduli space of the given open superstring background. Deriving manageable expressions for such constraints on marginal couplings, while keeping the setup as general as possible, is the main goal of this paper. As a concrete example of such a background, let us consider the system of superimposed stacks of $\text{D}(-1)$ and $\text{D}3$ branes in type IIB superstring. Here one can identify matter marginal operators associated both with the modes of open strings localized on the $\text{D}(-1)$ and the D$3$ brane stacks, as well as the modes of open strings stretched between the two stacks. We will see that imposing consistency of a classical solution of the OSFT equations of motion, which excites the vevs of these operators, yields the ADHM constraints on the moduli of $\mathcal{N}=4$ SYM instantons \cite{Atiyah:1978ri}. The consistent open superstring backgrounds provided by exactly marginal deformations in this system should therefore be generally identified with finite-size instantons living on the $\text{D}3$ worldvolume \cite{Witten:1995gx,Billo:2002hm}. The discussion in this paper will, however, demonstrate that our results are applicable to a much wider variety of backgrounds.



While it is well-known that there is an algorithmic way of writing down the classical equations of motion order by order in the deformation parameter, non-trivial conditions for the existence of a solution arise at each order. Namely, we have to require that the obstruction to inverting the BRST operator at each order in the deformation parameter vanishes. These obstructions can be identified as obstructions to exact marginality of the deformation at each order. We will first present conditions which the deformation needs to satisfy at first order in order for the obstruction appearing at second order to be absent. We will argue that these conditions are automatically satisfied whenever the background which we start with has at least $N=(2,0)$ supersymmetry in two non-compact target dimensions. On the other hand, a non-trivial constraint will be obtained by requiring that the obstruction at the third order vanishes (as first observed by \cite{Mattiello:2019gxc}). While this obstruction can be in general evaluated using the techniques of \cite{Berkovits:2003ny}, we will see that substantial simplifications will arise in situations where all marginal operators $\mathbb{V}_{1/2}$ in the NS sector decompose as $\mathbb{V}_{1/2}^++\mathbb{V}_{1/2}^-$ into states with charge $\pm 1$ under the $U(1)$ $R$-current of a global $\mathcal{N}=2$ worldsheet superconformal algebra \cite{Sen:2015uoa,Maccaferri:2018vwo}. Again, this will be argued to always be the case whenever our background has at least $N=(2,0)$ supersymmetry in two non-compact target dimensions. We will show that in such cases, the third-order obstruction localizes on the boundary of the worldsheet moduli space and that it vanishes if and only if the auxiliary fields
(as introduced in \cite{Maccaferri:2018vwo,Maccaferri:2019ogq})
\begin{subequations}
\label{eq:HsIntro}
\begin{align}
 \mathbb{H}_1^\pm &=\lim_{z\to 0}\left[\mathbb{V}^\pm_\frac{1}{2}(z)\mathbb{V}^\pm_\frac{1}{2}(-z)\right] \,,\\
\mathbb{H}_0 &=\lim_{z\to 0}\left[2z\left(\mathbb{V}^-_\frac{1}{2}(z)\mathbb{V}^+_\frac{1}{2}(-z)-\mathbb{V}^+_\frac{1}{2}(z)\mathbb{V}^-_\frac{1}{2}(-z)\right)\right]   \,,
\end{align}
\end{subequations}
are set to zero. Noting that \eqref{eq:HsIntro} give rise to algebraic (quadratic) constraints on the marginal couplings, we observe that this procedure provides a general, yet very simple prescription for extracting the geometry of the moduli space for any background where the worldsheet SCFT description is known and where our assumptions are valid. Borrowing the terminology of \cite{Maccaferri:2019ogq}, we shall call the constraints $\mathbb{H}_1^\pm = \mathbb{H}_0=0$ the \emph{generalized ADHM equations}. Also note that these conditions were identified by \cite{Maccaferri:2018vwo,Maccaferri:2019ogq} as the flatness conditions for the quartic effective potential as derived in the context of both Berkovits and $A_\infty$ OSFT. We can therefore conclude that (at least in the cases with the above-described enhanced $\mathcal{N}=2$ worldsheet superconformal symmetry) the notion of exactness of a marginal deformation up to third order coincides with the notion of flatness of the quartic potential of classical effective action.

We will first derive \eqref{eq:HsIntro} starting with the classical equations of motion of the $A_\infty$ open superstring field theory. We will also show that the localization property of the third-order obstruction persists if we deform the theory by adding stubs: we will see that the additional terms arising due to non-associativity of the star product with stubs exactly compensate the addition of fundamental bosonic 4-string vertex, thus avoiding the need to integrate over the corresponding bosonic moduli.
This provides a strong indication that obstructions arising at third order for marginal deformations of closed string backgrounds with an enhanced worldsheet superconformal symmetry will be amenable to a similar localization procedure (in the context an NS heterotic string field theory or an NSNS type II closed superstring theory; see also the Conclusions section of \cite{Maccaferri:2019ogq} for a discussion). 
We also show that the third-order obstruction derived in the Berkovits open superstring field theory is identical to the one derived in the $A_\infty$ theory provided that one assumes that the obstruction at second order vanishes. 


We will then go on to demonstrate the utility of the generalized ADHM equations $\mathbb{H}_1^\pm = \mathbb{H}_0 =0$ for deriving constraints on moduli in several cases of relevant backgrounds. Starting with the superposition of a stack of D$(-1)$ branes with a stack of (euclidean) D3 branes, we show that the usual ADHM equations \cite{Atiyah:1978ri,Corrigan:1978ce} (see \cite{Dorey:2002ik} for a review) are reproduced upon substituting the boundary marginal vertex operators appearing in the system into \eqref{eq:HsIntro}. In particular, we will see that the vanishing of $\mathbb{H}_1^\pm$ implies vanishing of the complex hyper-K\"{a}hler moment map $\mu^\mathbb{C}$ \cite{Hitchin:1986ea} (that is, the D-term of the corresponding 4d ${N}=2$ low-energy effective action of the brane configuration) while the vanishing of $\mathbb{H}_0$ implies vanishing of the real hyper-K\"{a}hler moment map $\mu^\mathbb{R}$ (the F-term). This shows that at least up to the third order in the marginal deformation parameter, it is possible to construct solutions to the classical equations of motion of open superstring field theory which correspond to non-trivial objects such as finite-size instantons. Put in different words, we show (up to third order in the deformation parameter) that finite-size instantons give rise to consistent open superstring backgrounds. Using the example of the D$(-1)$/D3 system, we also check that evaluating the third-order obstruction directly (by introducing Schwinger parametrization for the propagator in the spirit of \cite{Berkovits:2003ny}) yields the same constraints on the marginal couplings as setting the localized auxiliary fields \eqref{eq:HsIntro} to zero. We will then consider examples of more complicated backgrounds. Starting with the D$(-1)$/D3 system sitting at an unresolved $\mathbb{C}^2/\mathbb{Z}_n$ singularity, we recover the complex and real hyper-K\"{a}hler moment maps which first appeared in \cite{kronheimer1990yang} and were further discussed in the string theory context by \cite{Douglas:1996sw}. Finally, by considering general systems of D$(-1)$, D3 and D7 branes containing stretched string sectors with four Neumann-Dirichlet directions, we manage to reproduce some of the results of \cite{Nekrasov:2016qym,Nekrasov:2016gud} for the equations governing the moduli spaces of crossed and folded instantons at zero $B$-field. Note that our discussion will involve moduli for all marginal boundary fields which are present in the theory for the given background (including e.g.\ strings stretched between two D3 brane stacks spanning different complex 2-planes).

The paper is organized as follows. In section \ref{sec:obstAinf} we begin with a general discussion of marginal deformations in the $A_\infty$ formulation of open superstring field theory: we will first focus on deriving conditions which need to be satisfied in order for the second-order obstruction to vanish (subsection \ref{subsec:margDefAinf}), then discuss the structure of the obstruction arising at the third order (subsection \ref{subsec:ainf}) and finally briefly describe the changes which need to be put in place when we render the bosonic 2-product non-associative by adding stubs to the Witten star product (subsection \ref{subsec:Ainfstubs}). We will briefly repeat this discussion in section \ref{sec:berk} in the context of the Berkovits (WZW-like) formulation of open superstring theory, showing that give the deformation is chosen to make the second-order obstruction vanishing, then the third-order obstructions arising in the two theories are equivalent. In section \ref{sec:eval} we will exhibit three methods for obtaining explicit expressions for the third-order obstruction written directly in terms of the marginal NS boundary fields $\mathbb{V}_{1/2}$. The first two methods (subsections \ref{subsec:locBerk} and \ref{subsec:locX2}) will assume presence of a global $\mathcal{N}=2$ superconformal symmetry of the worldsheet theory and will yield a simple algebraic expression for the third-order obstruction. The third method (subsection \ref{subsec:direct}), while being available for more general backgrounds, will only yield expressions containing integral over a Schwinger parameter. In section \ref{sec:ex} we consider three examples of open superstring backgrounds (D$(-1)$/D3 system in flat space in subsection \ref{subsec:N4flat}, D$(-1)$/D3 system at $\mathbb{C}^2/\mathbb{Z}_n$ orbifold singularity in subsection \ref{subsec:N4orb} and spiked instantons at zero $B$-field in subsection \ref{subsec:spiked}) where we apply our results to compute algebraic constraints on moduli. Finally, in section \ref{sec:disc} we point out the main contributions of this paper, put them into the context of recent developments and offer a brief discussion of possible future directions.
We provide three appendices where we collect our conventions for spinors in 4d (appendix \ref{app:spin}), our conventions for the $A_\infty$ formulation of open superstring field theory (appendix \ref{app:conv}) and also some OPE and correlation functions to be used in the paper (appendix~\ref{app:ope}).

\section{Analysis of the third-order obstruction in $A_\infty$ OSFT}
\label{sec:obstAinf}

In this section we will analyse in detail the structure of the obstruction to marginal deformations in the $A_\infty$ OSFT \cite{Erler:2013xta} (both with and without stubs) which arises at third order in the deformation parameter.
We collect our conventions for $A_\infty$ OSFT in Appendix~\ref{app:conv}. As the discussion which is to follow is somewhat technical, let us now briefly summarize the main points. Considering a marignal deformation $\Psi(\lambda)$ whose leading order $\Psi_1$ is given by a $h=1/2$ zero-momentum NS Grassmann-odd state $\mathbb{V}_{1/2}$ as $\Psi_1 = c\mathbb{V}_{1/2}e^{-\phi}$, we will first show that the deformation is unobstructed at second order if and only if the projector condition 
\begin{align}
    P_0 M_2(\Psi_1,\Psi_1)=0
    \label{eq:introProjCon}
\end{align}
holds. Here $P_0$ projects on the kernel of the zero-mode of the total worldsheet stress-energy tensor (see subsection \ref{subsec:margDefAinf} for details) and $M_2$ is the 2-product of the $A_\infty$ OSFT. 
The projector condition \eqref{eq:introProjCon} holds both in the case with and without stubs. 
In the case without stubs, we will denote by $m_2$ the bare 2-string product, which can be defined in terms of the usual Witten's star product as $m_2(A,B)=(-1)^{d(A)}A\ast B$ (where $d(A)$ denotes the degree of $A$). Adding stubs of length $w$, the bare 2-string product gets deformed to $M_2^{0}(A,B) = (-1)^{d(A)}e^{-wL_0}[(e^{-wL_0} A)\ast(e^{-wL_0}B)]$.
Evaluating the obstruction at third order against arbitrary test state $e$, we will then obtain expressions
\begingroup\allowdisplaybreaks
\begin{subequations}
\begin{align}
\mathcal{O} &=  -\omega_S\left[ X^2 e, m_2\left[\frac{b_0}{L_0}\overline{P}_0 m_2(\Psi_1,\Psi_1),\Psi_1\right]\right] +\nonumber\\
&\hspace{3cm}-  \omega_S\left[X^2 e, m_2\left[\Psi_1,\frac{b_0}{L_0}\overline{P}_0 m_2(\Psi_1,\Psi_1)\right] \right]+\mathcal{O}_3\label{eq:introX2}\\
&=  \frac{1}{2}\omega_S \left[b_2(\Psi_1,Xe),\frac{b_0}{L_0} \overline{P}_0 b_2(\Psi_1,X\Psi_1) \right]+\frac{1}{6}\omega_L[b_2(Xe,\xi \Psi_1),b_2(\xi \Psi_1,\Psi_1)]\,,\label{eq:introBer}
\end{align}
\end{subequations}
\endgroup
in the case without stubs (see subsection \ref{subsec:ainf}), and analogous expressions 
\begingroup\allowdisplaybreaks
\begin{subequations}
\begin{align}
\mathcal{O}&= -\omega_S\left[X^2  e, M_2^{(0)}\bigg[\frac{b_0}{L_0}\overline{P}_0 M_2^{(0)}(\Psi_1,\Psi_1),\Psi_1\bigg]\right] +\nonumber\\
&\hspace{3.3cm}-  \omega_S\left[X^2 e, M_2^{(0)}\bigg[\Psi_1,\frac{b_0}{L_0}\overline{P}_0 M_2^{(0)}(\Psi_1,\Psi_1)\bigg] \right]+\nonumber\\
&\hspace{6.3cm}+\omega_S\left[X^2 e,M_3^{(0)}(\Psi_1,\Psi_1,\Psi_1)\right]+\mathcal{O}_3\label{eq:introX2stub}\\
&= \frac{1}{2}\omega_S\left[B_2^{(0)}(\Psi_1,Xe),\frac{b_0}{L_0} \overline{P}_0 B_2^{(0)}(\Psi_1,X\Psi_1)\right]+\nonumber\\
&\hspace{+1.3cm}+\frac{1}{6}\left\{\omega_L\left[B_2^{(0)}(Xe,\xi \Psi_1),B_2^{(0)}(\xi \Psi_1,\Psi_1)\right]+\omega_S\left[Xe,B_3^{(0)}(X\Psi_1,\Psi_1,\Psi_1)\right]\right\}\,,\label{eq:introBerstub}
\end{align}
\end{subequations}
\endgroup
in the case with stubs (see subsection \ref{subsec:Ainfstubs}). Here $\mathcal{O}_3$ cancels the anomalous terms arising in the evaluation of star products due to the non-primary nature of the state $X^2 e$ (where $X$ is the PCO) and $b_2$ denotes the degree-graded commutator based on $m_2$ (analogously for the products $B_2^{(0)}$ and $M_2^{(0)}$ deformed by adding stubs). Finally, $M_3^{(0)}$ is the bosonic 3-product arising in the presence of stubs and $B_3^{(0)}$ is its degree-graded symmetrization. We will refer to \eqref{eq:introX2} and \eqref{eq:introX2stub} as the $X^2$-form of the obstruction (due to the appearance of $X^2 e$) and to \eqref{eq:introBer} and \eqref{eq:introBerstub} as the Berkovits-like form (due to its similarity to the quartic vertex of classical effective action derived in Berkovits OSFT -- see \cite{Maccaferri:2018vwo,Maccaferri:2019ogq}). Evaluating these expressions for $\mathcal{O}$ (that is, deriving explicit expressions for $\mathcal{O}$ in terms of $\mathbb{V}_{1/2}$ and the test state, with the ghost structure stripped away) using various methods will be the subject of section \ref{sec:eval}.

\subsection{Marginal deformations in the $A_\infty$ OSFT}
\label{subsec:margDefAinf}

The purpose of this subsection will be to set the scene for the detailed analysis of the obstruction to exact marginality which arises at third order in the deformation parameter. In subsection \ref{subsubsec:prel} we will discuss the range of validity of the projector condition \eqref{eq:introProjCon}. In particular, we will use the results of \cite{Banks:1987cy,Melnikov:2017yvz} to argue that it holds for all $h=1/2$ NS states whenever the initial open superstring background conserves at least two spacetime supercharges with the same chirality in two non-compact dimensions. We will then show in subsection \ref{subsub:MargDef23} that the projector condition \eqref{eq:introProjCon} guarantees vanishing of the obstruction at second order and we will derive an expression for the third-order obstruction, which will be used as a starting point for the manipulations in subsection \ref{subsec:ainf}.

\subsubsection{Projector condition}
\label{subsubsec:prel}

Let us denote by $P_0$ the projector on $\mathrm{ker}\,L_0$. Defining $\overline{P}_0\equiv 1-P_0$ we have $P_0P_0 =P_0$, $\overline{P}_0\overline{P}_0=\overline{P}_0$, $P_0\overline{P}_0 = \overline{P}_0 P_0=0$ together with
\begin{align}
Q\frac{b_0}{L_0}\overline{P}_0 + \frac{b_0}{L_0}\overline{P}_0Q = \overline{P}_0\,.
\end{align} 
Let us now consider any two NS states\footnote{We will assume that the matter part of the NS sector only contains states with non-negative conformal weight and that matter vacuum is the unique state with $h=0$. This is true for superstring compactifications at zero momentum. The matter part will be allowed to carry Chan-Paton factors.} $V=c\mathbb{V}_{1/2}e^{-\phi}$, $W=c\mathbb{W}_{1/2}e^{-\phi}$, where $\mathbb{V}_\frac{1}{2}$, $\mathbb{W}_\frac{1}{2}$ are $h=1/2$ Grassmann-odd matter primaries. Note that this always needs to be the case if $V$ is to be identified with the leading order term $\Psi_1$ in the open superstring field theory marginal deformation, because the string field $\Psi$ needs to be Grassmann-odd. This is automatic in the case of the $\text{GSO}(+)$ projection (which we will focus on in the following), while in the case of the $\text{GSO}(-)$ projection this would necessitate inclusion of internal Chan-Paton factors. Let us also define $\mathbb{V}_1 = G_{-\frac{1}{2}}\mathbb{V}_\frac{1}{2}$, $\mathbb{W}_1 = G_{-\frac{1}{2}}\mathbb{W}_\frac{1}{2}$ and denote by $\{\mathbb{V}_\frac{1}{2}\mathbb{W}_\frac{1}{2}\}_n$ coefficient of $(2z)^{-n}$ in the OPE of $\mathbb{V}_\frac{1}{2}(+z)$ and $\mathbb{W}_\frac{1}{2}(-z)$ (that is, $\{\mathbb{V}_\frac{1}{2}\mathbb{W}_\frac{1}{2}\}_1$ is proportional to the identity). Here $G_r$ are the Laurent modes of the $\mathcal{N}=1$ worldsheet matter supercurrent $G(z)$. Let us now consider the following two assumptions:
\begin{enumerate}
    \item $\{\mathbb{V}_\frac{1}{2}\mathbb{W}_\frac{1}{2}\}_0$ has pole of order at most 1 in the OPE with $G(z)$, that is 
    \begin{align}
     G_{+\frac{1}{2}}\{\mathbb{V}_\frac{1}{2}\mathbb{W}_\frac{1}{2}\}_0=0\,,   
    \end{align}
    \item the OPE of $\mathbb{V}_1$ with $\mathbb{W}_\frac{1}{2}$, and, $\mathbb{V}_\frac{1}{2}$ with $\mathbb{W}_1$ do not have poles of integral order, that is
    \begin{align}
        \{\mathbb{V}_1\mathbb{W}_\frac{1}{2}\}_1=\{\mathbb{V}_\frac{1}{2}\mathbb{W}_1\}_1=0\,.
    \end{align}
\end{enumerate}
First, we note that using the formula (6.206) of \cite{DiFrancesco:1997nk} (the generalized Wick theorem) we have $G_{+1/2}\{\mathbb{V}_{1/2}\mathbb{W}_{1/2}\}_0=\{\mathbb{V}_1\mathbb{W}_{1/2}\}_1$, so that it follows that Assumption 2 implies Assumption 1. Also, if Assumption 1 holds for \emph{any} two $h=1/2$ NS states, then Assumption 2 is implied.\footnote{If Assumption 2 was to fail, there would need to be a $h=1/2$ NS state $\mathbb{Y}_{1/2}\equiv \{\mathbb{V}_1\mathbb{W}_{1/2}\}_1\neq 0$. 
That is
 $\langle   (G_{-1/2}\mathbb{V}_{1/2})(z_1)\mathbb{W}_{1/2}(z_2)(\mathbb{Y}_{1/2})^\dagger(z_3)\rangle\neq 0$,
where $(\mathbb{Y}_{1/2})^\dagger$ is the conjugate of $\mathbb{Y}_{1/2}$. But this correlator is non-zero if only if the two-point function
   $\langle   (G_{-1/2}\mathbb{V}_{1/2})(z_1)\{\mathbb{W}_{1/2}(\mathbb{Y}_{1/2})^\dagger\}_0 (z_2)\rangle$
is non-zero. 
However, this vanishes if Assumption 1 holds for any two $h=1/2$ states in the NS sector because then we have
\begin{align}
    \big\langle G_{-\frac{1}{2}}\mathbb{V}_\frac{1}{2}\big|\{\mathbb{W}_\frac{1}{2}(\mathbb{Y}_\frac{1}{2})^\dagger
    \}_0\big\rangle = \big\langle \mathbb{V}_\frac{1}{2}\big|G_{+\frac{1}{2}}\{\mathbb{W}_\frac{1}{2}(\mathbb{Y}_\frac{1}{2})^\dagger
    \}_0\big\rangle=0\,.
\end{align}
It therefore follows that if we adopt Assumption 1 for any two states in the theory, then Assumption 2 follows, as claimed.}
Given the two assumptions, we will now show that we have properties
\begin{subequations}
\label{eq:Xvan}
\begin{align}
P_0 m_2 (XV,W)+P_0m_2(V,XW)&=0\,,\label{eq:Xvan1}\\
P_0X m_2(V,W)&=0\,,\label{eq:Xvan2}
\end{align}
\end{subequations}
which together imply the \emph{projector condition}
\begin{align}
P_0 M_2 (V,W)=\frac{1}{3}[P_0 Xm_2(V,W)+P_0m_2(XV,W)+P_0 m_2(V,XW)]=0\,.
\label{eq:ProjCon}
\end{align}
To see this, note that using formula (3.9) of \cite{Schnabl:2002gg}, we obtain
\begin{subequations}
\label{eq:expRef1}
\begin{align}
P_0 m_2(XV,W) &= + \eta c\{\mathbb{V}_\frac{1}{2}\mathbb{W}_\frac{1}{2}\}_1 -c\p c \{\mathbb{V}_1\mathbb{W}_\frac{1}{2}\}_1 e^{-\phi}\,,\\
P_0 m_2(V,XW) &= -\eta c\{\mathbb{V}_\frac{1}{2}\mathbb{W}_\frac{1}{2}\}_1 -c\p c \{\mathbb{V}_\frac{1}{2}\mathbb{W}_1\}_1 e^{-\phi}\,,
\end{align}
\end{subequations}
where used that $XV=c\mathbb{V}_1-e^\phi \eta \mathbb{V}_{1/2}$ (and similarly for $XW$). This establishes \eqref{eq:Xvan1} provided that Assumption 2 holds. We also have
\begin{align}
P_0 X m_2(V,W)=P_0 Q\xi m_2(V,W) = Q\xi  c\p c \{\mathbb{V}_\frac{1}{2}\mathbb{W}_\frac{1}{2}\}_0  e^{-2\phi}\,,
\end{align}
where the only contribution comes from the supercurrent term, that is
\begin{align}
P_0 X m_2(V,W)=\oint\frac{dz}{2\pi i} \eta e^\phi G(z) \xi c\p c \{\mathbb{V}_\frac{1}{2}\mathbb{W}_\frac{1}{2}\}_0  e^{-2\phi}= -c\p c\, G_{+\frac{1}{2}}\{\mathbb{V}_\frac{1}{2}\mathbb{W}_\frac{1}{2}\}_0 e^{-\phi}\,,
\label{eq:expRef2}
\end{align}
which, however, vanishes, as long as Assumption 1 holds (i.e.\ that the OPE of $G$ with $\{\mathbb{V}_\frac{1}{2}\mathbb{W}_\frac{1}{2}\}_0$ does not contain higher-than-simple poles). This establishes \eqref{eq:Xvan2} and therefore \eqref{eq:ProjCon}. Alternatively, it is possible to prove \eqref{eq:Xvan} by showing that the expressions on the l.h.s.\ evaluate to zero against an arbitrary test state $e$: since we are working at zero momentum, we can conclude that the states $P_0 m_2(XV,W)$, $P_0 m_2(V,XW)$, $P_0 Xm_2(V,W)$ (which all have ghost number $+2$, picture number $-1$ and conformal weight $0$) can each be expanded as $\tilde{B} c\eta+c\p c \tilde{\mathbb{V}}_{1/2}e^{-\phi}$, where $\tilde{B}$ is some number and $\tilde{\mathbb{V}}_{1/2}$ is some $h=1/2$ zero-momentum matter operator (as manifested by the above-derived expressions).
That is, the expressions on the l.h.s.\ of \eqref{eq:Xvan} vanish if and only if they vanish when evaluated in the BPZ product against the dual basis of test states $e_\text{g} = c\tilde{\mathbb{V}}_{1/2}e^{-\phi}$ (gluon-like vertex) or $e_\text{NL} = c\p c\p \xi e^{-2\phi}$ (Nakanishi-Lautrup vertex) at ghost number $+1$, picture number $-1$ and $h=0$. We would therefore need to show that
\begin{subequations}
\begin{align}
    \omega_S\big[e,P_0 m_2(XV,W)+P_0 m_2(V,XW)\big]&=0\,,\\
    \omega_S\big[e,P_0 X m_2(V,W)\big]&=0\,,
\end{align}
\end{subequations}
for both $e=e_\text{g}$ and $e=e_\text{NS}$. Furthermore, noting that $P_0 M_2 (V,W)$ is actually proportional to $c\p c\tilde{\mathbb{V}}_{1/2} e^{-\phi}$ only (see \eqref{eq:expRef1} and \eqref{eq:expRef2}), we can conclude that
\begin{align}
    P_0 M_2 (V,W)=0\quad \Longleftrightarrow\quad \omega_S(e_\text{g},P_0 M_2(V,W))=0
    \label{eq:StrongWeakEquiv}
\end{align}
for any two states $V=c\mathbb{V}_{1/2}e^{-\phi}$, $W=c\mathbb{W}_{1/2}e^{-\phi}$ which are present in the theory. 

Finally, let us briefly discuss additional constraints imposed on the boundary (i.e.\ chiral) worldsheet theory in the cases when our background conserves some number of spacetime supercharges. For compactifications\footnote{Here, the notion of ``compactification'' is taken to include also the brane configuration. That is, we consider spacetime supersymmetries of the theory living on the component of the worldvolume common to all branes constituting our configuration.} down to four spacetime dimensions, it was argued long ago \cite{Banks:1987cy} that requiring $N=1$ spacetime supersymmetry necessitates that the local RNS $\mathcal{N}=1$ worldsheet superconformal symmetry enhances to a global $\mathcal{N}=2$ superconformal symmetry. For compactifications to dimensions higher than four, it automatically follows that spacetime supersymmetry implies extended worldsheet superconformal symmetry, as one can always dimensionally reduce back to four dimensions. Results for compactifications to dimensions lower than four, which appeared only recently \cite{Melnikov:2017yvz} (for the heterotic worldsheet), seem to suggest that the boundary worldsheet theory has a global $\mathcal{N}=2$ superconformal symmetry as long as the background conserves at least two spacetime supercharges with the same chirality in two non-compact dimensions (i.e.\ $N=(2,0)$ supersymmetry in 2d -- we are going to discuss a concrete example of this minimal setting in subsection \ref{subsec:spiked}). Furthermore, recalling the unitarity bound $h\geqslant |q|/2$ for two-dimensional $\mathcal{N}=2$ superconformal theories (where $q$ denotes the charge under the $U(1)$ $R$-current $J$) and noting that the GSO projection is by the construction of \cite{Banks:1987cy,Melnikov:2017yvz} implemented by projecting onto states with $q\in 2\mathbb{Z}+1$, we conclude that the matter primaries $\mathbb{V}_{1/2}$ with $h=1/2$ can be all chosen to carry charges either $q=+1$ or $q=-1$ under $J$. 
Matter primaries $\mathbb{V}_{1/2}^\pm$ with $(h,q)=(1/2,\pm 1)$ belong to the (anti-)chiral ring of the theory and they satisfy
\begin{subequations}
\begin{align}
    G^\pm(z) \mathbb{V}_{\frac{1}{2}}^\mp(0) &=\frac{1}{z}\mathbb{V}_1^\mp(0)+\mathrm{reg.}\,,\\
    G^\pm(z) \mathbb{V}_{\frac{1}{2}}^\pm(0) &= \mathrm{reg.}\,,
\end{align}
\end{subequations}
where $\mathbb{V}_1^\pm$ are $(h,q)=(1,0)$ matter fields. We also have
\begin{subequations}
\begin{align}
    G^\pm(z) \mathbb{V}_{1}^\pm(0) &=\frac{1}{z^2}\mathbb{V}_{\frac{1}{2}}^\pm(0)+\frac{1}{z}\p\mathbb{V}_{\frac{1}{2}}^\pm(0)+\mathrm{reg.}\,,\\
    G^\pm(z) \mathbb{V}_{1}^\mp(0) &= \mathrm{reg.}
\end{align}
\end{subequations}
Using these properties, we will now show that Assumptions 1 and 2 hold for any two $h=1/2$ fields in the matter sector. We first note that for any states $\mathbb{V}_{1/2}^\pm$ and $\mathbb{W}_{1/2}^\pm$ in the (anti-)chiral ring, we have (using the generalized Wick theorem)
\begin{subequations}
\label{eq:N2step1}
\begin{align}
    G^+_{+\frac{1}{2}}\{\mathbb{V}_\frac{1}{2}^+\mathbb{W}_\frac{1}{2}^\pm\}_0 &=0\,,\\
    G^-_{+\frac{1}{2}}\{\mathbb{V}_\frac{1}{2}^-\mathbb{W}_\frac{1}{2}^\pm\}_0 &=0\,.
\end{align}
\end{subequations}
Using similar ideas to those which we have employed above when discussing the relation between Assumptions 1 and 2, we can show that it follows from \eqref{eq:N2step1} that\footnote{For instance, if $\mathbb{Y}_{1/2}^+\equiv \{\mathbb{V}_1^-\mathbb{W}_{1/2}^+\}_1\neq 0$, then $\langle (G^+_{-1/2}\mathbb{V}_{1/2}^-)(z_1)\mathbb{W}_{1/2}^+(z_2)(\mathbb{Y}_{1/2}^+)^\dagger(z_3)\rangle\neq 0$ which would mean that $G^+_{+1/2}\{\mathbb{W}_{1/2}^+(\mathbb{Y}_{1/2})^\dagger\}_0\neq 0$, contradicting \eqref{eq:N2step1}.} 
\begin{align}
    \{\mathbb{V}^\mp_1\mathbb{W}^\pm_\frac{1}{2}\}_1 &= 0\label{eq:N2step2}
\end{align}
But then, \eqref{eq:N2step2} and the generalized Wick theorem give that
\begin{subequations}
\label{eq:N2step3}
\begin{align}
    G^+_{+\frac{1}{2}}\{\mathbb{V}_\frac{1}{2}^-\mathbb{W}_\frac{1}{2}^+\}_0 &=0\,,\\
    G^-_{+\frac{1}{2}}\{\mathbb{V}_\frac{1}{2}^+\mathbb{W}_\frac{1}{2}^-\}_0 &=0\,.
\end{align}
\end{subequations}
Finally, \eqref{eq:N2step3} then implies
\begin{align}
    \{\mathbb{V}_1^\pm\mathbb{W}_\frac{1}{2}^\pm\}_1 &=0.
\end{align}
which in turn gives that
\begin{subequations}
\begin{align}
    G^+_{+\frac{1}{2}}\{\mathbb{V}_\frac{1}{2}^-\mathbb{W}_\frac{1}{2}^-\}_0 &=0\,,\\
    G^-_{+\frac{1}{2}}\{\mathbb{V}_\frac{1}{2}^+\mathbb{W}_\frac{1}{2}^+\}_0 &=0\,.
\end{align}
\end{subequations}
We have therefore shown that Assumptions 1 and 2 (and therefore the projector conditions \eqref{eq:Xvan} and \eqref{eq:ProjCon}) hold for all states in a theory with $\mathcal{N}=2$ global worldsheet superconformal symmetry where all $h=1/2$ states can be chosen to carry $R$-charge $q=\pm 1$. As per the discussion above, this should always be the case when the background preserves at least $N=(2,0)$ supersymmetry in two non-compact dimensions.

\subsubsection{Marginal deformations in $A_\infty$ OSFT at second and third order}
\label{subsub:MargDef23}

Writing down the $A_\infty$ OSFT action up to quartic order, we obtain
\begin{align}
S_{A_\infty}[\Psi] = \frac{1}{2}\omega_S(\Psi,Q\Psi)+\frac{1}{3}\omega_S(\Psi,M_2(\Psi,\Psi))+\frac{1}{4}\omega_S(\Psi,M_3(\Psi,\Psi,\Psi))+\ldots\,,
\end{align}
so that varying this action with respect to $\Psi$, we get the equations of motion
\begin{align}
Q\Psi + M_2(\Psi,\Psi)+M_3(\Psi,\Psi,\Psi)+\ldots=0\,.
\label{eq:eomAI}
\end{align}
Note that $\Psi$ carries picture number $-1$ and ghost number $+1$. 
We want to construct a continuous family of classical solutions $\Psi(\lambda)$, such that $\Psi(0)=0$ and such that the leading term in $\lambda$ is given by a $Q$-closed state $\Psi_1= c \mathbb{V}_\frac{1}{2}e^{-\phi}$ where $\mathbb{V}_\frac{1}{2}$ is a zero-momentum Grassmann-odd $h=1/2$ matter primary. 
Writing the classical solution $\Psi(\lambda)$ as a perturbative expansion
\begin{align}
\Psi(\lambda)=\sum_{k=1}^\infty\lambda^k \Psi_k =\lambda\Psi_1+\lambda^2\Psi_2+\lambda^3 \Psi_3+\ldots\,,
\label{eq:margAn}
\end{align}
and substituting \eqref{eq:margAn} into \eqref{eq:eomAI}, we obtain, order by order in $\lambda$,
\begingroup
\allowdisplaybreaks
\begin{subequations}
\begin{align}
0&= Q\Psi_1\,,\\
0&= Q\Psi_2 +M_2(\Psi_1,\Psi_1)\,,\label{eq:secOrd}\\
0&= Q\Psi_3 +M_2(\Psi_1,\Psi_2)+M_2(\Psi_2,\Psi_1)+M_3(\Psi_1,\Psi_1,\Psi_1)\,.\label{eq:thirdOrd}\\
&\hspace{0.2cm}\vdots\nonumber
\end{align}
\end{subequations}
\endgroup
At second order, we have to satisfy the equation \eqref{eq:secOrd}. This is clearly integrable because 
\begin{align}
QM_2(\Psi_1,\Psi_1)=-M_2(Q\Psi_1,\Psi_1)-M_2(\Psi_1,Q\Psi_1)=0\,.
\end{align}
A putative solution in Siegel gauge reads
\begin{align}
    \Psi_2 = -\frac{b_0}{L_0}\overline{P}_0 M_2(\Psi_1,\Psi_1)+\psi_2\,,
    \label{eq:solSecOrd}
\end{align}
where $\psi_2$ is a ghost number $+1$, picture number $-1$ string field with $\eta_0 \psi_2=0$.
However, in order for \eqref{eq:solSecOrd} to actually solve \eqref{eq:secOrd}, we need
\begin{align}
Q\Psi_2 
= \left(\frac{b_0}{L_0}Q -\overline{P}_0\right)\overline{P}_0 M_2(\Psi_1,\Psi_1)+Q\psi_2 =-\overline{P}_0 M_2(\Psi_1,\Psi_1)+Q\psi_2
\end{align}
to be equal to $ -M_2(\Psi_1,\Psi_1)$. That is, we need the second order obstruction
\begin{align}
O_2=P_0 M_2(\Psi_1,\Psi_1)+Q\psi_2\,,
\label{eq:SecObst}
\end{align}
to vanish. Put in different words, in order for the solution \eqref{eq:solSecOrd} to be consistent, we need $P_0 M_2(\Psi_1,\Psi_1)$ to vanish up to $Q$-exact terms. But since $P_0M_2(\Psi_1,\Psi_1)$ is a zero-momentum state in $\text{ker}\, L_0$ at ghost number $+2$ and picture number $-1$, it has to be equal to a linear combination of $c\p c \tilde{\mathbb{V}}_{1/2}e^{-\phi}$ and $\eta c = Q(\frac{1}{2}c\p c\p \xi e^{-2\phi})$, where $\tilde{\mathbb{V}}_{1/2}$ is an arbitrary NS state with $h=1/2$. Consistency therefore requires that $P_0 M_2(\Psi_1,\Psi_1)$ does not contain the state $c\p c \tilde{\mathbb{V}}_{1/2}e^{-\phi}$, so that it is necessary and sufficient to check that\footnote{Note that we can actually always set $\psi_2 = c\hat{\mathbb{V}}_{\frac{1}{2}}e^{-\phi}$
where $\hat{\mathbb{V}}_{1/2}$ is some $h=1/2$ state in the NS sector (so that $Q\psi_2=0$) because as per our discussion in subsection \ref{subsubsec:prel}, the state $P_0 M_2(\Psi_1,\Psi_1)$ can never contain $\eta c$.}
\begin{align}
    \omega_S(e_\text{g}, P_0 M_2(\Psi_1,\Psi_1))=0\,,\label{eq:testProjCond}
\end{align}
where $e_\text{g}$ is a zero-momentum test state in $\text{ker}\, L_0$ at ghost number $+1$ and picture number $-1$ of the form $e_\text{g}=c\tilde{\mathbb{V}}_{1/2}e^{-\phi}$. Thus, recalling \eqref{eq:StrongWeakEquiv}, we can conclude that the necessary and sufficient condition for the vanishing of $O_2$ is actually $P_0 M_2(\Psi_1,\Psi_1)=0$ (which is a special case of the projector condition \eqref{eq:ProjCon} with $V=W=\Psi_1$). 

Assuming from now on that $P_0M_2(\Psi_1,\Psi_1)=0$ and proceeding to the third order, we need to solve the equation \eqref{eq:thirdOrd} for $\Psi_3$. We have integrability condition
\begin{align}
QM_2(\Psi_1,\Psi_2)+QM_2(\Psi_2,\Psi_1)+QM_3(\Psi_1,\Psi_1,\Psi_1)=0\,,
\end{align}
which is satisfied provided that $\Psi_2$ solves the equation of motion at second order. Indeed, we have
\begingroup
\allowdisplaybreaks
\begin{subequations}
\label{eq:Ainf}
\begin{align}
   &QM_2(\Psi_1,\Psi_2)+QM_2(\Psi_2,\Psi_1)+QM_3(\Psi_1,\Psi_1,\Psi_1)\\
  &\hspace{+1.5cm}=  -M_2(Q\Psi_1,\Psi_2)-M_2(\Psi_1,Q\Psi_2)-M_2(Q\Psi_2,\Psi_1)-M_2(\Psi_2,Q\Psi_1)+\nonumber\\
  &\hspace{10cm}+QM_3(\Psi_1,\Psi_1,\Psi_1)\\
    &\hspace{+1.5cm}=  M_2(\Psi_1,M_2(\Psi_1,\Psi_1))+M_2(M_2(\Psi_1,\Psi_1),\Psi_1)+QM_3(\Psi_1,\Psi_1,\Psi_1)\\
    &\hspace{+1.5cm}=0\,,
\end{align}
\end{subequations}
\endgroup
where in the second equality we have assumed that $\Psi_1$ is a consistent solution of \eqref{eq:secOrd} and the third equality follows by one of the $A_\infty$ relations
\begin{align}
[Q,M_3]+\frac{1}{2}[M_2,M_2]=0\,.
\end{align}
A putative solution of \eqref{eq:thirdOrd} can be written as
\begin{align}
    \Psi_3 =-\frac{b_0}{L_0}\overline{P}_0[M_2(\Psi_2,\Psi_1) + M_2(\Psi_1,\Psi_2) +M_3(\Psi_1,\Psi_1,\Psi_1)]+\psi_3\,,
    \label{eq:solThirdOrd}
\end{align}
where $\psi_3$ is a ghost number $+1$, picture number $-1$ string field with $\eta_0\psi_3=0$.
Again, this only solves the equation \eqref{eq:thirdOrd} provided that
\begin{align}
    Q\Psi_3 &= \left(\frac{b_0}{L_0}Q-\overline{P}_0\right)\overline{P}_0[M_2(\Psi_2,\Psi_1) + M_2(\Psi_1,\Psi_2) +M_3(\Psi_1,\Psi_1,\Psi_1)]+Q\psi_3
\end{align}
is equal to $ -M_2(\Psi_2,\Psi_1) - M_2(\Psi_1,\Psi_2) -M_3(\Psi_1,\Psi_1,\Psi_1)$, that is, provided that the third-order obstruction 
\begin{align}
    O_3 &=P_0\bigg\{M_2\left[\frac{b_0}{L_0}\overline{P}_0M_2(\Psi_1,\Psi_1),\Psi_1\right] +\nonumber\\
    &\hspace{4cm}+ M_2\left[\Psi_1,\frac{b_0}{L_0}\overline{P}_0 M_2(\Psi_1,\Psi_1)\right]-M_3(\Psi_1,\Psi_1,\Psi_1)\bigg\}-Q\psi_3
    \label{eq:obstrDef}
\end{align}
vanishes.\footnote{Note that in general $\psi_2$ also enters $O_3$ and, in some cases, it may be possible that it can be fine-tuned so as to make $O_3$ vanish. However, in most cases of interest, the projector condition \eqref{eq:ProjCon} will give $P_0 M_2 (\psi_2,\Psi_1)=P_0 M_2(\Psi_1,\psi_2)=0$ so that $\psi_2$ does not contribute to $O_3$.} 
Again, we therefore need to ensure that the projector part $O_3^\text{proj}=P_0\{\ldots\}$ of $O_3$ vanishes up to $Q$-exact terms. Since $O_3^\text{proj}\in\mathrm{ker}\,L_0$ at ghost number $+2$ and picture number $-1$, the only states which it can be proportional to (and which are not $Q$-exact) are of the form $c\p c\tilde{\mathbb{V}}_{1/2}e^{-\phi}$. The necessary and sufficient condition for the obstruction to vanish is therefore
\begin{align}
 \label{eq:curlyO}
    \mathcal{O}\equiv  -\omega_S(e_\text{g},O_3^\text{proj})=\omega_L(e_\text{g},\xi O_3^\text{proj})=0\,.
\end{align}
From now on, let us drop the lower index on 
$e_\text{g}$, as for the rest of the paper we will work only with the test state 
$e\equiv c\tilde{\mathbb{V}}_{1/2}e^{-\phi}$. Also note that upon identifying $e=\Psi_1$, the expression $\omega_L(e,\xi O_3^\text{proj})$ becomes proportional to the quartic part of the classical effective action of \cite{Maccaferri:2018vwo,Maccaferri:2019ogq}. More precisely, we obtain
\begin{align}
\mathcal{O} = -4S_\text{eff}^{(4)}\,.
\label{eq:ObstActRel}
\end{align}
The necessity of existence of such a relation was already proven in \cite{Maccaferri:2019ogq} where it is also noted that this relation implies that all marginal deformations which are unobstructed at third order automatically give rise to flat directions of the quartic effective action. In fact, we shall see in section \ref{sec:eval} that under certain assumptions, the converse appears to be true as well.

\subsection{Simplifying the third-order obstruction}
\label{subsec:ainf}

We will now expose algebraic manipulations whose aim will be to simplify $\mathcal{O}$ into a computable form. Although we have checked that it is in principle possible to proceed by generalizing the calculations of \cite{Maccaferri:2019ogq} and keep all intermediate expressions manifestly in the small Hilbert space, we found it much more economic to perform the computations in the large Hilbert space. Bearing in mind that our main goal is to provide a practical expression for the obstruction, we will therefore adopt a pragmatic approach and expose here a relatively short path the main results which leads through the large Hilbert space. 
In subsection \ref{subsub:X2body}, we will derive the $X^2$-form \eqref{eq:introX2} for $\mathcal{O}$, while in subsection \ref{subsub:Berkbody}, we will derive the Berkovits-like form \eqref{eq:introBer}.

\subsubsection{$X^2$ form}
\label{subsub:X2body}

Proceeding along the lines of \cite{Mattiello:2019gxc}, we will first show that $\mathcal{O}$ can be rewritten as 
\begingroup
\allowdisplaybreaks
\begin{align}
\mathcal{O} &=  -\omega_S\left[ X^2 e, m_2\left[\frac{b_0}{L_0}\overline{P}_0 m_2(\Psi_1,\Psi_1),\Psi_1\right]\right] +\nonumber\\
&\hspace{3cm}-  \omega_S\left[X^2 e, m_2\left[\Psi_1,\frac{b_0}{L_0}\overline{P}_0 m_2(\Psi_1,\Psi_1)\right] \right]+\mathcal{O}_3\,,
\label{eq:OX2}
\end{align}
\endgroup
where $\mathcal{O}_3$ (to be defined below) consists only of terms which are localized on the boundary of the worldsheet moduli space and which are zero up to contributions which cancel the anomalous terms which appear due to the non-primary nature of the state $X^2 e$. 

\subsubsection*{Quartic vertex}

Focusing on the quartic vertex term $\mathcal{O}^{(4)}\equiv -\omega_L[ e,\xi M_3(\Psi_1,\Psi_1,\Psi_1)]$ first, we obtain
\begin{align}
   \mathcal{O}^{(4)}
     &=-\frac{1}{2}\bigg\{\omega_L \left[ e, \xi M_2(\overline{M}_2(\Psi_1,\Psi_1),\Psi_1\right]-\xi\overline{M}_2\left[M_2(\Psi_1,\Psi_1),\Psi_1)\right]+\nonumber\\
    &\hspace{+2.5cm}+\omega_L\left[e,\xi M_2[\Psi_1,\overline{M}_2(\Psi_1,\Psi_1)\right]-\xi\overline{M}_2\left[\Psi_1,M_2(\Psi_1,\Psi_1)]\right]+\nonumber\\
    &\hspace{+8cm}+\omega_L\left[e,X\overline{M}_3(\Psi_1,\Psi_1,\Psi_1)\right]\bigg\}\,,\label{eq:Bfin}
\end{align}
where we have used that $Qe=0$.

\subsubsection*{Cubic vertex}

The cubic vertex terms  
\begingroup\allowdisplaybreaks
\begin{subequations}
\begin{align}
\mathcal{O}_1^{(3)} &\equiv \omega_L\left[ e,\xi M_2\left[\frac{b_0}{L_0}\overline{P}_0 M_2(\Psi_1,\Psi_1),\Psi_1\right] \right]\,,\\
  \mathcal{O}^{(3)}_2&\equiv \omega_L\left[ e,\xi M_2\left[\Psi_1,\frac{b_0}{L_0}\overline{P}_0 M_2(\Psi_1,\Psi_1)\right]\right]\,,
\end{align}
\end{subequations}
\endgroup
yield
\begingroup
\allowdisplaybreaks
\begin{subequations}
\begin{align}
    \mathcal{O}_1^{(3)} 
%
   &=\frac{1}{2}\Bigg\{\omega_L\left[ e, X\overline{M}_2\left[\frac{b_0}{L_0}\overline{P}_0 M_2(\Psi_1,\Psi_1),\Psi_1\right]\right]-\omega_L\left[e,\xi\overline{M}_2\left( M_2(\Psi_1,\Psi_1),\Psi_1\right)\right]+\nonumber\\
    &\hspace{0.3cm}+\omega_L\left[e,X M_2\left[\frac{b_0}{L_0}\overline{P}_0  \overline{M}_2(\Psi_1,\Psi_1),\Psi_1\right] \right] +\omega_L\left[e,\xi M_2\left( \overline{M}_2(\Psi_1,\Psi_1),\Psi_1\right) \right] +\nonumber\\
   &\hspace{6.8cm}-\omega_L\left[e,\xi M_2\left({P}_0  \overline{M}_2(\Psi_1,\Psi_1),\Psi_1\right) \right]
 \Bigg\}\label{eq:A1fin}\,,\\
    \mathcal{O}^{(3)}_2
    &=\frac{1}{2}\Bigg\{\omega_L\left[ e, X\overline{M}_2\left[\Psi_1,\frac{b_0}{L_0}\overline{P}_0 M_2(\Psi_1,\Psi_1)\right]\right]-\omega_L\left[e,\xi\overline{M}_2\left(\Psi_1, M_2(\Psi_1,\Psi_1)\right)\right]+\nonumber\\
    &\hspace{0.3cm}+\omega_L\left[e,X M_2\left[\Psi_1,\frac{b_0}{L_0}\overline{P}_0  \overline{M}_2(\Psi_1,\Psi_1)\right] \right] +\omega_L\left[e,\xi M_2\left(\Psi_1, \overline{M}_2(\Psi_1,\Psi_1)\right) \right] +\nonumber\\
   &\hspace{6.8cm}-\omega_L\left[e,\xi M_2\left(\Psi_1,{P}_0  \overline{M}_2(\Psi_1,\Psi_1)\right) \right]
 \Bigg\}
   \,.\label{eq:A2fin}
\end{align}
\end{subequations}
\endgroup
Here we note that the second and fourth terms in \eqref{eq:A1fin} and \eqref{eq:A2fin} cancel with the first four terms in \eqref{eq:Bfin}. 
Also, note that we have
\begin{align}
\omega_L\left[ e, X\overline{M}_2\left[\frac{b_0}{L_0}\overline{P}_0M_2(\Psi_1,\Psi_1),\Psi_1\right] \right] = \omega_L\left[ e, X\xi m_2\left[\frac{b_0}{L_0}\overline{P}_0M_2(\Psi_1,\Psi_1),\Psi_1\right] \right] \,,
\end{align}
because the difference of the second insertions on the l.h.s.\ and the r.h.s.\ would lie in the small Hilbert space.
We then have
\begingroup
\allowdisplaybreaks
\begin{align}
  &\hspace{-0.4cm}  \omega_L\left[ e, X\overline{M}_2\left[\frac{b_0}{L_0}\overline{P}_0M_2(\Psi_1,\Psi_1),\Psi_1\right] \right]=\nonumber\\
     &=\omega_L\left[ e, X^2 \xi m_2\left[\frac{b_0}{L_0}\overline{P}_0 m_2(\Psi_1,\Psi_1),\Psi_1\right] \right]
     +\nonumber\\
     &\hspace{1.5cm}+\omega_L\left[ e, X\xi m_2\left(\overline{M}_2(\Psi_1,\Psi_1),\Psi_1\right) \right]-
     \omega_L\left[ e, X\xi m_2\left(P_0\overline{M}_2(\Psi_1,\Psi_1),\Psi_1\right) \right]
     \,,
\end{align}
\label{eq:flag1}
\endgroup
together with
\begingroup
\allowdisplaybreaks
\begin{align}
      &\hspace{-0.4cm}\omega_L\left[ e,XM_2\left[\frac{b_0}{L_0}\overline{P}_0\overline{M}_2(\Psi_1,\Psi_1),\Psi_1\right] \right]= \nonumber\\
&=\omega_L\left[ e,X^2\xi m_2\left[\frac{b_0}{L_0}\overline{P}_0 m_2(\Psi_1,\Psi_1),\Psi_1\right] \right]+\nonumber\\
&\hspace{1.5cm}-\omega_L\left[e,X\xi \overline{M}_2\left( m_2(\Psi_1,\Psi_1),\Psi_1\right) \right]+\omega_L\left[ e,X\xi \overline{M}_2\left(P_0 m_2(\Psi_1,\Psi_1),\Psi_1\right) \right]
\,.
\end{align}
\endgroup
Altogether we obtain
{\begin{align}
    \mathcal{O}=\mathcal{O}_1^{(3)}+\mathcal{O}_2^{(3)}+\mathcal{O}^{(4)} &= \mathcal{O}_1+\mathcal{O}_2+\mathcal{O}_3 \,,
\end{align}}
where we define
\begingroup
\allowdisplaybreaks
\begin{subequations}
\begin{align}
\mathcal{O}_1 &=  \frac{1}{2}\bigg\{-\omega_L\left[ e, X\overline{M}_3(\Psi_1,\Psi_1,\Psi_1)\right] +\nonumber\\
&\hspace{1cm}+ \omega_L\left[ e, X\xi m_2[\overline{M}_2(\Psi_1,\Psi_1),\Psi_1] \right]-\omega_L\left[ e,X\xi \overline{M}_2[ m_2(\Psi_1,\Psi_1),\Psi_1] \right]\nonumber\\
&\hspace{1cm}+  \omega_L\left[ e, X\xi m_2[\Psi_1,\overline{M}_2(\Psi_1,\Psi_1)]\right]-\omega_L\left[ e,X\xi \overline{M}_2[\Psi_1, m_2(\Psi_1,\Psi_1)] \right]\bigg\}\,,\\[+2mm]
\mathcal{O}_2&= \omega_L\left[ e,X^2\xi m_2\left[\frac{b_0}{L_0}\overline{P}_0 m_2(\Psi_1,\Psi_1),\Psi_1\right]\right] +\nonumber\\
&\hspace{6cm}+  \omega_L\left[ e,X^2\xi m_2\left[\Psi_1,\frac{b_0}{L_0}\overline{P}_0 m_2(\Psi_1,\Psi_1)\right] \right]\,,\\[+2mm]
\mathcal{O}_3&= \frac{1}{2}\bigg\{-\omega_L\left[ e, X\xi m_2[P_0\overline{M}_2(\Psi_1,\Psi_1),\Psi_1]\right]+\omega_L\left[ e,X\xi \overline{M}_2[P_0 m_2(\Psi_1,\Psi_1),\Psi_1]\right]+\nonumber\\
&\hspace{7.3cm}-\omega_L\left[ e,\xi M_2[P_0\overline{M}_2(\Psi_1,\Psi_1),\Psi_1]\right]\nonumber\\[+2.1mm]
&\hspace{1cm}-\omega_L\left[ e, X\xi m_2[\Psi_1,P_0\overline{M}_2(\Psi_1,\Psi_1)]\right]+\omega_L\left[e,X\xi \overline{M}_2[\Psi_1,P_0 m_2(\Psi_1,\Psi_1)]\right]+\nonumber\\
&\hspace{7.3cm}-\omega_L\left[ e,\xi M_2[\Psi_1,P_0\overline{M}_2(\Psi_1,\Psi_1)]\right] \bigg\}\,.
\end{align}
\end{subequations}
\endgroup
First, $\mathcal{O}_1$ clearly vanishes: to see this, we note that $m_3=[\eta ,\overline{M}_3]$, so that
\begin{align}
\omega_L\left[ e, X\overline{M}_3(\Psi_1,\Psi_1,\Psi_1)\right]=\omega_L\left[ e, X\xi m_3(\Psi_1,\Psi_1,\Psi_1)\right]
\end{align}
and then we use $m_3=[m_2,\overline{M}_2]$. Second, $\mathcal{O}_2$ consists of terms containing single propagator. Finally, $\mathcal{O}_3$ contains only terms with $P_0$ and it is therefore completely localised on the boundary of the worldsheet moduli space. Using the cyclic property \eqref{eq:cyc} of $m_2$ and \eqref{eq:BPZpar}, it can be rewritten as
\begin{align}
    \mathcal{O}_3 &= \frac{1}{6}\bigg\{4\omega_L \left[P_0\overline{M}_2(\Psi_1,\Psi_1),  m_2(\Psi_1, X\xi e) +m_2(X\xi e,\Psi_1)\right]+\nonumber\\
    &\hspace{3.0cm}+\omega_L\left[ P_0\overline{M}_2(\Psi_1,\Psi_1), m_2(X\Psi_1,\xi e) +m_2(\xi e,X\Psi_1)\right] +\nonumber\\[+2mm]
    &\hspace{1.2cm}+\omega_L\left[ \xi P_0 m_2(\Psi_1,\Psi_1), m_2(\Psi_1,X\xi e)+m_2(X\xi e,\Psi_1) \right]+\nonumber\\[+2mm]
    &\hspace{3.0cm}+\omega_L\left[ P_0 m_2(\Psi_1,\Psi_1), m_2(\xi \Psi_1,X\xi e)-m_2(X\xi e,\xi \Psi_1) \right]+\nonumber\\
    &\hspace{4.0cm}+\omega_L\left[\xi P_0 Xm_2(\Psi_1,\Psi_1), m_2(\Psi_1,\xi e)+ m_2(\xi e,\Psi_1)\right]
    \bigg\}\,.
    \label{eq:711}
\end{align}
First, note that the last line in \eqref{eq:711} can be dropped even without assuming the condition \eqref{eq:Xvan2} because the matter part of $P_0 Xm_2(\Psi_1,\Psi_1)$ can only be proportional to $G_{+\frac{1}{2}}\{\mathbb{V}_\frac{1}{2}\mathbb{V}_\frac{1}{2}\}_0$, which is a $h=1/2$ state, so that it gives zero when inserted in the symplectic form against
\begin{align}
  P_0  \xi m_2(\Psi_1,\xi e)+P_0\xi m_2(\xi e,\Psi_1)
\end{align}
which, by \eqref{eq:PVW}, is proportional to identity in the matter sector.
The rest of the expression \eqref{eq:711} can be evaluated as well and turns out to give zero up to terms which arise due to anomalous transformation properties of the non-primary state $\xi X e$ (see \cite{Mattiello:2019gxc} for details). Below these will be shown to cancel with anomalous contributions to $\mathcal{O}_2$. 

\subsubsection{Berkovits-like form}
\label{subsub:Berkbody}

We will now show that the $X^2$-form \eqref{eq:OX2} can be recast in the Berkovits-like form
\begin{align}
    \mathcal{O} &=  \frac{1}{2}\omega_S \left[b_2(\Psi_1,Xe),\frac{b_0}{L_0} \overline{P}_0 b_2(\Psi_1,X\Psi_1) \right]+\frac{1}{6}\omega_L[b_2(Xe,\xi \Psi_1),b_2(\xi \Psi_1,\Psi_1)]\,, \label{eq:O2propcon}
\end{align}
where we have defined $b_2(A,B)\equiv m_2(A,B)+(-1)^{d(A)d(B)}m_2(B,A)$. Note that only primary insertions appear in \eqref{eq:O2propcon}. We will show in section \ref{sec:berk} that exactly the same expression is obtained by analyzing the third-order obstruction which arises in the Berkovits open superstring field theory provided that we assume that the deformation is unobstructed at second order. 

\subsubsection*{$\mathcal{O}_3$ terms}

Let us start with analyzing $\mathcal{O}_3$. 
To this end, note that it is possible show that
\begin{align}
-\omega_L\left[ P_0 b_2(\Psi_1,\Psi_1),\xi b_2(X\Psi_1,\xi e )\right]
&=-\omega_L\left[ P_0 b_2 (\Psi_1,\Psi_1),\xi b_2(\xi \Psi_1,X e )\right]\,.
\end{align}
Indeed, we have
\begingroup\allowdisplaybreaks
\begin{subequations}
\begin{align}
 -\omega_L\left[ P_0 b_2(\Psi_1,\Psi_1),\xi b_2(X\Psi_1,\xi e )\right]   
 &=-\omega_L\left[ P_0 \xi b_2(\Psi_1,\Psi_1),  Qb_2(\xi \Psi_1,\xi e )\right] +\nonumber\\
 &\hspace{2cm}+\omega_L\left[ P_0 \xi b_2(\Psi_1,\Psi_1), b_2(\xi \Psi_1,X e )\right] \\
 &=+\omega_L\left[ P_0 X b_2(\Psi_1,\Psi_1),  b_2(\xi \Psi_1,\xi e )\right] +\nonumber\\
 &\hspace{2cm}-\omega_L\left[ P_0  b_2(\Psi_1,\Psi_1), \xi b_2(\xi \Psi_1,X e )\right] \\
 &=-\omega_L\left[ P_0  b_2(\Psi_1,\Psi_1), \xi b_2(\xi \Psi_1,X e )\right] \,,
\end{align}
\end{subequations}
\endgroup
where in order to write down the last equality, we have used \eqref{eq:xixiVW} to note that
\begin{align}
    b_2(\xi \Psi_1,\xi e)=m_2(\xi \Psi_1,\xi e)-m_2(\xi e,\xi \Psi_1)
\end{align}
is proportional to identity in the matter sector so that it gives zero in the symplectic form against $P_0 Xb_2(\Psi_1,\Psi_1)$ which is proportional to a $h=1/2$ state in the matter sector.\footnote{$P_0 Xb_2(\Psi_1,\Psi_1)$ itself vanishes if we assume the projector condition \eqref{eq:Xvan2}.} We therefore end up with
\begin{align}
    \mathcal{O}_3 &= -\frac{1}{12}\bigg\{5\omega_L\left[ P_0 b_2(\Psi_1,\Psi_1),\xi b_2(\Psi_1, X\xi e)\right] +\omega_L\left[ P_0 b_2(\Psi_1,\Psi_1),\xi b_2(\xi \Psi_1, X e)\right]
   + \nonumber\\
    &\hspace{+7.4cm}
-\omega_L\left[ P_0 b_2(\Psi_1,\Psi_1),b_2(\xi \Psi_1, X\xi e)\right]\bigg\}\,.
    \label{eq:711witten}
\end{align}
Although it is straightforward to explicitly evaluate \eqref{eq:711witten}, we do not need to do so at this point, as we will soon show that it is exactly cancelled by a $P_0$ term which we pick up when we move one of the PCOs in $\mathcal{O}_2$. 

\subsubsection*{$\mathcal{O}_2$ terms}

Next, let us analyze $\mathcal{O}_2$. Reabsorbing $\xi$, we obtain
\begin{align}
    \mathcal{O}_2 
    &= \frac{1}{2}\omega_S\left[b_2(\Psi_1,X^2  e) ,\frac{b_0}{L_0}\overline{P}_0 b_2(\Psi_1,\Psi_1)  \right]\,.
    \label{eq:O2pre}
\end{align}
%
In order to avoid the appearance of non-primary fields during the explicit evaluation of \eqref{eq:O2pre} (these would arise due to the $X^2 e$ insertion), let us move one of the two PCOs sitting on $e$ inside the $\overline{P}_0b_2(\Psi_1,\Psi_1)$ part of \eqref{eq:O2pre}. We end up with
\begin{align}
    \mathcal{O}_2 
       &= \frac{1}{2}\omega_S\left[b_2( \Psi_1, X e),\frac{b_0}{L_0}\overline{P}_0 b_2(\Psi_1,X \Psi_1)\right]+\frac{1}{2}\omega_L\left[b_2 (\Psi_1,\xi X e),\overline{P}_0 b_2(\Psi_1,\xi \Psi_1)\right]
       \,,\label{eq:refxxx}
\end{align}
with the $P_0$ part of the second term in \eqref{eq:refxxx} satisfying
\begin{align}
   -\frac{1}{2} \omega_L\left[ b_2(\Psi_1,\xi X e),P_0b_2(\Psi_1,\xi \Psi_1)\right]=    +\frac{1}{2}\omega_L\left[P_0b_2(\Psi_1, \Psi_1),\xi b_2(\Psi_1,\xi X e) \right]\,,
\end{align}
where we have used \eqref{eq:PVW}. 
Note, however, that this (localised) contribution will precisely cancel with $\mathcal{O}_3$: introducing the string field
\begin{align}
 \Xi&\equiv  \xi  b_2(\Psi_1, X\xi e)-\xi  b_2(\xi \Psi_1, X e)+ b_2(\xi \Psi_1, X\xi e)\,,
\end{align}
and using \eqref{eq:711witten}, it can be shown that
\begin{subequations}
\begin{align}
\omega_L\big[P_0b_2(\Psi_1,\Psi_1),\Xi\big]&= -12\left(\mathcal{O}_3 +\frac{1}{2}\omega_L\left[P_0 b_2(\Psi_1, \Psi_1),\xi b_2(\Psi_1,\xi X e) \right]\right)\,.
\end{align}
\end{subequations}
However, it can be also shown that $\eta \Xi =0$, 
which in turn gives that $\omega_L[P_0b_2(\Psi_1,\Psi_1),\Xi]=0$ and therefore 
\begin{align}
\mathcal{O}_3 +\frac{1}{2}\omega_L\left[P_0b_2(\Psi_1, \Psi_1),\xi b_2(\Psi_1,\xi X e) \right]=0\,,
\end{align}
that is, $\mathcal{O}_3$ is completely canceled by the $P_0$ part of the second term in \eqref{eq:refxxx}.
Finally, in order to rid ourselves of the non-primary insertion $\xi X e$ in the identity part of the second term in \eqref{eq:refxxx}, we can use the super-Jacobi identity and the fact that the string field
\begin{align}
b_2\left[\Psi_1,b_2(\xi X e,\xi \Psi_1)\right]-b_2\left[\xi \Psi_1,b_2(\xi X e, \Psi_1)\right]+b_2\left[\xi \Psi_1,b_2( X e,\xi \Psi_1)\right]\,,
\end{align}
lies in the small Hilbert space to show that
\begin{align}
\frac{1}{2}\omega_L\left[b_2(\Psi_1,\xi X e),b_2(\Psi_1,\xi \Psi_1)]\right] =&+ \frac{1}{6}\omega_L\left[b_2( X e,\xi \Psi_1),b_2(\xi \Psi_1,\Psi_1)\right]\,.
\end{align}
Putting our results together, we therefore recover the Berkovits-like expression \eqref{eq:O2propcon} for the obstruction. For the sake of the discussion which is to follow in Section \ref{sec:eval}, we introduce the notation
\begin{subequations}
\begin{align}
    \mathcal{O}^\text{prop} &=  \frac{1}{2}\omega_S\left[b_2( \Psi_1, X e),\frac{b_0}{L_0}\overline{P}_0 b_2(\Psi_1,X \Psi_1)\right]\,,\\
        \mathcal{O}' &=  \frac{1}{6}\omega_L\left[b_2( X e,\xi \Psi_1),b_2(\xi \Psi_1,\Psi_1)\right]\,,
\end{align}
\end{subequations}
so that $\mathcal{O}=\mathcal{O}^\text{prop}+\mathcal{O}'$. We can also check the validity of the intermediate manipulations we have performed so far by comparing \eqref{eq:O2propcon} with the Berkovits-like form of the quartic part of the classical effective action of \cite{Maccaferri:2018vwo,Maccaferri:2019ogq}. Indeed, we again recover the relation \eqref{eq:ObstActRel}. 

\subsection{$A_\infty$ OSFT with stubs}
\label{subsec:Ainfstubs}

Let us now consider $A_\infty$ OSFT with stubs.\footnote{I thank Ashoke Sen for this suggestion.} We will show that apart from deforming the star product $m_2$ into a non-associative product $M_2^{(0)}$, adding stubs introduces additional term into both the $X^2$-form (subsection \ref{subsub:stubX23}) and the Berkovits-like form (subsection \ref{subsub:stubBer3}) as a consequence of the appearance of the bosonic 3-product $M_3^{(0)}$. 
See subsection \ref{subsub:stubPrel} for our conventions for $A_\infty$ OSFT with stubs.

\subsubsection{Preliminaries}
\label{subsub:stubPrel}

Here we largely follow the notation of \cite{Erler:2014eba}. Denoting the picture by a superscript in the round brackets, we define the bosonic products $M_1^{(0)}=Q$,
\begin{align}
M_2^{(0)}(A,B)=(-1)^{d(A)}e^{-w L_0}[(e^{-w L_0}A)\ast(e^{-w L_0}B)]
\end{align}
and higher products $M_3^{(0)},\ldots$ so as to cover the missing regions of the bosonic moduli space. The superstring products are then defined similarly to the case without stubs by suitably distributing PCO charges among the insertions. For instance, the superstring 2-product then reads
\begin{align}
M_2^{(1)}(A,B)=\frac{1}{3}\left[XM_2^{(0)}(A,B)+M_2^{(0)}(XA,B)+M_2^{(0)}(A,XB)\right]\,.
\end{align}
In the spirit of the case without stubs, we introduce the gauge 2-product $\mu_2^{(1)}$, so that when acting on the states in the small Hilbert space, the superstring 2-product can be computed as
\begin{subequations}
\begin{align}
M_2^{(1)} &=[Q,\mu_2^{(1)}]\,,\\
\mu_2^{(1)}(A,B) &= \frac{1}{3}\left[\xi M_2^{(0)}(A,B)- M_2^{(0)}(\xi A, B)-(-1)^{d(A)}M_2^{(0)}(A,\xi B)\right]\,.
\end{align}
\end{subequations}
We also have the property that $M_2^{(0)} = [\eta, \mu_2^{(1)}]$. The superstring 3-product can be defined in terms of the following tower of products
\begingroup
\allowdisplaybreaks
\begin{subequations}
\begin{align}
M_3^{(2)} &= \frac{1}{2}([Q,\mu_3^{(2)}]+[M_2^{(1)},\mu_2^{(1)}])\,,\\
\mu_3^{(2)}(A,B,C) &=\frac{1}{4}\left[\xi M_3^{(1)}(A,B,C)-M_3^{(1)}(\xi A,B,C)\right.\nonumber\\
&\hspace{+1cm}\left.-(-1)^{d(A)}M_3^{(1)}( A,\xi B,C)-(-1)^{d(A)+d(B)}M_3^{(1)}(A,B,\xi C)\right]\,,\\
M_3^{(1)} &= [Q,\mu_3^{(1)}]+[M_2^{(0)},\mu_2^{(1)}]\,,\\
\mu_3^{(1)}(A,B,C)&=\frac{1}{2}\left[\xi M_3^{(0)}(A,B,C)-M_3^{(0)}(\xi A,B,C)\right.\nonumber\\
&\hspace{+1cm}\left.-(-1)^{d(A)} M_3^{(0)}(A,\xi B,C)-(-1)^{d(A)+d(B)}M_3^{(0)}(A,B,\xi C)]
\right]\,,
\end{align}
\end{subequations}
\endgroup
with the properties $M_3^{(1)} = [\eta,\mu_3^{(2)}]$ and $2M_3^{(0)}=[\eta, \mu_3^{(1)}]$. The equation of motion is written in terms of the multi-superstring products $M_n^{(n-1)}$, which satisfy a cyclic $A_\infty$ algebra, as
\begin{align}
\sum_{n=1}^\infty M_n^{(n-1)}(\Psi^n) = Q\Psi+ M_2^{(1)}(\Psi,\Psi)+M_3^{(2)}(\Psi,\Psi,\Psi)+\ldots =0\,.
\end{align}
It is straightforward to check that the relations \eqref{eq:ProjCon}, \eqref{eq:Xvan} and \eqref{eq:xiCon} continue to be satisfied under the same assumptions as before if we replace $m_2$ by $M_2^{(0)}$ (this is due to the presence of the projector $P_0$ and the fact that the insertions have $L_0=0$). 
It is also easy to see that we again have that the second-order obstruction to exact marginality vanishes if and only if the projector condition $P_0 M_2^{(1)}(\Psi_1,\Psi_1)=0$ holds, which we shall from now on assume.

\subsubsection{Third-order obstruction with stubs: $X^2$ form}

\label{subsub:stubX23}

It is straightforward to check that the computation goes through mostly along the lines of the case without stubs with only a couple of minor changes wherever we encounter $M_3$ or make use the associativity of $m_2$. The integrability of the equation of motion at third order in $\lambda$ follows again straightforwardly by using the fact that $[Q,M_2^{(1)}]=0$, the fact that $\Psi_2$ solves the equation of motion at second order and also the $A_\infty$ relation
\begin{align}
[Q,M_3^{(2)}]+\frac{1}{2}[M_2^{(1)},M_2^{(1)}]=0\,.
\end{align}
In order for a consistent solution to exist, we need the obstruction
\begin{align}
    O_3^\text{proj} &= P_0\bigg\{M_2^{(1)}\left[\frac{b_0}{L_0}M_2^{(1)}(\Psi_1,\Psi_1),\Psi_1\right] +\nonumber\\
    &\hspace{4cm}+ M_2^{(1)}\left[\Psi_1,\frac{b_0}{L_0}M_2^{(1)}(\Psi_1,\Psi_1)\right]-M_3^{(2)}(\Psi_1,\Psi_1,\Psi_1)\bigg\}\,.
    \label{eq:OStubs}
\end{align}
to be vanishing up to $Q$-exact terms. Going through identical steps as in the case without stubs, we can show that it is necessary and sufficient to require vanishing of $\mathcal{O}\equiv -\omega_S(e,O^\text{proj}_3)=\mathcal{O}_1+\mathcal{O}_2+\mathcal{O}_3$, where $e=c\tilde{\mathbb{V}}_{1/2}e^{-\phi}$ and $\mathcal{O}_1$, $\mathcal{O}_2$, 
$\mathcal{O}_3$ will now be described. First, we have
\begingroup\allowdisplaybreaks
\begin{align}
\mathcal{O}_1 &=  \frac{1}{2}\bigg\{-\omega_L\left[ e, X\mu_3^{(2)}(\Psi_1,\Psi_1,\Psi_1)\right]+\nonumber\\
&\hspace{1cm}+ \omega_L\left[ e, X\xi M_2^{(0)}[\mu_2^{(1)}(\Psi_1,\Psi_1),\Psi_1]\right]-\omega_L\left[ e,X\xi \mu_2^{(1)}[M_2^{(0)}(\Psi_1,\Psi_1),\Psi_1]\right]\nonumber\\
&\hspace{1cm}+  \omega_L\left[ e, X\xi M_2^{(0)}[\Psi_1,\mu_2^{(1)}(\Psi_1,\Psi_1)]\right]-\omega_L\left[ e,X\xi \mu_2^{(1)}[\Psi_1,M_2^{(0)}(\Psi_1,\Psi_1)]\right]\bigg\}\,,
\end{align}
\endgroup
where we note that the string field $X\mu_3^{(2)}(\Psi_1,\Psi_1,\Psi_1)-X\xi M_3^{(1)}(\Psi_1,\Psi_1,\Psi_1)$ lies in the small Hilbert space and therefore $-\omega_L[e,X\mu_3^{(2)}(\Psi_1,\Psi_1,\Psi_1)]=-\omega_L[e,X\xi M_3^{(1)}(\Psi_1,\Psi_1,\Psi_1)]$ so that we can use the relation $M_3^{(1)}=[Q,\mu_3^{(1)}]+[M_2^{(0)},\mu_2^{(1)}]$ (see \cite{Erler:2014eba}) to write
\begin{align}
\mathcal{O}_1 = -\frac{1}{2}\omega_L\left[e,X\xi Q\mu_3^{(1)}(\Psi_1,\Psi_1,\Psi_1)\right]\,.
\end{align}
Finally, using the relation $[\eta,\mu_3^{(1)}]=2M_3^{(0)}$, we conclude that 
\begin{align}
\mathcal{O}_1 = \omega_S\left[X^2 e,M_3^{(0)}(\Psi_1,\Psi_1,\Psi_1)\right]\,.
\end{align}
Note that this is different compared to the case without stubs where we had $\mathcal{O}_1$ vanishing. As for the remaining two contributions to $\mathcal{O}$, we again obtain
\begin{align}
\mathcal{O}_2&= \omega_L\left[ e,X^2\xi M_2^{(0)}\left[\frac{b_0}{L_0}\overline{P}_0 M_2^{(0)}(\Psi_1,\Psi_1),\Psi_1\right]\right] +\nonumber\\
&\hspace{3cm}+  \omega_L\left[ e,X^2\xi M_2^{(0)}\left[\Psi_1,\frac{b_0}{L_0}\overline{P}_0 M_2^{(0)}(\Psi_1,\Psi_1)\right] \right]\,,
\end{align}
together with
\begingroup\allowdisplaybreaks
\begin{align}
    \mathcal{O}_3 =& \frac{1}{6}\bigg\{+4\omega_L \left[P_0\mu_2^{(1)}(\Psi_1,\Psi_1),  M_2^{(0)}(\Psi_1, X\xi e) +M_2^{(0)}(X\xi e,\Psi_1)\right]+\nonumber\\
    &\hspace{2.0cm}+\omega_L\left[ P_0\mu_2^{(1)}(\Psi_1,\Psi_1),M_2^{(0)}(X\Psi_1,\xi e) +M_2^{(0)}(\xi e,X\Psi_1)\right] +\nonumber\\
    &\hspace{0.2cm}+\omega_L\left[ \xi P_0 M_2^{(0)}(\Psi_1,\Psi_1), M_2^{(0)}(\Psi_1,X\xi e)+M_2^{(0)}(X\xi e,\Psi_1)\right]+\nonumber\\
    &\hspace{2.0cm}+\omega_L\left[ P_0 M_2^{(0)}(\Psi_1,\Psi_1), M_2^{(0)}(\xi \Psi_1,X\xi e)-M_2^{(0)}(X\xi e,\xi \Psi_1) \right]\bigg\}\,.
\end{align}
\endgroup
Altogether, the $X^2$ form of the obstruction in the case with stubs therefore reads
\begin{align}
\mathcal{O}&= -\omega_S\left[X^2  e, M_2^{(0)}[\frac{b_0}{L_0}\overline{P}_0 M_2^{(0)}(\Psi_1,\Psi_1),\Psi_1]\right] +\nonumber\\
&\hspace{3.3cm}-  \omega_S\left[X^2 e, M_2^{(0)}[\Psi_1,\frac{b_0}{L_0}\overline{P}_0 M_2^{(0)}(\Psi_1,\Psi_1)] \right]+\nonumber\\
&\hspace{6.3cm}+\omega_S\left[X^2 e,M_3^{(0)}(\Psi_1,\Psi_1,\Psi_1)\right]+\mathcal{O}_3\,.
\end{align}

\subsubsection{Third-order obstruction with stubs: Berkovits-like form}

\label{subsub:stubBer3}

We will now show that the obstruction can be rewritten in terms of the product $B_2^{(0)}$ (see \eqref{eq:B2def} for definition) as
\begin{align}
\mathcal{O} &= \frac{1}{2}\omega_S\left[B_2^{(0)}(\Psi_1,Xe),\frac{b_0}{L_0} \overline{P}_0 B_2^{(0)}(\Psi_1,X\Psi_1)\right]+\nonumber\\
&\hspace{+1.3cm}+\frac{1}{6}\left\{\omega_L\left[B_2^{(0)}(Xe,\xi \Psi_1),B_2^{(0)}(\xi \Psi_1,\Psi_1)\right]+\omega_S\left[Xe,B_3^{(0)}(X\Psi_1,\Psi_1,\Psi_1)\right]\right\}\,,
\label{eq:BerStub}
\end{align}
so that all insertions are primary.

\subsubsection*{Non-associative commutator algebra}

Let us define the degree-graded commutator based on the Witten star product with stubs as
\begin{align}
B_2^{(0)}(A,B) \equiv M_2^{(0)}(A,B)+(-1)^{d(A)d(B)}M_2^{(0)}(B,A)\,.
\label{eq:B2def}
\end{align}
Denoting
\begin{align}
[A,B]_\text{st}\equiv e^{-w L_0}[(e^{-w L_0}A)\ast (e^{-w L_0}B)]-(-1)^{|A||B|}e^{-w L_0}[(e^{-w L_0}B)\ast (e^{-w L_0}A)]\,,
\end{align}
we therefore have $B_2^{(0)}(A,B) = (-1)^{d(A)}[A,B]_\text{st}$. Clearly we have 
\begin{align}
B_2^{(0)}(A,B) = (-1)^{d(A)d(B)}B_2^{(0)}(B,A)
\end{align}
and it can be shown that cyclicity of the symplectic form w.r.t.\ $M_2^{(0)}$ implies 
\begin{align}
\omega(B_2^{(0)}(A,B),C) = (-1)^{d(A)+1}\omega(A,B_2^{(0)}(B,C))\,.
\end{align}
We also have the generalized super-Jacobi identity
\begingroup
\begin{align}
&(-1)^{d(A)(d(C)+1)}B_2^{(0)}\left[A,B_2^{(0)}(B,C)\right]+(-1)^{d(B)(d(A)+1)}B_2^{(0)}\left[B,B_2^{(0)}(C,A)\right]
\nonumber\\
&\hspace{7cm}+(-1)^{d(C)(d(B)+1)}B_2^{(0)}\left[C,B_2^{(0)}(A,B)\right]\nonumber\\
&\hspace{+2cm}=-(-1)^{d(A)d(C)}[Q,B_3^{(0)}](A,B,C)
\label{eq:GenSuperJac}
\end{align}
\endgroup
where we have defined
\begin{align}
B_3^{(0)}(A,B,C) \equiv &\, M_3^{(0)}(A,B,C)+(-1)^{d(A)(d(B)+d(C))}M_3^{(0)}(B,C,A)+\nonumber\\
&\hspace{3.5cm}+(-1)^{d(C)(d(A)+d(B))}M_3^{(0)}(C,A,B)+\nonumber\\
&+(-1)^{d(A)d(B)}M_3^{(0)}(B,A,C)+(-1)^{d(B)d(C)}M_3^{(0)}(A,C,B)\nonumber\\
&\hspace{3.5cm}+(-1)^{d(A)(d(B)+d(C))+d(B)d(C)}M_3^{(0)}(C,B,A)\,.
\end{align}
In particular, we obtain
\begin{align}
B_2^{(0)}\left[\xi \Psi_1,B_2^{(0)}( \Psi_1, \Psi_1)\right]- 2B_2^{(0)}\left[\Psi_1,B_2^{(0)}( \Psi_1,\xi \Psi_1)\right]=[Q,B_3^{(0)}](\Psi_1,\xi \Psi_1,\Psi_1)\,,
\label{eq:SuperJacSxi}
\end{align}
where
\begin{align}
\frac{1}{2}[Q,B_3^{(0)}](\Psi_1,\xi \Psi_1,\Psi_1)&=QM_3^{(0)}(\Psi_1,\xi \Psi_1, \Psi_1)+QM_3^{(0)}(\xi \Psi_1, \Psi_1,\Psi_1)+\nonumber\\
&\hspace{2.0cm}+QM_3^{(0)}(\Psi_1, \Psi_1,\xi \Psi_1)+\frac{1}{2}B_3^{(0)}(X\Psi_1,\Psi_1,\Psi_1)\,.
\end{align}

\subsubsection*{$\mathcal{O}_3$ terms}

It is straightforward to see that we obtain
\begin{align}
    \mathcal{O}_3 &= -\frac{1}{12}\bigg\{5\omega_L\left[ P_0 B_2^{(0)}(\Psi_1,\Psi_1),\xi B_2^{(0)}(\Psi_1, X\xi e)\right] +\nonumber\\
    &\hspace{3.5cm}+\omega_L\left[ P_0B_2^{(0)}(\Psi_1,\Psi_1),\xi B_2^{(0)}(\xi \Psi_1, X e)\right]+\nonumber\\
&\hspace{6cm}-\omega_L\left[ P_0 B_2^{(0)}(\Psi_1,\Psi_1),B_2^{(0)}(\xi \Psi_1, X\xi e)\right]\bigg\}\,.
\end{align}

\subsubsection*{$\mathcal{O}_2$ terms}

First, using cyclicity of the simplectic form and the definition of $B_2^{(0)}$, we obtain
\begin{align}
\mathcal{O}_2 &=  \frac{1}{2}\omega_S\left[ B_2^{(0)}(\Psi_1,X^2  e),\frac{b_0}{L_0}\overline{P}_0 B_2^{(0)}(\Psi_1,\Psi_1)\right]\,,
\end{align}
which we again rewrite as
\begin{align}
\mathcal{O}_2 &= \frac{1}{2}\omega_S\left[B_2^{(0)}( \Psi_1, X e),\frac{b_0}{L_0}\overline{P}_0 B_2^{(0)}(\Psi_1,X \Psi_1)\right]+\nonumber\\
&\hspace{6cm}+\frac{1}{2}\omega_L\left[B_2^{(0)}( \Psi_1,\xi X e),\overline{P}_0 B_2^{(0)}(\Psi_1,\xi \Psi_1)\right]\,.
\label{eq:O2xS}
\end{align}
It is easily checked that the $P_0$ part of the second term in \eqref{eq:O2xS} again cancels $\mathcal{O}_3$.
As for the identity part, defining the string field
\begin{align}
\Upsilon &\equiv B_2^{(0)}\left[\Psi_1,B_2^{(0)}(\xi X e,\xi \Psi_1)\right]-B_2^{(0)}\left[\xi \Psi_1,B_2^{(0)}(\xi X e, \Psi_1)\right]+\nonumber\\
&\hspace{7cm}+B_2^{(0)}\left[\xi \Psi_1,B_2^{(0)}( X e,\xi \Psi_1)\right]\,,
\end{align}
which satisfies $\eta \Upsilon=0$, so that $\omega_L(\Psi_1,\Upsilon)=0$, we can use the generalized super-Jacobi identity \eqref{eq:SuperJacSxi} and cyclicity of the symplectic form to show that
\begin{align}
\frac{1}{2}\omega_L\left[B_2^{(0)}(\Psi_1,\xi X e),B_2^{(0)}(\Psi_1,\xi \Psi_1)]\right] =&-\omega_S\left[X^2 e,M_3^{(0)}(\Psi_1,\Psi_1,\Psi_1)\right]+\nonumber\\
&\hspace{-5.0cm}+ \frac{1}{6}\left\{\omega_L\left[B_2^{(0)}( X e,\xi \Psi_1),B_2^{(0)}(\xi \Psi_1,\Psi_1)\right]+\omega_S\left[Xe,B_3^{(0)}(X\Psi_1,\Psi_1,\Psi_1)\right]\right\}\,.
\end{align}
Substituting back into \eqref{eq:O2xS}, we recover the Berkovits-like form \eqref{eq:BerStub}.

\section{Equivalence of the $A_\infty$ and Berkovits obstructions at third order}

\label{sec:berk}

Here we show that the third-order obstruction arising from the reduced Berkovits open superstring field theory (i.e.\ $\xi_0 \Phi=0$) is identical to the one derived in the $A_\infty$ OSFT without stubs. After setting up the stage by reviewing the machinery of marginal deformations in the Berkovits theory in subsection \ref{subsec:margBerk}, we will evaluate the third-order obstruction against arbitrary test states in \ref{subsec:simpleBerk}, recovering the Berkovits-like form \eqref{eq:introBer} for the obstruction which we derived in the context of $A_\infty$ OSFT in the previous section.

\subsection{Marginal deformations in Berkovits open superstring field theory}
\label{subsec:margBerk}

Expanding the Berkovits action up to quartic order, we obtain
\begin{align}
S_\text{Ber}[\Phi] = -\frac{1}{2}\mathrm{Tr}_L [\eta \Phi Q\Phi]+ \frac{1}{6}\mathrm{Tr}_L[\eta \Phi[\Phi,Q\Phi]]-\frac{1}{24}\mathrm{Tr}_L[\eta\Phi[\Phi,[\Phi,Q\Phi]]]+\ldots
\end{align}
The equation of motion which we obtain by varying this action reads
\begin{align}
Q\eta  \Phi + \frac{1}{2}[\eta \Phi,Q\Phi]+\frac{1}{12}([\eta \Phi,[Q\Phi,\Phi]]+[\Phi,[\Phi,Q\eta \Phi]]+[Q\Phi,[\Phi,\eta \Phi]])+\ldots=0\,.
\end{align}
We partially fix gauge as $\xi_0 \Phi = 0$, meaning that we can write $\Phi = \xi_0 \Psi$ where $\Psi$ is in picture $-1$ with $\eta_0 \Psi =0$. Again, we want to find a continuous family $\Psi(\lambda)=\sum_{k=1}^\infty\lambda^k \Psi_k$ of classical solutions with $\Psi_1 = c\mathbb{V}_{1/2}e^{-\phi}$. Order by order in $\lambda$, we obtain conditions
\begingroup
\allowdisplaybreaks
\begin{subequations}
\begin{align}
0&= Q\Psi_1 \,,\\[+1.1mm]
0&=Q\Psi_2 + \frac{1}{2}[\Psi_1,X\Psi_1]\,,\\
0&=Q\Psi_3 + \frac{1}{2}[\Psi_1,Q\xi\Psi_2]+\frac{1}{2}[\Psi_2,X\Psi_1]+\frac{1}{12}([\Psi_1,[X\Psi_1,\xi\Psi_1]]+[X\Psi_1,[\xi\Psi_1,\Psi_1]])\,.\label{eq:Cc}\\
&\hspace{0.2cm}\vdots\nonumber
\end{align}
\end{subequations}
\endgroup
The equation arising at second order in $\lambda$ is clearly integrable because $Q[\Psi_1,X\Psi_1]=0$. A putative solution for $\Psi_2$ then reads
\begin{align}
\Psi_2 = -\frac{1}{2}\frac{b_0}{L_0}\overline{P}_0[\Psi_1,X\Psi_1]+\psi_2\,.
\label{eq:35Ber}
\end{align}
In order for \eqref{eq:35Ber} to actually solve the equation of motion at $\mathcal{O}(\lambda^2)$, we need
\begin{align}
Q\Psi_2 = \frac{1}{2}\left(\frac{b_0}{L_0}Q-\overline{P}_0\right)\overline{P}_0[\Psi_1,X\Psi_1] +Q\psi_2
\end{align}
to be equal to $-\frac{1}{2}[\Psi_1,X\Psi_1]$, that is, we need the second order obstruction
\begin{align}
O_2^\text{Ber} =\frac{1}{2} P_0[\Psi_1,X\Psi_1]+Q\psi_2
\end{align}
to vanish. Analogously to the $A_\infty$ case, one can show that the necessary and sufficient condition for $O_2^\text{Ber}$ to vanish is $P_0[\Psi_1,X\Psi_1]=0$ with $Q\psi_2=0$ (so that again, we need to take $\psi_2$ to be of the form $c\hat{\mathbb{V}}_{1/2}e^{-\phi}$). Proceeding to the third order, integrability requires that
\begin{align}
\frac{1}{2}Q[\Psi_1,Q\xi\Psi_2]+\frac{1}{2}Q[\Psi_2,X\Psi_1]+\frac{1}{12}(Q[\Psi_1,[X\Psi_1,\xi\Psi_1]]+Q[X\Psi_1,[\xi\Psi_1,\Psi_1]])=0\,.
\label{eq:lineBer}
\end{align}
Assuming that $\Psi_2$ solves the second order equation of motion, \eqref{eq:lineBer} can be straightforwardly shown to hold as a consequence of the super-Jacobi identity
\begin{align}
2[X \Psi_1 ,[ \Psi_1 ,X \Psi_1 ]]
+[ \Psi_1 ,[X \Psi_1 ,X \Psi_1 ]]=0\,.
\end{align}
A putative solution for $\Psi_3$ then reads
\begin{align}
\Psi_3 &= -\frac{b_0}{L_0}\overline{P}_0\left\{\frac{1}{2}\left[ \Psi_1 ,Q\xi \Psi_2\right]+\frac{1}{2}\left[\Psi_2,X \Psi_1 \right]\right.+\nonumber\\
&\hspace{5cm}+\left.\frac{1}{12}([ \Psi_1 ,[X \Psi_1 ,\xi \Psi_1 ]]+[X \Psi_1 ,[\xi \Psi_1 , \Psi_1 ]])\right\}+\psi_3
\label{eq:Sol2}
\end{align}
and the corresponding obstruction can be readily seen to be equal to\footnote{Similarly to the $A_\infty$ case, we ignore any potential contributions of $\psi_2$ to $O_3^\text{Ber}$ because in most cases of interest we will have $P_0[\Psi_1,X\psi_2]+P_0[\psi_2,X\Psi_1]=0$ by the projector condition \eqref{eq:Xvan}.}
\begin{align}
O_3^\text{Ber} &= {P}_0\left\{\frac{1}{4}\left[ \Psi_1 ,Q\xi \frac{b_0}{L_0}\overline{P}_0[ \Psi_1 ,X \Psi_1]\right]+\frac{1}{4}\left[\frac{b_0}{L_0}\overline{P}_0[ \Psi_1 ,X \Psi_1 ],X \Psi_1 \right]+\right.\nonumber\\
&\hspace{4.0cm}\left.-\frac{1}{12}([ \Psi_1 ,[X \Psi_1 ,\xi \Psi_1 ]]+[X \Psi_1 ,[\xi \Psi_1 , \Psi_1 ]])\right\}-Q\psi_3\,.
\label{eq:O3DefBerk}
\end{align}
Recalling our discussion of the $A_\infty$ case, it is clear that necessary and sufficient condition for the vanishing of $O_3^\text{Ber}$ is that
\begin{align}
    \mathcal{O}_\text{Ber} &\equiv -\mathrm{Tr}_S[e_\text{g} O_3^\text{Ber,proj}] =0\,,
\end{align}
where $O_3^\text{Ber,proj}=P_0\{\ldots\}$ is the projector part of \eqref{eq:O3DefBerk} and $e_\text{g} = c\tilde{\mathbb{V}}_{1/2}e^{-\phi}$. 

\subsection{Simplifying the third-order obstruction}
\label{subsec:simpleBerk}
Evaluating $O_3^\text{Ber,proj}$ against all possible test states of the form $e=c\tilde{\mathbb{V}}_{1/2}e^{-\phi}$ (again, we drop the lower index 'g' from $e$), we have
\begin{align}
\mathcal{O}_\text{Ber}
&= -\frac{1}{4}\mathrm{Tr}_S\left[[e,X \Psi_1 ]\frac{b_0}{L_0}\overline{P}_0[ \Psi_1 , X \Psi_1 ]\right]-\frac{1}{4}\mathrm{Tr}_S\left[[ \Psi_1 ,X e] \frac{b_0}{L_0}\overline{P}_0[ \Psi_1 , X \Psi_1 ]\right]+\nonumber\\
&\hspace{3cm}+\frac{1}{12}\mathrm{Tr}_L\left[\xi e[ \Psi_1 ,[X \Psi_1 ,\xi \Psi_1 ]]\right]+\frac{1}{12}\mathrm{Tr}_L\left[\xi e[X \Psi_1 ,[\xi \Psi_1 , \Psi_1 ]]\right]\,.\label{eq:largeTr}
\end{align}
Finally, to make contact with the $A_\infty$ obstruction derived in Section \ref{subsec:ainf}, we can first use the vanishing of the second-order obstruction (i.e.\ that $P_0[\Psi_1,X\Psi_1]=0$) to establish that
\begingroup
\allowdisplaybreaks
\begin{subequations}
\begin{align}
\frac{1}{4}\mathrm{Tr}_S\left[[e,X\Psi_1]\frac{b_0}{L_0}\overline{P}_0 [\Psi_1,X\Psi_1]\right]
&=-\frac{1}{4}\mathrm{Tr}_L\left[[\xi e,\xi \Psi_1][\Psi_1,X\Psi_1]\right]+\nonumber\\
&\hspace{2cm}+\frac{1}{4}\mathrm{Tr}_S\left[[ \Psi_1,X e]\frac{b_0}{L_0}\overline{P}_0[\Psi_1,X\Psi_1]\right]
\end{align}
\end{subequations}
\endgroup
so that the expression \eqref{eq:largeTr} for $\mathcal{O}_\text{Ber}$ can be rewritten as
\begingroup
\allowdisplaybreaks
\begin{subequations}
\begin{align}
\mathcal{O}_\text{Ber}
&=-\frac{1}{2}\mathrm{Tr}_S\left[[ \Psi_1 ,X e] \frac{b_0}{L_0}\overline{P}_0[ \Psi_1 , X \Psi_1 ]\right]+\frac{1}{4}\mathrm{Tr}_L\left[[\xi e,\xi \Psi_1][\Psi_1,X\Psi_1]\right]\nonumber\\
&\hspace{3cm}+\frac{1}{12}\mathrm{Tr}_L[\xi e[ \Psi_1 ,[X \Psi_1 ,\xi \Psi_1 ]]]+\frac{1}{12}\mathrm{Tr}_L[\xi e[X \Psi_1 ,[\xi \Psi_1 , \Psi_1 ]]]\\
&=-\frac{1}{2}\mathrm{Tr}_S\left[[ \Psi_1 ,X e]\frac{b_0}{L_0}\overline{P}_0[ \Psi_1 , X \Psi_1 ]\right]-\frac{1}{6}\mathrm{Tr}_L[[X e, \xi \Psi_1][ \xi \Psi_1,\Psi_1  ]]\,,
\end{align}
\end{subequations}
\endgroup
where, in the second step, we used a super-Jacobi identity. That is
\begin{align}
\mathcal{O}_\text{Ber}&=\frac{1}{2}\omega_S \left[b_2(\Psi_1,Xe),\frac{b_0}{L_0} \overline{P}_0 b_2(\Psi_1,X\Psi_1) \right]+\frac{1}{6}\omega_L[b_2(Xe,\xi \Psi_1),b_2(\xi \Psi_1,\Psi_1)]\,.\label{eq:OBerInf}
\end{align}
Since \eqref{eq:OBerInf} is identical with \eqref{eq:O2propcon}, we have shown that the third-order obstructions arising in the $A_\infty$ and Berkovits open superstring field theories are equal when evaluated against a test state $c\tilde{\mathbb{V}}_{1/2}e^{-\phi}$. Note that strictly speaking this is only true provided that the obstruction at second order in both theories was already arranged to vanish.

\section{Evaluation of the obstruction}
\label{sec:eval}

Here we will present three ways of evaluating the obstruction at third order. Here, by ``evaluating'' the obstruction, we will mean deriving an explicit expression for $\mathcal{O}$ in terms of $\mathbb{V}_{1/2}$ and $\tilde{\mathbb{V}}_{1/2}$ which would be suitable for practical applications.
The first two methods will rely on the presence of a global $\mathcal{N}=2$ worldsheet superconformal symmetry for the given background with only $\pm 1$ $R$-charge marginal fields appearing in the boundary spectrum: in subsection \ref{subsec:locBerk} we will use the Berkovits-like form \eqref{eq:introBer} as our starting point, while in subsection \ref{subsec:locX2} we will start from the $X^2$-form \eqref{eq:introX2}. Both methods will lead to the same result
\begin{align}
\mathcal{O}  = -\langle \tilde{\mathbb{H}}_1^+|\mathbb{H}_1^-\rangle-\langle \tilde{\mathbb{H}}_1^-|\mathbb{H}_1^+\rangle-\frac{1}{2}\langle \tilde{\mathbb{H}}_0 |\mathbb{H}_0\rangle\,,
\label{eq:obstrIntro}
\end{align}
where the auxilliary fields $\mathbb{H}_1^\pm$, $\mathbb{H}_0$ and $\tilde{\mathbb{H}}_1^\pm$, $\tilde{\mathbb{H}}_0$ will be defined in terms of $\mathbb{V}_{1/2}^\pm$ and $\tilde{\mathbb{V}}_{1/2}^\pm$ in \eqref{eq:Hs} and \eqref{eq:Es}. Note that no integration over worldsheet moduli appears in \eqref{eq:obstrIntro} -- the result localised on the boundary of the worldsheet moduli space. Upon identifying $e=\Psi_1$, the expression \eqref{eq:obstrIntro} for the obstruction becomes proportional to the localized quartic part of the classical effective action, in accordance with the prediction of \cite{Maccaferri:2019ogq}. Also, for both methods, we will show that the final result is unaffected by adding stubs: that is, adding the bosonic 4-string vertex $M_3^{(0)}$, which inherently comes with integration over a bosonic modulus, does not seem to spoil the localization property of the third-order obstruction. Finally, in subsection \ref{subsec:direct}, we will present a method for evaluating $\mathcal{O}$ directly along the lines of \cite{Berkovits:2003ny} -- this will work also for more general setups. 

As we have hinted at above, the core of the first two methods will be the recipe of \cite{Sen:2015uoa,Maccaferri:2018vwo,Maccaferri:2019ogq}, that is, we will assume that we can decompose the string fields $\Psi_1$ and $e$ into eigenstates of the $R$-current $J$ of an $\mathcal{N}=2$ worldsheet superconformal algebra $\{T,J,G^\pm\}$ with $R$-charge $\pm 1$. In particular, we will assume that the theory contains only such NS marginal operators $V$, which satisfy $V=V^++V^-$, where $V^\pm$ carry charge $\pm$ under the $R$-current. 
Writing $V^\pm=c \mathbb{V}_{1/2}^\pm e^{-\phi}$, we have 
\begin{align}
XV^\pm = c\mathbb{V}_1^\pm-e^\phi\eta \mathbb{V}_{1/2}^\pm\,,
\end{align}
where we assume
\begin{subequations}
\begin{align}
    G^\pm(z) \mathbb{V}_{\frac{1}{2}}^\mp(0) &=\frac{1}{z}\mathbb{V}_1^\mp(0)+\mathrm{reg.}\,,\\
    G^\pm(z) \mathbb{V}_{\frac{1}{2}}^\pm(0) &= \mathrm{reg.}
\end{align}
\end{subequations}
While the $h=\frac{1}{2}$ matter fields are charged under $J$, their $h=1$ counterparts are neutral
\begin{subequations}
\begin{align}
    J_0 \mathbb{V}_{\frac{1}{2}}^\pm &= \pm \mathbb{V}_{\frac{1}{2}}^\pm\,,\\
    J_0 \mathbb{V}_{1}^\pm &= 0\,.
\end{align}
\end{subequations}
Note that as per our discussion at the end of subsection \ref{subsubsec:prel}, these assumptions hold automatically if we assume that the background at hand conserves at least two spacetime supercharges with the same chirality in two non-compact dimensions. 

Finally, we note that first two methods for evaluating $\mathcal{O}$ will allow also for slightly more general setups then we described above.\footnote{We would like to thank Luca Mattiello and Ivo Sachs for a useful discussion on this point.} Namely, we will be allowed to assume that the worldsheet theory may contain also NS marginal fields in the matter sector with charge zero under the localising $R$-current: while we will always deform our theory by a subsector of marginal operators which can be decomposed as $V^++V^-$, the obstruction has to be always computed against all possible test states. And these we will allow to include also marginal operators with zero $R$-charge. That is, we will allow for test states $e = e^+ +e^- + e^0$, where $e^\pm=c\tilde{\mathbb{V}}_{1/2}^\pm e^{-\phi}$ together with $e^0=c\tilde{\mathbb{V}}_{1/2}^0 e^{-\phi}$. We then have
\begin{subequations}
\begin{align}
    G^\pm(z) \tilde{\mathbb{V}}_{\frac{1}{2}}^\mp(0) &=\frac{1}{z}\tilde{\mathbb{V}}_1^\mp(0)+\mathrm{reg.}\,,\\
G(z)\tilde{\mathbb{V}}_\frac{1}{2}^0(0) &=\frac{1}{z}\tilde{\mathbb{V}}_1^0(0)+\text{reg.}\,,\\[1.8mm]
    G^\pm(z) \tilde{\mathbb{V}}_{\frac{1}{2}}^\pm(0) &= \mathrm{reg.}\,,
\end{align}
\end{subequations}
where we note that
\begin{align}
\tilde{\mathbb{V}}^0_1 = G^+_{-\frac{1}{2}}\tilde{\mathbb{V}}_\frac{1}{2}^0+G^-_{-\frac{1}{2}}\tilde{\mathbb{V}}_\frac{1}{2}^0 \equiv (\tilde{\mathbb{V}}^0_1)^+ + (\tilde{\mathbb{V}}^0_1)^-\,,
\end{align}
where $(\tilde{\mathbb{V}}^0_1)^\pm$ carry charge $\pm 1$ under $J$. We also have $Xe^0 = c\tilde{\mathbb{V}}_1^0 - e^\phi\eta \tilde{\mathbb{V}}^0_{1/2}$.

\subsection{Localization: Berkovits-like form}

\label{subsec:locBerk}

We will now show that starting with the Berkovits-like form \eqref{eq:O2propcon} of the obstruction $\mathcal{O} = \mathcal{O}^\text{prop}+\mathcal{O}'$ and exploiting the virtues of the $\mathcal{N}=2$ $R$-charge decomposition of $\Psi_1$, one can write down an expression for $\mathcal{O}$ which does not contain integration over the worldsheet moduli. We will first show (subsection \ref{subsub:locBerProp}) that the propagator term of the Berkovits-like form decomposes into a localized part $\mathcal{O}^\text{loc}$ and a contact part, which will be then shown (subsection \ref{subsub:locBerCon}) to exactly cancel with $\mathcal{O}'$. Finally, in subsection \ref{subsec:elt} we will evaluate the OPE in $\mathcal{O}^\text{loc}$ to derive the result \eqref{eq:obstrIntro} which is suitable for applications. In subsection \ref{subsub:locBerStubs}, we will shortly discuss that adding stubs, while introducing an additional term into the Berkovits-like form (see \eqref{eq:BerStub}), it leaves the final result \eqref{eq:obstrIntro} unchanged as the appearance of the bosonic 3-product $M_3^{(0)}$ is exactly compensated by the associator of $M_2^{(0)}$.

\subsubsection{Propagator term}

\label{subsub:locBerProp}

Focusing on the propagator term of \eqref{eq:O2propcon} first, we use the $R$-charge conservation and $c$-ghost saturation to write
\begin{align}
   \mathcal{O}^\text{prop} = \mathcal{O}^{\pm \pm}+\mathcal{O}^{\pm \mp}\,,
\end{align}
where we have defined
\begingroup\allowdisplaybreaks
\begin{subequations}
\begin{align}
     \mathcal{O}^{\pm \pm} &=  \frac{1}{2}\omega_S \left[b_2(\Psi_1^-,X\Psi_1^-),\frac{b_0}{L_0} \overline{P}_0 b_2(\Psi_1^+,Xe) \right]+\nonumber\\
     &\hspace{5cm}+\frac{1}{2}\omega_S \left[b_2(\Psi_1^+,X\Psi_1^+),\frac{b_0}{L_0} \overline{P}_0 b_2(\Psi_1^-,Xe) \right]\,,\\
     \mathcal{O}^{\pm \mp} &=  \frac{1}{2}\omega_S \left[b_2(\Psi_1^-,X\Psi_1^+),\frac{b_0}{L_0}\overline{P}_0 b_2(\Psi_1^+,Xe) \right]+\nonumber\\
     &\hspace{5cm}+\frac{1}{2}\omega_S \left[b_2(\Psi_1^+,X\Psi_1^-),\frac{b_0}{L_0}\overline{P}_0 b_2(\Psi_1^-,Xe) \right]\,.
\end{align}
\end{subequations}
\endgroup
Here we have used the fact that $Xe^\pm$ and $X\Psi_1^\pm$ are $R$-neutral in the above correlators, because the Siegel gauge propagator provides a $b$-ghost, so that we need to take the $c$-ghost parts for all insertions. 
For the same reason, we have that
\begin{align}
    \omega_S \left[b_2(\Psi_1^\pm,X\Psi_1),\frac{b_0}{L_0}\overline{P}_0b _2(\Psi_1^\pm,Xe) \right]=0\,.
\end{align}
Also, note that any potential $R$-neutral part $e_0$ of $e$ can never contribute to the propagator term, because $c$-ghost saturation tells us that $Xe_0$ carries charge $\pm 1$. 
We will now remove the propagators by moving the PCO which does not sit on the test state $e$ onto a $\Psi_1^\pm$ insertion with $R$-charge different from the remaining two. This can be done by first going to the large Hilbert space by placing $\xi$ on the insertion where we want the PCO to be moved, then writing $X\Psi_1^\pm = Q\xi \Psi_1^\pm$ and finally moving $Q$ onto the insertion with $\xi$. In particular, starting with $\mathcal{O}^{\pm\pm}$, we get
\begingroup\allowdisplaybreaks
\begin{align}
 \mathcal{O}^{\pm \pm} 
     =&\,  +\frac{1}{2}\omega_L \left[ b_2(\Psi_1^-,\xi \Psi_1^-), \overline{P}_0 b_2(\xi \Psi_1^+,Xe) \right]+\nonumber\\
     &\hspace{5cm}+\frac{1}{2}\omega_L \left[ b_2(\Psi_1^+,\xi \Psi_1^+), \overline{P}_0 b_2(\xi \Psi_1^-,Xe) \right]+\nonumber\\
      &\,+\frac{1}{2}\omega_S \left[ b_2(\Psi_1^-, \Psi_1^-),\frac{b_0}{L_0} \overline{P}_0 b_2(X \Psi_1^+,Xe) \right]+\nonumber\\
     &\hspace{5cm}+\frac{1}{2}\omega_S \left[ b_2(\Psi_1^+, \Psi_1^+),\frac{b_0}{L_0} \overline{P}_0 b_2(X \Psi_1^-,Xe) \right]\,,
\end{align}
\endgroup
where the last two terms vanish by the $R$-charge conservation. We therefore end up with
\begin{align}
 \mathcal{O}^{\pm \pm} 
     =&\,  +\frac{1}{2}\omega_L \left[ b_2(\Psi_1^-,\xi \Psi_1^-), \overline{P}_0 b_2(\xi \Psi_1^+,Xe) \right]+\nonumber\\
     &\hspace{5cm}+\frac{1}{2}\omega_L \left[ b_2(\Psi_1^+,\xi \Psi_1^+), \overline{P}_0 b_2(\xi \Psi_1^-,Xe) \right]\,.
\end{align}
Similarly for $\mathcal{O}_2^{\pm \mp}$, where we get
\begingroup\allowdisplaybreaks
\begin{align}
  \mathcal{O}^{\pm \mp} =&\,+  \frac{1}{2}\omega_S \left[b_2(X \Psi_1^-, \Psi_1^+),\frac{b_0}{L_0}\overline{P}_0 b_2(\Psi_1^+,Xe) \right]+\nonumber\\
     &\hspace{5cm}+\frac{1}{2}\omega_S \left[b_2(X \Psi_1^+, \Psi_1^-),\frac{b_0}{L_0}\overline{P}_0 b_2(\Psi_1^-,Xe) \right]\nonumber\\
     &\,+  \frac{1}{2}\omega_L \left[b_2(\xi \Psi_1^-,\xi \Psi_1^+),\overline{P}_0 b_2(\Psi_1^+,Xe) \right]+\nonumber\\
     &\hspace{5cm}+\frac{1}{2}\omega_L \left[b_2(\xi \Psi_1^+,\xi \Psi_1^-),\overline{P}_0 b_2(\Psi_1^-,Xe) \right]\,.
\end{align}
\endgroup
with the first two terms vanishing by $R$-charge conservation, that is
\begin{align}
  \mathcal{O}^{\pm \mp} =  &\,+  \frac{1}{2}\omega_L \left[b_2(\xi \Psi_1^-,\xi \Psi_1^+),\overline{P}_0 b_2(\Psi_1^+,Xe) \right]+\nonumber\\
     &\hspace{5cm}+\frac{1}{2}\omega_L \left[b_2(\xi \Psi_1^+,\xi \Psi_1^-),\overline{P}_0 b_2(\Psi_1^-,Xe) \right]\,.
\end{align}
Finally, we note that the  $\mathcal{O}'$ contribution to \eqref{eq:O2propcon} decomposes into the $R$-charge eigenstates as
\begingroup
\allowdisplaybreaks
\begin{align}
\mathcal{O}' &= \frac{1}{6}\bigg\{\omega_L\left[b_2(Xe,\xi \Psi_1^+),b_2(\xi \Psi_1^-,\Psi_1^-)\right]+\omega_L\left[b_2(Xe,\xi \Psi_1^-),b_2(\xi \Psi_1^+,\Psi_1^+)\right]+\nonumber\\    
&\hspace{1.0cm}+\omega_L\left[b_2(Xe,\xi \Psi_1^-),b_2(\xi \Psi_1^+,\Psi_1^-)\right]+\omega_L\left[b_2(Xe,\xi \Psi_1^+),b_2(\xi \Psi_1^-,\Psi_1^+)\right]+\nonumber\\    
&\hspace{1.5cm}+\omega_L\left[b_2(Xe,\xi \Psi_1^-),b_2(\xi \Psi_1^-,\Psi_1^+)\right]+\omega_L\left[b_2(Xe,\xi \Psi_1^+),b_2(\xi \Psi_1^+,\Psi_1^-)\right]\bigg\}\,.
\end{align}
\endgroup
Again, we note that $c$-ghost saturation requires that we take the $e^\phi \eta$ part of $Xe$. For $e_\pm$, the corresponding correlators are generally non-zero because the $e^\phi \eta$ part of $Xe_\pm$ carries $R$-charge $\pm 1$. On the other hand the $e^\phi \eta$ part of $Xe_0$ is always $R$-neutral so that it always give zero and can be ignored. Summarizing our results up to this point, we have shown that $\mathcal{O}$ can be written as a sum of localized and contact terms
\begin{align}
    \mathcal{O} = \mathcal{O}^\text{loc} + \mathcal{O}^\text{con}\,,
\end{align}
where 
\begingroup
\allowdisplaybreaks
\begin{subequations}
\begin{align}
   \mathcal{O}^\text{loc} =& \,-\frac{1}{2}\bigg\{\omega_L \left[ b_2(\Psi_1^-,\xi \Psi_1^-),P_0 b_2(\xi \Psi_1^+,Xe^+) \right]+\nonumber\\
     &\hspace{5cm}+\omega_L \left[ b_2(\Psi_1^+,\xi \Psi_1^+), P_0 b_2(\xi \Psi_1^-,Xe^-) \right]\nonumber\\[+3mm]
     &\hspace{3cm} +\omega_L \left[b_2(\xi \Psi_1^-,\xi \Psi_1^+),P_0 b_2(\Psi_1^+,Xe^{-}) \right]+\nonumber\\
     &\hspace{5cm}+\omega_L \left[b_2(\xi \Psi_1^+,\xi \Psi_1^-), P_0 b_2(\Psi_1^-,Xe^{+}) \right]\bigg\}\,\,,\\
    \mathcal{O}^\text{con} =&\,\frac{1}{6}\bigg\{\omega_L\left[b_2(Xe,\xi \Psi_1^+),b_2(\xi \Psi_1^-,\Psi_1^-)\right]+\omega_L\left[b_2(Xe,\xi \Psi_1^-),b_2(\xi \Psi_1^+,\Psi_1^+)\right]+\nonumber\\    &\hspace{1cm}+\omega_L\left[b_2(Xe,\xi \Psi_1^-),b_2(\xi \Psi_1^+,\Psi_1^-)\right]+\omega_L\left[b_2(Xe,\xi \Psi_1^+),b_2(\xi \Psi_1^-,\Psi_1^+)\right]+\nonumber\\[2mm]    &\hspace{1cm}+\omega_L\left[b_2(Xe,\xi \Psi_1^-),b_2(\xi \Psi_1^-,\Psi_1^+)\right]+\omega_L\left[b_2(Xe,\xi \Psi_1^+),b_2(\xi \Psi_1^+,\Psi_1^-)\right]+\nonumber\\[2mm]
     &\hspace{1cm}+3\omega_L \left[ b_2(\Psi_1^-,\xi \Psi_1^-), b_2(\xi \Psi_1^+,Xe) \right]+\nonumber\\[2mm]
     &\hspace{5cm}+3\omega_L \left[ b_2(\Psi_1^+,\xi \Psi_1^+), b_2(\xi \Psi_1^-,Xe) \right]\nonumber\\[2mm]
     &\hspace{1cm} +3\omega_L \left[b_2(\xi \Psi_1^-,\xi \Psi_1^+), b_2(\Psi_1^+,Xe) \right]+\nonumber\\
     &\hspace{5cm}+3\omega_L \left[b_2(\xi \Psi_1^+,\xi \Psi_1^-),b_2(\Psi_1^-,Xe) \right]\bigg\}\,.
   \label{eq:O2con}
\end{align}
\end{subequations}
\endgroup
We will now show that $\mathcal{O}^\text{con}=0$ while $\mathcal{O}^\text{loc}$ is generally non-zero. Requiring that $\mathcal{O}^\text{loc}$ (and therefore the whole obstruction) vanishes will yield a non-trivial constraint on $\mathbb{V}_\frac{1}{2}$. 

\subsubsection{Cancellation of contact terms}
\label{subsub:locBerCon}

Let us first use cyclicity to absorb all terms shown in \eqref{eq:O2con} inside one simplectic form taken against $Xe$. We obtain
\begin{align}
  \mathcal{O}^\text{con} =\omega_L(Xe,Y^++Y^-)\,,
\end{align}
where we define
\begingroup
\begin{align}
Y^\pm &= -\frac{1}{6}\bigg\{b_2\left[\xi \Psi_1^\mp,b_2(\xi \Psi_1^\pm,\Psi_1^\pm)\right]+b_2\left[\xi \Psi_1^\pm,b_2(\xi \Psi_1^\mp,\Psi_1^\pm)\right]+\nonumber\\&\hspace{+3.0cm}+b_2\left[\xi \Psi_1^\pm,b_2(\xi \Psi_1^\pm,\Psi_1^\mp)\right]-3b_2\left[\xi \Psi_1^\mp,b_2(\Psi_1^\pm,\xi \Psi_1^\pm)\right]+\nonumber\\
&\hspace{8.5cm}+3b_2\left[\Psi_1^\pm,b_2(\xi \Psi_1^\mp,\xi \Psi_1^\pm)\right]\bigg\}\,.
\end{align}
\allowdisplaybreaks
\endgroup
Let us now show that $\eta  Y^\pm=0$. We have
\begingroup\allowdisplaybreaks
\begin{subequations}
\begin{align}
\eta  Y^\pm =&\,  \frac{1}{6}\bigg\{b_2\left[ \Psi_1^\mp,b_2(\xi \Psi_1^\pm,\Psi_1^\pm)\right]+b_2\left[ \Psi_1^\pm,b_2(\xi \Psi_1^\mp,\Psi_1^\pm)\right]+\nonumber\\
&\hspace{+3.0cm}+b_2\left[ \Psi_1^\pm,b_2(\xi \Psi_1^\pm,\Psi_1^\mp)\right]-3b_2\left[ \Psi_1^\mp,b_2(\Psi_1^\pm,\xi \Psi_1^\pm)\right]+\nonumber\\[+2.2mm]
&\hspace{8cm}-3b_2\left[\Psi_1^\pm,b_2( \Psi_1^\mp,\xi \Psi_1^\pm)\right]\nonumber \\[+2mm]
&\, +b_2\left[\xi \Psi_1^\mp,b_2( \Psi_1^\pm,\Psi_1^\pm)\right]+b_2\left[\xi \Psi_1^\pm,b_2( \Psi_1^\mp,\Psi_1^\pm)\right]+\nonumber\\[+2.2mm]
&\hspace{+3.0cm}+b_2\left[\xi \Psi_1^\pm,b_2( \Psi_1^\pm,\Psi_1^\mp)\right]-3b_2\left[\xi \Psi_1^\mp,b_2(\Psi_1^\pm, \Psi_1^\pm)\right]+\nonumber\\
&\hspace{8cm}+3b_2\left[\Psi_1^\pm,b_2(\xi \Psi_1^\mp, \Psi_1^\pm)\right]\bigg\}\nonumber\\
=&\,-\frac{1}{3}\bigg\{b_2\left[ \Psi_1^\mp,b_2(\Psi_1^\pm,\xi \Psi_1^\pm)\right]+b_2\left[\Psi_1^\pm,b_2(\xi \Psi_1^\pm, \Psi_1^\mp)\right]+\nonumber\\
&\hspace{+3.0cm}-b_2\left[\xi \Psi_1^\pm,b_2( \Psi_1^\mp,\Psi_1^\pm)\right]+b_2\left[\xi \Psi_1^\mp,b_2(\Psi_1^\pm, \Psi_1^\pm)\right]+\nonumber\\
&\hspace{8cm}-2b_2\left[\Psi_1^\pm,b_2(\xi \Psi_1^\mp, \Psi_1^\pm)\right]\bigg\}\nonumber\,,
\end{align}
\end{subequations}
\endgroup
where the last equality is easily seen to vanish due to the super-Jacobi identity. It follows that $\omega_L(Xe,Y^\pm)=0$ and therefore $\mathcal{O}^\text{con}=0$.

\subsubsection{Evaluation of localized terms}
\label{subsec:elt}

Let us finally evaluate the localized terms. For our convenience, we will do so in the large Hilbert space. We have
\begin{align}
\mathcal{O}^\text{loc}=
     &\,-\frac{1}{2}\bigg\{\omega_L \left[ b_2(\Psi_1^+,Xe^{-})-b_2(\Psi_1^-,Xe^{+}),P_0 b_2(\xi \Psi_1^-,\xi \Psi_1^+) \right]\nonumber\\
&\,\hspace{3cm}+\omega_L \left[ b_2(\xi \Psi_1^-,\Psi_1^-),P_0 b_2(\xi \Psi_1^+,Xe^+) \right]+\nonumber\\
     &\hspace{5cm}+\omega_L \left[ b_2(\xi \Psi_1^+,\Psi_1^+), P_0 b_2(\xi \Psi_1^-,Xe^-) \right]\bigg\}
\,.
\end{align}
It is then straightforward to compute that
\begingroup
\allowdisplaybreaks
\begin{subequations}
\begin{align}
P_0 b_2 (\xi \Psi_1^{-},\xi \Psi_1^{+})&= -\xi\p\xi c\p c \mathbb{H}_0 e^{-2\phi}\,,\\
P_0[b_2( \Psi_1^+,X e^-)-b_2 (\Psi_1^-,X e^+)]&= \,\eta c\, \tilde{\mathbb{H}}_0+\ldots\,,\label{eq:cE0}\\
P_0b_2(\xi \Psi_1^\pm,\Psi_1^\pm) &= -2\xi c\p c \mathbb{H}_1^\pm e^{-2\phi}\,,\\
P_0b_2(\xi \Psi_1^\pm, Xe^\pm) &= c\, \tilde{\mathbb{H}}_1^\pm+\ldots\,,\label{eq:1pa}
\end{align}
\end{subequations}
\endgroup
where we have denoted
\begin{subequations}
\label{eq:Hs}
\begin{align}
\lim_{z\to 0}\left[\mathbb{V}^\pm_\frac{1}{2}(z)\mathbb{V}^\pm_\frac{1}{2}(-z)\right] &= \mathbb{H}_1^\pm\,,\\
\lim_{z\to 0}\left[2z\left(\mathbb{V}^-_\frac{1}{2}(z)\mathbb{V}^+_\frac{1}{2}(-z)-\mathbb{V}^+_\frac{1}{2}(z)\mathbb{V}^-_\frac{1}{2}(-z)\right)\right] &= \mathbb{H}_0 \,,
\end{align}
\end{subequations}
and
\begin{subequations}
\label{eq:Es}
\begin{align}
\lim_{z\to 0}\left[\mathbb{V}^\pm_\frac{1}{2}(z)\tilde{\mathbb{V}}^\pm_\frac{1}{2}(-z)+\tilde{\mathbb{V}}^\pm_\frac{1}{2}(z)\mathbb{V}^\pm_\frac{1}{2}(-z)\right] &= \tilde{\mathbb{H}}_1^\pm \,,\\
\lim_{z\to 0}\left[2z\left(\mathbb{V}_{\frac{1}{2}}^-(z) \tilde{\mathbb{V}}_{\frac{1}{2}}^+(-z)\!-\!\tilde{\mathbb{V}}_{\frac{1}{2}}^+(z)\mathbb{V}_{\frac{1}{2}}^-(-z)\!-\!\mathbb{V}_{\frac{1}{2}}^+(z) \tilde{\mathbb{V}}_{\frac{1}{2}}^-(-z)\!+\!\tilde{\mathbb{V}}_{\frac{1}{2}}^-(z)\mathbb{V}_{\frac{1}{2}}^+(-z)\right)\right]  &=\tilde{\mathbb{H}}_0 \,.
\end{align}
\end{subequations}
Using \eqref{eq:ghostCorr} we finally obtain
\begin{align}
\mathcal{O} = \mathcal{O}^\text{loc} = -\langle \tilde{\mathbb{H}}_1^+|\mathbb{H}_1^-\rangle-\langle \tilde{\mathbb{H}}_1^-|\mathbb{H}_1^+\rangle-\frac{1}{2}\langle \tilde{\mathbb{H}}_0 |\mathbb{H}_0\rangle\,.
\label{eq:obstr}
\end{align}
Note that the overlaps in \eqref{eq:obstr} may potentially include a Chan-Paton trace. As the test state $e$ can be chosen arbitrarily, \eqref{eq:obstr} makes it clear that the necessary and sufficient conditions for the third-order obstruction to vanish are
\begin{subequations}
\begin{align}
    \mathbb{H}_1^\pm &=0\,,\\
    \mathbb{H}_0 &=0\,.
\end{align}
\end{subequations}
These are what \cite{Maccaferri:2019ogq} call the generalized ADHM equations. At the same time, they should be viewed as only necessary conditions in order for the deformation to be exactly marginal to all orders as one cannot exclude possible corrections potentially arising at higher orders in the deformation parameter. 
Also note that setting $e=\Psi_1$, we have $\tilde{\mathbb{H}}_1^\pm=2\mathbb{H}_1^\pm$ and $\tilde{\mathbb{H}}_0=2\mathbb{H}_0$ so that the obstruction becomes proportional to the localized quartic effective action of \cite{Maccaferri:2018vwo,Maccaferri:2019ogq}
\begin{align}
    \mathcal{O}\big|_{e=\Psi_1} = -4\,\left(\langle \mathbb{H}_1^+|\mathbb{H}_1^-\rangle+ \frac{1}{4}\langle \mathbb{H}_0|\mathbb{H}_0\rangle\right) = -4S_\text{eff}^{(4)}\,.
    \label{eq:O2eff}
\end{align}
This serves as a check of consistency of our manipulations, as we have shown that the relation \eqref{eq:ObstActRel}, which we have noted at the beginning (and which is originally due to \cite{Maccaferri:2019ogq}), continued to hold throughout our analysis. Ref.\ \cite{Maccaferri:2018vwo} also notes that the generalized ADHM equations $\mathbb{H}_1^\pm = \mathbb{H}_0=0$ are the flatness conditions for the quartic effective potential (as it is clear from the form of \eqref{eq:O2eff}). While it is clear from \eqref{eq:O2eff} that any marginal deformation which has vanishing third-order obstruction has to give rise to a flat direction of $S_\text{eff}^{(4)}$ (this was noted already by \cite{Maccaferri:2019ogq}), we observe that our analysis therefore also shows that provided that the third-order obstruction is given by the expression \eqref{eq:obstr} (that is, provided that the worldsheet theory admits an extended global $\mathcal{N}=2$ superconformal algebra with all marginal operators carrying $R$-charge $\pm 1$), then any flat direction of the quartic effective action gives rise to a marginal deformation which is exact up to third order in $\lambda$. This is a non-trivial result because vanishing of the obstructions to exact marginality against all possible test states could in principle be more restrictive than flatness of the effective potential.

\subsubsection{Adding stubs}
\label{subsub:locBerStubs}

In the case with stubs, the computation goes along similar lines as in the case without stubs (replacing $b_2$ by $B_2^{0}$), except for the fact that there is the additional fundamental 4-vertex term
\begin{align}
\mathcal{O}^\text{fund} =\frac{1}{6}\omega_S\left[Xe, B_3^{(0)}(X\Psi_1,\Psi_1,\Psi_1)\right]\,.
\end{align}
We also have to use the generalized super-Jacobi identity \eqref{eq:GenSuperJac} when manipulating the identity part of the $\overline{P}_0$ terms picked when moving one of the PCOs during the localization procedure. We will now show that these two modifications exactly compensate each other. Let us therefore write
\begin{align}
\mathcal{O} = \mathcal{O}^\text{loc}+\mathcal{O}^\text{con}+\mathcal{O}^\text{fund}\,.
\end{align}
For $\mathcal{O}^\text{loc}$ we again obtain
\begin{align}
\mathcal{O}^\text{loc}=
     &\,-\frac{1}{2}\bigg\{\omega_L \left[ B_2^{(0)}(\Psi_1^+,Xe^{-})-B_2^{(0)}(\Psi_1^-,Xe^{+}),P_0 B_2^{(0)}(\xi \Psi_1^-,\xi \Psi_1^+) \right]\nonumber\\
&\,\hspace{3cm}+\omega_L \left[ B_2^{(0)}(\xi \Psi_1^-,\Psi_1^-),P_0 B_2^{(0)}(\xi \Psi_1^+,Xe^+) \right]+\nonumber\\
     &\hspace{5cm}+\omega_L \left[ B_2^{(0)}(\xi \Psi_1^+,\Psi_1^+), P_0 B_2^{(0)}(\xi \Psi_1^-,Xe^-) \right]\bigg\}
\,,
\end{align}
which evaluates to the same expression \eqref{eq:obstr} as we have found in the case without stubs. For $\mathcal{O}^\text{con}$, we obtain
\begin{align}
\mathcal{O}^\text{con}=\omega_L(Xe,Y^++Y^-)\,,
\label{eq:YpYm}
\end{align}
where
\begingroup
\begin{align}
Y^\pm &= -\frac{1}{6}\bigg\{B_2^{(0)}\left[\xi \Psi_1^\mp,B_2^{(0)}(\xi \Psi_1^\pm,\Psi_1^\pm)\right]+B_2^{(0)}\left[\xi \Psi_1^\pm,B_2^{(0)}(\xi \Psi_1^\mp,\Psi_1^\pm)\right]+\nonumber\\
&\hspace{+7cm}+B_2^{(0)}\left[\xi \Psi_1^\pm,B_2^{(0)}(\xi \Psi_1^\pm,\Psi_1^\mp)\right]+\nonumber\\&\hspace{+2.0cm}-3B_2^{(0)}\left[\xi \Psi_1^\mp,B_2^{(0)}(\Psi_1^\pm,\xi \Psi_1^\pm)\right]+3B_2^{(0)}\left[\Psi_1^\pm,B_2^{(0)}(\xi \Psi_1^\mp,\xi \Psi_1^\pm)\right]\bigg\}\,.
\end{align}
\allowdisplaybreaks
\endgroup
This time, however, we obtain a non-zero answer when acting with $\eta $ on $Y^++Y^-$, because $B_2^{(0)}$ does not associate. In fact, we have
\begin{subequations}
\begin{align}
\eta  Y^\pm =&\,-\frac{1}{3}\bigg\{\eta  Q \xi B_3^{(0)}(\Psi_1^\mp,\Psi_1^\pm,\xi \Psi_1^\pm)-\eta  Q\xi B_3^{(0)}(\xi \Psi_1^\mp, \Psi_1^\pm, \Psi_1^\pm) \nonumber\\
&\hspace{3cm}-\eta \xi B_3^{(0)}(\Psi_1^\mp,\Psi_1^\pm,X \Psi_1^\pm)+\eta \xi B_3^{(0)}(X \Psi_1^\mp, \Psi_1^\pm, \Psi_1^\pm) \bigg\}\,,\label{eq:430a}\\
\equiv &\, \eta D^\pm
\end{align}
\end{subequations}
where we have used the generalized super-Jacobi identity \eqref{eq:GenSuperJac} and also inserted $1=\eta \xi_0+ \xi_0\eta $ to write the result as manifestly $\eta $-exact with 
\begin{align}
     D^\pm&=-\frac{1}{3}\bigg\{ Q \xi B_3^{(0)}(\Psi_1^\mp,\Psi_1^\pm,\xi \Psi_1^\pm)-  Q\xi B_3^{(0)}(\xi \Psi_1^\mp, \Psi_1^\pm, \Psi_1^\pm) \nonumber\\
&\hspace{4cm}- \xi B_3^{(0)}(\Psi_1^\mp,\Psi_1^\pm,X \Psi_1^\pm)+ \xi B_3^{(0)}(X \Psi_1^\mp, \Psi_1^\pm, \Psi_1^\pm) \bigg\}\,.
\end{align}
That is, the difference between $Y^\pm$ and $D^\pm$ will necessarily lie in the small Hilbert space so that we can replace $Y^\pm$ with $D^\pm$ inside \eqref{eq:YpYm}. This means that we can write
\begin{align}
\mathcal{O}^\text{con}=&\,-\frac{1}{3}\bigg\{\omega_S\left[Xe, B_3^{(0)}(\Psi_1^+,\Psi_1^-,X\Psi_1^-)\right]-\omega_S\left[Xe, B_3^{(0)}(X\Psi_1^+,\Psi_1^-,\Psi_1^-)\right]+\nonumber\\
&\hspace{1.5cm} +\omega_S\left[Xe, B_3^{(0)}(\Psi_1^-,\Psi_1^+,X\Psi_1^+)\right]-\omega_S\left[Xe, B_3^{(0)}(X\Psi_1^-,\Psi_1^+,\Psi_1^+)\right]\bigg\}\,.
\label{eq:refloc}
\end{align}
Here we note that $B_3^{(0)}$ provides a $b$-ghost, so that the $c$-ghost part of $Xe$ and $X\Psi_1^\pm$ is selected. However, recalling that the $c$-ghost part of $Xe$ and $X\Psi_1$ is $R$-neutral, we conclude that the second and the fourth term in \eqref{eq:refloc} are zero by $R$-charge conservation. Using the symmetry of the $B_3^{(0)}$ product, we therefore end up with
\begin{align}
\mathcal{O}^\text{con}=-\frac{1}{3}\omega_S\left[Xe, B_3^{(0)}(X\Psi_1^-,\Psi_1^+,\Psi_1^-)\right]-\frac{1}{3}\omega_S\left[Xe, B_3^{(0)}(X\Psi_1^+,\Psi_1^-,\Psi_1^+)\right]\,.
\end{align}
However, decomposing the fundamental bosonic 4-vertex term $\mathcal{O}^\text{fund}$ into $R$-charge eigenstates, we obtain
\begin{align}
\mathcal{O}^\text{fund}=&\,+\frac{1}{6}\omega_S\left[Xe, B_3^{(0)}(X\Psi_1^+,\Psi_1^+,\Psi_1^-)\right]+\frac{1}{6}\omega_S\left[Xe, B_3^{(0)}(X\Psi_1^+,\Psi_1^-,\Psi_1^+)\right]+\nonumber\\
&\,\hspace{+1cm}+\frac{1}{6}\omega_S\left[Xe, B_3^{(0)}(X\Psi_1^-,\Psi_1^+,\Psi_1^-)\right]+\frac{1}{6}\omega_S\left[Xe, B_3^{(0)}(X\Psi_1^-,\Psi_1^-,\Psi_1^+)\right]\nonumber\\
=&\,+\frac{1}{3}\omega_S\left[Xe, B_3^{(0)}(X\Psi_1^+,\Psi_1^-,\Psi_1^+)\right]+\frac{1}{3}\omega_S\left[Xe, B_3^{(0)}(X\Psi_1^-,\Psi_1^+,\Psi_1^-)\right]\,,
\end{align}
where in the last step we have again made us of the symmetry of $B_3^{(0)}$. We therefore obtain that $\mathcal{O}^\text{con}+\mathcal{O}^\text{fund}=0$. Altogether, we conclude that the obstruction is again given by \eqref{eq:obstr}.

\subsection{Localization: $X^2$ form}
\label{subsec:locX2}

Here we will localize the obstruction starting from the $X^2$ form \eqref{eq:OX2} (see also \cite{Mattiello:2019gxc}). Below (subsection \ref{subusub:locX2stubs}) we also show what changes need to be made when working with stubs.
It is clear that the only terms in $X^2 e$ which will contribute into the obstruction are those containing a single $c$ ghost insertion. It is straightforward to show that these are precisely
\begin{align}
X^2 e &= \frac{1}{2}c\p^2 (e^\phi \tilde{\mathbb{V}}_\frac{1}{2})-\frac{1}{2}(\p^2 c) e^\phi \tilde{\mathbb{V}}_\frac{ 1}{2}+c :\p \xi \eta : e^\phi \tilde{\mathbb{V}}_{\frac{1}{2}}+ce^\phi :G\tilde{\mathbb{V}}_1:-\frac{1}{2}ce^\phi \p^2\tilde{\mathbb{V}}_{\frac{1}{2}}+\ldots
\end{align}
It is therefore clear that we can write $X^2 e= (X^2 e)^++(X^2 e)^-$, where the two states
\begin{align}
(X^2 e)^\pm &= \frac{1}{2}c\p^2 (e^\phi \tilde{\mathbb{V}}_\frac{1}{2}^\pm)-\frac{1}{2}(\p^2 c) e^\phi \tilde{\mathbb{V}}^\pm_\frac{ 1}{2}+c :\p \xi \eta : e^\phi \tilde{\mathbb{V}}_{\frac{1}{2}}^\pm+ce^\phi :G^\pm\tilde{\mathbb{V}}_1:-\frac{1}{2}ce^\phi \p^2\tilde{\mathbb{V}}^\pm_{\frac{1}{2}}
\end{align}
carry charge $\pm 1$ under the localising $R$-current. This means that the $X^2$ propagator term in \eqref{eq:OX2} can be rewritten as
\begin{align}
&-\omega_S\left[X^2 e, m_2\left[\frac{b_0}{L_0}\overline{P}_0 m_2(\Psi_1^+,\Psi_1^+),\Psi_1^-\right]\right]- \omega_S\left[X^2 e, m_2\left[\Psi_1^+,\frac{b_0}{L_0}\overline{P}_0 m_2(\Psi_1^+,\Psi_1^-)\right] \right]\nonumber\\
&-\omega_S\left[X^2 e, m_2\left[\frac{b_0}{L_0}\overline{P}_0 m_2(\Psi_1^+,\Psi_1^-),\Psi_1^+\right]\right]- \omega_S\left[X^2 e, m_2\left[\Psi_1^+,\frac{b_0}{L_0}\overline{P}_0 m_2(\Psi_1^-,\Psi_1^+)\right] \right]\nonumber\\
&-\omega_S\left[X^2 e, m_2\left[\frac{b_0}{L_0}\overline{P}_0 m_2(\Psi_1^-,\Psi_1^+),\Psi_1^+\right]\right]- \omega_S\left[X^2 e, m_2\left[\Psi_1^-,\frac{b_0}{L_0}\overline{P}_0 m_2(\Psi_1^+,\Psi_1^+)\right] \right]\nonumber\\
&-\omega_S\left[X^2 e, m_2\left[\frac{b_0}{L_0}\overline{P}_0 m_2(\Psi_1^-,\Psi_1^-),\Psi_1^+\right]\right]- \omega_S\left[X^2 e, m_2\left[\Psi_1^-,\frac{b_0}{L_0}\overline{P}_0 m_2(\Psi_1^-,\Psi_1^+)\right] \right]\nonumber\\
&-\omega_S\left[X^2 e, m_2\left[\frac{b_0}{L_0}\overline{P}_0 m_2(\Psi_1^-,\Psi_1^+),\Psi_1^-\right]\right]- \omega_S\left[X^2 e, m_2\left[\Psi_1^-,\frac{b_0}{L_0}\overline{P}_0 m_2(\Psi_1^+,\Psi_1^-)\right] \right]\nonumber\\
&-\omega_S\left[X^2 e, m_2\left[\frac{b_0}{L_0}\overline{P}_0 m_2(\Psi_1^+,\Psi_1^-),\Psi_1^-\right]\right]- \omega_S\left[X^2 e, m_2\left[\Psi_1^+,\frac{b_0}{L_0}\overline{P}_0 m_2(\Psi_1^-,\Psi_1^-)\right] \right]\,.\nonumber
\end{align}
In each of the above terms, we will now move one of the PCOs from the test state on the insertion with opposite $R$-charge than the remaining two. That is, we will first go to the large Hilbert space by placing $\xi$ on the insertion where we want to move the PCO, then 
write $X^2 e = Q\xi Xe$ and finally move $Q$ onto the insertion with $\xi$. We obtain
\begingroup
\allowdisplaybreaks
\begin{align}
&\!\!\!\!\!\!+\!\omega_L\!\left[\xi X e,  m_2\left[\overline{P}_0 m_2(\Psi_1^+,\Psi_1^+),\xi \Psi_1^-\right]\right]\!+\! \omega_L\!\left[\xi X e,   m_2\left[\Psi_1^+,\overline{P}_0 m_2(\Psi_1^+,\xi \Psi_1^-)\right] \right]\nonumber\\
&\!\!\!\!\!\!+\!\omega_L\!\left[\xi X e,   m_2\left[\overline{P}_0 m_2(\Psi_1^+,\xi \Psi_1^-),\Psi_1^+\right]\right]\!+\! \omega_L\!\left[\xi X e,   m_2\left[\Psi_1^+,\overline{P}_0 m_2(\xi \Psi_1^-,\Psi_1^+)\right] \right]\nonumber\\
&\!\!\!\!\!\!+\!\omega_L\!\left[\xi X e, m_2\left[\overline{P}_0 m_2(  \xi \Psi_1^-,\Psi_1^+),\Psi_1^+\right]\right]\!-\! \omega_L\!\left[\xi X e,  m_2\left[\xi \Psi_1^-,\overline{P}_0 m_2(\Psi_1^+,\Psi_1^+)\right] \right]\nonumber\\
&\!\!\!\!\!\!+\!\omega_L\!\left[\xi X e,  m_2\left[\overline{P}_0 m_2(\Psi_1^-,\Psi_1^-),\xi \Psi_1^+\right]\right]\!+\! \omega_L\!\left[\xi X e,   m_2\left[\Psi_1^-,\overline{P}_0 m_2(\Psi_1^-,\xi \Psi_1^+)\right] \right]\nonumber\\
&\!\!\!\!\!\!+\!\omega_L\!\left[\xi X e,  m_2\left[\overline{P}_0 m_2(\Psi_1^-,\xi \Psi_1^+),\Psi_1^-\right]\right]\!+\! \omega_L\!\left[\xi X e,   m_2\left[\Psi_1^-,\overline{P}_0 m_2(\xi \Psi_1^+,\Psi_1^-)\right] \right]\nonumber\\
&\!\!\!\!\!\!+\!\omega_L\!\left[\xi X e,  m_2\left[\overline{P}_0 m_2(\xi \Psi_1^+,\Psi_1^-),\Psi_1^-\right]\right]\!-\! \omega_L\!\left[\xi X e,   m_2\left[\xi \Psi_1^+,\overline{P}_0 m_2(\Psi_1^-,\Psi_1^-)\right] \right]\nonumber\\
&\!\!\!\!\!\!+\!\omega_L\!\left[\xi X e,  m_2\left[\frac{b_0}{L_0}\overline{P}_0 m_2(\Psi_1^+,\Psi_1^+),X \Psi_1^-\right]\right]\!+\! \omega_L\!\left[\xi X e,  m_2\left[\Psi_1^+,\frac{b_0}{L_0}\overline{P}_0 m_2(\Psi_1^+,X \Psi_1^-)\right] \right]\nonumber\\
&\!\!\!\!\!\!+\!\omega_L\!\left[\xi X e,   m_2\left[\frac{b_0}{L_0}\overline{P}_0 m_2(\Psi_1^+,X \Psi_1^-),\Psi_1^+\right]\right]\!+\! \omega_L\!\left[\xi X e,  m_2\left[\Psi_1^+,\frac{b_0}{L_0}\overline{P}_0 m_2(X \Psi_1^-,\Psi_1^+)\right] \right]\nonumber\\
&\!\!\!\!\!\!+\!\omega_L\!\left[\xi X e, m_2\left[\frac{b_0}{L_0}\overline{P}_0 m_2(  X\Psi_1^-,\Psi_1^+),\Psi_1^+\right]\right]\!+\! \omega_L\!\left[\xi X e,  m_2\left[X \Psi_1^-,\frac{b_0}{L_0}\overline{P}_0 m_2(\Psi_1^+,\Psi_1^+)\right] \right]\nonumber\\
&\!\!\!\!\!\!+\!\omega_L\!\left[\xi X e,  m_2\left[\frac{b_0}{L_0}\overline{P}_0 m_2(\Psi_1^-,\Psi_1^-),X \Psi_1^+\right]\right]\!+\! \omega_L\!\left[\xi X e,  m_2\left[\Psi_1^-,\frac{b_0}{L_0}\overline{P}_0 m_2(\Psi_1^-,X \Psi_1^+)\right] \right]\nonumber\\
&\!\!\!\!\!\!+\!\omega_L\!\left[\xi X e,  m_2\left[\frac{b_0}{L_0}\overline{P}_0 m_2(\Psi_1^-,X \Psi_1^+),\Psi_1^-\right]\right]\!+\! \omega_L\!\left[\xi X e,  m_2\left[\Psi_1^-,\frac{b_0}{L_0}\overline{P}_0 m_2(X \Psi_1^+,\Psi_1^-)\right] \right]\nonumber\\
&\!\!\!\!\!\!+\!\omega_L\!\left[\xi X e,  m_2\left[\frac{b_0}{L_0}\overline{P}_0 m_2(X \Psi_1^+,\Psi_1^-),\Psi_1^-\right]\right]\!+\! \omega_L\!\left[\xi X e,   m_2\left[X \Psi_1^+,\frac{b_0}{L_0}\overline{P}_0 m_2(\Psi_1^-,\Psi_1^-)\right] \right]\label{eq:refAss}
\end{align}
\endgroup
We note that all propagator terms now vanish due to $R$-charge conservation (we are forced to take the $c$-ghost part in both $Xe$ and $X\Psi_1^\pm$, which is $R$-neutral) and the identity parts of the $\overline{P}_0 = 1-P_0$ terms cancel by associativity of $m_2$. We are therefore left with 
\begingroup
\allowdisplaybreaks
\begin{align}
&-\omega_L\left[{P}_0 m_2(\Psi_1^+,\Psi_1^+),  m_2(\xi \Psi_1^-,\xi X e^-)\right]+ \omega_L\left[{P}_0 m_2(\Psi_1^+,\xi \Psi_1^-),   m_2(\xi X e^-,\Psi_1^+)\right]\nonumber\\
&+\omega_L\left[{P}_0 m_2(\Psi_1^+,\xi \Psi_1^-),   m_2(\Psi_1^+,\xi X e^-)\right]+ \omega_L\left[{P}_0 m_2(\xi \Psi_1^-,\Psi_1^+),   m_2(\xi X e^-,\Psi_1^+)\right]\nonumber\\
&+\omega_L\left[{P}_0 m_2(  \xi \Psi_1^-,\Psi_1^+), m_2(\Psi_1^+,\xi X e^-)\right]+ \omega_L\left[{P}_0 m_2(\Psi_1^+,\Psi_1^+),  m_2(\xi X e^-,\xi \Psi_1^-) \right]\nonumber\\
&-\omega_L\left[{P}_0 m_2(\Psi_1^-,\Psi_1^-),  m_2(\xi \Psi_1^+,\xi X e^+)\right]+\omega_L\left[{P}_0 m_2(\Psi_1^-,\xi \Psi_1^+),   m_2(\xi X e^+,\Psi_1^-) \right]\nonumber\\
&+\omega_L\left[{P}_0 m_2(\Psi_1^-,\xi \Psi_1^+),  m_2(\Psi_1^-,\xi X e^+)\right]+\omega_L\left[{P}_0 m_2(\xi \Psi_1^+,\Psi_1^-),   m_2(\xi X e^+,\Psi_1^-) \right]\nonumber\\
&+\omega_L\left[{P}_0 m_2(\xi \Psi_1^+,\Psi_1^-),  m_2(\Psi_1^-,\xi X e^+)\right]+ \omega_L\left[{P}_0 m_2(\Psi_1^-,\Psi_1^-),   m_2(\xi X e^+,\xi \Psi_1^+) \right]\nonumber\,,
\end{align}
\endgroup
which can be rewritten in terms of the $b_2$ product as
\begin{align}
&-\frac{1}{2}\omega_L\left[ b_2(\xi \Psi_1^-,\xi X e^-),{P}_0 b_2(\Psi_1^+,\Psi_1^+) \right]- \omega_L\left[b_2(\Psi_1^+,\xi X e^-),{P}_0 b_2(\Psi_1^+,\xi \Psi_1^-)\right]\nonumber\\
&-\frac{1}{2}\omega_L\left[ b_2(\xi \Psi_1^+,\xi X e^+),{P}_0 b_2(\Psi_1^-,\Psi_1^-)\right]-\omega_L\left[b_2(\Psi_1^-,\xi X e^+),{P}_0 b_2(\Psi_1^-,\xi \Psi_1^+) \right]\,.
\end{align}
We are now ready to evaluate the obstruction. We will treat all insertions as if they were primary, because contributions coming from the anomalous transformation properties can be shown to exactly cancel with $\mathcal{O}_3$ (see \cite{Mattiello:2019gxc} for details).
It can then be shown that
\begin{subequations}
\begin{align}
P_0b_2(\Psi_1^\pm,\Psi_1^\pm) &= +2 c\p c\,\mathbb{H}_1^\pm e^{-2\phi}\,,\\
P_0b_2 (\Psi_1^\pm,\xi \Psi_1^\mp) &= - \xi c\p c\,\mathbb{H}_1 e^{-2\phi}\mp (1/2) \p\xi c\p c\,\mathbb{H}_0 e^{-2\phi}
\end{align}
\end{subequations}
where the auxiliary fields $\mathbb{H}_1^\pm$, $\mathbb{H}_0$ are as in \eqref{eq:Hs} and we define
\begin{subequations}
\begin{align}
\lim_{z\to 0}[\mathbb{V}^-_\frac{1}{2}(z)\mathbb{V}^+_\frac{1}{2}(-z)+\mathbb{V}^+_\frac{1}{2}(z)\mathbb{V}^-_\frac{1}{2}(-z)] &= \mathbb{H}_1 \,.
\end{align}
\end{subequations}
Keeping only the contributions containing exactly one $c$-ghost and neglecting anomalous terms in the OPEs, we further have
\begin{subequations}
\begin{align}
-P_0b_2(\Psi_1^+,\xi Xe^-)+P_0b_2(\Psi_1^-,\xi Xe^+) &=  c :\!\xi\eta\!:\mathbb{\tilde{H}}_0\,,\\
P_0b_2(\xi\Psi_1^\pm,\xi Xe^\pm)&= \xi c \mathbb{\tilde{H}}_1^\pm\,,
\end{align}
\end{subequations}
where the test-state auxiliary fields $\tilde{\mathbb{H}}_1^\pm$, $\tilde{\mathbb{H}}_0$ are as in \eqref{eq:Es}.
Using these results, it is then straightforward to establish that we recover expression \eqref{eq:obstr}, that is
\begin{align}
\mathcal{O} = -\langle \mathbb{\tilde{H}}_1^+|\mathbb{H}_1^-\rangle-\langle \mathbb{\tilde{H}}_1^-|\mathbb{H}_1^+\rangle-\frac{1}{2}\langle \mathbb{\tilde{H}}_0|\mathbb{H}_0\rangle\,.
\label{eq:obstruX2}
\end{align}

\subsubsection{Adding stubs}
\label{subusub:locX2stubs}

Two modifications of the above procedure are needed when working with stubs. First, as opposed to the case without stubs, the following terms in \eqref{eq:refAss}
\begingroup
\allowdisplaybreaks
\begin{align}
&+\omega_L\left[\xi X e^-,  M_2^{(0)}\left[ M_2^{(0)}(\Psi_1^+,\Psi_1^+),\xi \Psi_1^-\right]\right]+ \omega_L\left[\xi X e^-,   M_2^{(0)}\left[\Psi_1^+, M_2^{(0)}(\Psi_1^+,\xi \Psi_1^-)\right] \right]\nonumber\\
&+\omega_L\left[\xi X e^-,   M_2^{(0)}\left[ M_2^{(0)}(\Psi_1^+,\xi \Psi_1^-),\Psi_1^+\right]\right]+ \omega_L\left[\xi X e^-,   M_2^{(0)}\left[\Psi_1^+, M_2^{(0)}(\xi \Psi_1^-,\Psi_1^+)\right] \right]\nonumber\\
&+\omega_L\left[\xi X e^-,M_2^{(0)}\left[ M_2^{(0)}(  \xi \Psi_1^-,\Psi_1^+),\Psi_1^+\right]\right]- \omega_L\left[\xi X e^-,  M_2^{(0)}\left[\xi \Psi_1^-, M_2^{(0)}(\Psi_1^+,\Psi_1^+)\right] \right]\nonumber\\
&+\omega_L\left[\xi X e^+,  M_2^{(0)}\left[ M_2^{(0)}(\Psi_1^-,\Psi_1^-),\xi \Psi_1^+\right]\right]+ \omega_L\left[\xi X e^+,  M_2^{(0)}\left[\Psi_1^-, M_2^{(0)}(\Psi_1^-,\xi \Psi_1^+)\right] \right]\nonumber\\
&+\omega_L\left[\xi X e^+,  M_2^{(0)}\left[ M_2^{(0)}(\Psi_1^-,\xi \Psi_1^+),V^-\right]\right]+ \omega_L\left[\xi X e^+,  M_2^{(0)}\left[\Psi_1^-, M_2^{(0)}(\xi \Psi_1^+,\Psi_1^-)\right] \right]\nonumber\\
&+\omega_L\left[\xi X e^+,  M_2^{(0)}\left[ M_2^{(0)}(\xi \Psi_1^+,\Psi_1^-),\Psi_1^-\right]\right]- \omega_L\left[\xi X e^+,   M_2^{(0)}\left[\xi \Psi_1^+, M_2^{(0)}(\Psi_1^-,\Psi_1^-)\right] \right]\,,\nonumber
\end{align}
\endgroup
which arise when moving one of the PCOs in the propagator term, do not vanish, because the product $M_2^{(0)}$ does not associate. Instead, using the $A_\infty$ relation $[Q,M_3^{(0)}]+\frac{1}{2}[M_2^{(0)},M_2^{(0)}]=0$, these yield
\begingroup
\allowdisplaybreaks
\begin{align}
&-\omega_S\left[X^2 e^-, M_3^{(0)} ( \Psi_1^+,\Psi_1^+, \Psi_1^-)\right]+\omega_S\left[ X e^-, M_3^{(0)} ( \Psi_1^+,\Psi_1^+,X \Psi_1^-)\right]\nonumber\\
&-\omega_S\left[X^2 e^-, M_3^{(0)}(\Psi_1^+, \Psi_1^-,\Psi_1^+) \right]+\omega_S\left[ X e^-, M_3^{(0)}(\Psi_1^+,X \Psi_1^-,\Psi_1^+) \right]\nonumber\\
&-\omega_S\left[X^2 e^-, M_3^{(0)}(  \Psi_1^-,\Psi_1^+,\Psi_1^+) \right]+\omega_S\left[ X e^-, M_3^{(0)}(  X \Psi_1^-,\Psi_1^+,\Psi_1^+) \right]\nonumber\\
&-\omega_S\left[X^2 e^+, M_3^{(0)}(\Psi_1^-,\Psi_1^-,\Psi_1^+) \right]+\omega_S\left[ X e^+, M_3^{(0)}(\Psi_1^-,\Psi_1^-,X \Psi_1^+) \right]\nonumber\\
&-\omega_S\left[X^2 e^+, M_3^{(0)}(\Psi_1^-, \Psi_1^+,\Psi_1^-) \right]+\omega_S\left[ X e^+, M_3^{(0)}(\Psi_1^-,X \Psi_1^+,\Psi_1^-) \right]\nonumber\\
&-\omega_S\left[X^2 e^+, M_3^{(0)}( \Psi_1^+,\Psi_1^-,\Psi_1^-) \right]+\omega_S\left[ X e^+, M_3^{(0)}(X \Psi_1^+,\Psi_1^-,\Psi_1^-) \right]\,.\label{eq:X2associ}
\end{align}
\endgroup
Note that $M_3^{(0)}$ provides a $b$-ghost so that we are forced to take $c$-ghost terms in $X^2 e^{\pm}$, $Xe^\pm$ and also $X\Psi_1^\pm$. Since the $c$-ghost terms in $Xe^\pm$ and $X\Psi_1^\pm$ are $R$-neutral, the corresponding terms (second column of \eqref{eq:X2associ}) will vanish by $R$-charge conservation. Also, recall that the $c$-ghost terms in $X^2 e^\pm $ carry $R$-charge $\pm 1$ so that these will not in general vanish. This is, however, where the second modification comes into play: remember that the $X^2$ form of $\mathcal{O}$ with stubs contains, compared to the case without stubs, the term
\begin{align}
\omega_2\left[X^2 e, M_3^{(0)}(\Psi_1,\Psi_1,\Psi_1)\right]\,,
\end{align}
which, after the $R$-charge decomposition, precisely cancels with the $X^2$ terms in \eqref{eq:X2associ}. As the rest of the computation goes unchanged, we recover the result \eqref{eq:obstruX2}.
 
\subsection{Direct evaluation}
\label{subsec:direct}

Here we will use the strategy of \cite{Berkovits:2003ny} to evaluate the Berkovits-like form \eqref{eq:O2propcon} of the obstruction. While this will not put as strict requirements on the background as in the case of the previous two methods which were based on the $\mathcal{N}=2$ $R$-charge decomposition technique, we will not be able to express the obstruction as explicitly as we were able to in subsections \ref{subsec:locBerk} and \ref{subsec:locX2}. We will first deal with the propagator term $\mathcal{O}^\text{prop}$ in subsection
Let us define $a\equiv \sqrt{2}-1$. We will not consider stubs in this section. Proceeding along the lines of \cite{Berkovits:2003ny}, we can show that by introducing Schwinger parametrization for the Siegel-gauge propagator, it is possible to express  $\mathcal{O}^\text{prop}$ as
\begingroup
\allowdisplaybreaks
\begin{subequations}
\begin{align}
\mathcal{O}^\text{prop}&=-\frac{1}{2}\int_0^\infty dt\,\big\langle ( c\mathbb{V}_\frac{1}{2} e^{-\phi}(-a^{-1})c\tilde{\mathbb{V}}_1(+a^{-1})+c\tilde{\mathbb{V}}_1(-a^{-1})  c\mathbb{V}_\frac{1}{2} e^{-\phi}(+a^{-1}))\times\nonumber\\[-1.0mm]
&\hspace{2cm}\times b_0 e^{-tL_0}( c\mathbb{V}_\frac{1}{2} e^{-\phi}(+{a})c\mathbb{V}_1(-{a})+c\mathbb{V}_1(+{a})  c\mathbb{V}_\frac{1}{2} {e}^{-\phi}(-{a}))\big\rangle_S\\
&=-\frac{1}{2}\int_0^\infty dt\,\big\langle c(-a^{-1})c(+a^{-1}) b_0 c(e^{-t}{a})c(-e^{-t}{a})\big\rangle_S\times\nonumber\\[-1mm]
&\hspace{1.5cm}\times\big\langle( \mathbb{V}_\frac{1}{2} e^{-\phi}(-a^{-1})\tilde{\mathbb{V}}_1(+a^{-1})+\tilde{\mathbb{V}}_1(-a^{-1})  \mathbb{V}_\frac{1}{2} e^{-\phi}(+a^{-1}))\times\nonumber\\
&\hspace{2.0cm}\times( \mathbb{V}_\frac{1}{2} e^{-\phi}(+e^{-t}{a})\mathbb{V}_1(-e^{-t}{a})+\mathbb{V}_1(+e^{-t}{a})  \mathbb{V}_\frac{1}{2} {e}^{-\phi}(-e^{-t}{a}))\big\rangle_S\,.
\end{align}
\end{subequations}
\endgroup
Using the result
\begin{align}
\big\langle c(-a^{-1})c(+a^{-1}) b_0 c(e^{-t}{a})c(-e^{-t}{a})\big\rangle= -4e^{-t}(a^2 e^{-2t}-a^{-2})\,,
\end{align}
we eventually obtain
\begingroup\allowdisplaybreaks
\begin{align}
\mathcal{O}^\text{prop} &=2\int_0^\infty dt\, e^{-t}(a^2 e^{-2t}-a^{-2})\times \nonumber\\
&\hspace{0.9cm}\times\big[+ (a^{-1}+ae^{-t})^{-1}\big\langle\mathbb{V}_\frac{1}{2}(-a^{-1})\tilde{\mathbb{V}}_1(+a^{-1}) \mathbb{V}_\frac{1}{2}(+e^{-t}{a})\mathbb{V}_1(-e^{-t}{a})\big\rangle+\nonumber\\
&\hspace{1.5cm}-(a^{-1}-ae^{-t})^{-1}\big\langle \tilde{\mathbb{V}}_1(-a^{-1}) \mathbb{V}_\frac{1}{2} (+a^{-1}) \mathbb{V}_\frac{1}{2}(+e^{-t}{a})\mathbb{V}_1(-e^{-t}{a})\big\rangle+\nonumber\\
&\hspace{1.5cm}+(a^{-1}-ae^{-t})^{-1}\big\langle\mathbb{V}_\frac{1}{2} (-a^{-1}) \tilde{\mathbb{V}}_1(+a^{-1})\mathbb{V}_1(+e^{-t}{a}) \mathbb{V}_\frac{1}{2} (-e^{-t}{a})\big\rangle+\nonumber\\
&\hspace{1.5cm}-(a^{-1}+ae^{-t})^{-1} \big\langle \tilde{\mathbb{V}}_1(-a^{-1}) \mathbb{V}_\frac{1}{2} (+a^{-1}) \mathbb{V}_1(+e^{-t}{a}) \mathbb{V}_\frac{1}{2} (-e^{-t}{a}) \big\rangle\big]\,.
\label{eq:OpropDir}
\end{align}
\endgroup
For the contact term, we obtain
\begingroup
\allowdisplaybreaks
\begin{subequations}
\begin{align}
\mathcal{O}' &= -\frac{1}{6}\big\langle (\eta e^{\phi}\tilde{\mathbb{V}}_\frac{1}{2}(-a^{-1})\xi c\mathbb{V}_\frac{1}{2}e^{-\phi}(+a^{-1})-\xi c\mathbb{V}_\frac{1}{2}e^{-\phi}(-a^{-1})\eta e^{\phi}\tilde{\mathbb{V}}_\frac{1}{2}(+a^{-1}) )\times\nonumber\\
&\hspace{1.5cm}\times(\xi c\mathbb{V}_\frac{1}{2}e^{-\phi}(+a)c\mathbb{V}_\frac{1}{2}e^{-\phi}(-a)-c\mathbb{V}_\frac{1}{2}e^{-\phi}(a)\xi c\mathbb{V}_\frac{1}{2}e^{-\phi}(-a) )\big \rangle_L\\
&= +2 \big\langle \tilde{\mathbb{V}}_\frac{1}{2}(-a^{-1}) \mathbb{V}_\frac{1}{2}(+a^{-1}) \mathbb{V}_\frac{1}{2}(+a)\mathbb{V}_\frac{1}{2}(-a)\big\rangle+\nonumber\\
&\hspace{+4cm}+2\big\langle\mathbb{V}_\frac{1}{2}(-a^{-1}) \tilde{\mathbb{V}}_\frac{1}{2}(+a^{-1}) \mathbb{V}_\frac{1}{2}(+a)\mathbb{V}_\frac{1}{2}(-a) \big\rangle\,.
\label{eq:OprimeDir}
\end{align}
\end{subequations}
\endgroup

\section{Examples}
\label{sec:ex}

In this section we present a number of examples demonstrating the utility of the generalized ADHM equations $\mathbb{H}_1^\pm = \mathbb{H}_0=0$ in deriving algebraic constraints on moduli of various brane configurations. These will include the $\text{D}(-1)$/D3 brane system both in flat space (subsection \ref{subsec:N4flat}) and sitting at an orbifold singularity (subsection \ref{subsec:N4orb}), as well as a couple of more complicated brane configurations, some of which were discussed previously by \cite{Nekrasov:2015wsu,Nekrasov:2016qym,Nekrasov:2016gud} (subsection \ref{subsec:spiked}). In the case of the simple $\text{D}(-1)$/D3 brane system, we will explicitly verify validity of the localization technique by obtaining identical results using the direct evaluation method as outlined in subsection \ref{subsec:direct}.

\subsection{${N}=4$ SYM instantons}
\label{subsec:N4flat}
We will now apply our results on the system of superposed $k$ $\overline{\text{D}(-1)}$ branes and $N$ euclidean D3 branes which was in this context discussed by \cite{Witten:1995gx,Seiberg:1999xz,Billo:2002hm,Douglas:1995bn} and others and, most recently, by \cite{Mattiello:2019gxc,Maccaferri:2018vwo,Maccaferri:2019ogq}.
\begin{table}[htpb!]
\centering
\caption{${N}=4$ SYM instantons.}
\label{tab:N4sym}
\begin{tabular}{c||cc|cc|cc|cc||cc}
 & $X^1$ & $X^2$ & $X^3$ & $X^4$& $X^5 $& $X^6$& $X^7$& $X^8$& $X^9$& $X^0$\\
\hline
 $\overline{\text{D}(-1)}$ & $\bigcdot$ & $\bigcdot$ & $\bigcdot$ & $\bigcdot$ & $\bigcdot$ & $\bigcdot$ & $\bigcdot$ & $\bigcdot$ & $\bigcdot$& $\bigcdot$\\
 \hline
  D$3$ & $\times$ & $\times$ & $\times$ & $\times$ & $\bigcdot$ & $\bigcdot$ & $\bigcdot$ & $\bigcdot$ & $\bigcdot$& $\bigcdot$\\
\end{tabular}
\end{table}
We will complexify our target coordinates as $X^{r\pm} = (X^{2r-1}\pm i X^{2r})/\sqrt{2}$, where $r=1,\ldots,5$. The stack of D3 branes will be taken to span the complex coordinates $X^{1\pm},X^{2\pm}$ (see Table \ref{tab:N4sym}). These we may take to be toroidally compactified without changing the content of the discussion below. Such brane configuration preserves in total 8 spacetime supercharges, which give rise to $N=(1,0)$ supersymmetry in the six dimensions $X^5,X^6,X^7,X^8,X^9,X^0$ (that is $N=2$ in 4d and $N=(4,4)$ in 2d). Based on our discussion at the end of subsection \ref{subsubsec:prel}, we therefore expect to be able to extend the $\mathcal{N}=1$ worldsheet superconformal algebra to an $\mathcal{N}=2$ SCA with $R$-current $J$ with respect to which will all boundary marginal fields carry charges $\pm 1$. Focusing on the marginal operators along the Dirichlet-Neumann directions $X^1,X^2,X^3,X^4$, we will take
\begin{align}
\mathbb{V}_{\frac{1}{2}} &\equiv     \begin{pmatrix}
A_\mu \psi^\mu & w_{{\alpha}}\Delta S^{{\alpha}}\\
\bar{w}_{{\alpha}}\bar{\Delta} S^{{\alpha}} & a_\mu \psi^\mu   
    \end{pmatrix}\,,\qquad
    \tilde{\mathbb{V}}_{\frac{1}{2}} \equiv     \begin{pmatrix}
B_\mu \psi^\mu & v_{{\alpha}}\Delta S^{{\alpha}}\\
\bar{v}_{{\alpha}}\bar{\Delta} S^{{\alpha}} & b_\mu  \psi^\mu   
    \end{pmatrix}\,.
    \label{eq:VVtN4}
\end{align}
The Chan-Paton sectors explicitly displayed in \eqref{eq:VVtN4} therefore describe the strings localized on the D3 branes (upper-left corner), strings localized on the $\overline{\text{D}(-1)}$ branes (lower-right corner) and the strings stretched between the two brane stacks. Moreover, each of the four entries in \eqref{eq:VVtN4} is itself a matrix as we assume that the stacks of the two kinds of branes may consist of multiple branes. The $\mu=1,\ldots,4$ indices therefore run over the four (euclidean) D3 directions, $\psi^\mu$ are the $h=1/2$ worldsheet fermions and, $A$ are $N\times N$ matrix-valued $SO(4)$ vectors, $a_\mu$ are $k\times k$ matrix-valued $SO(4)$ vectors, 
$w_\alpha$ are $N\times k$ matrix-valued $SO(4)$ spinors (where $\alpha \in \{+,-\}$ is the chiral Weyl spinor index) and $\bar{w}_\alpha$ are $k\times N$ matrix-valued $SO(4)$ spinors. Also, $\Delta,\bar{\Delta}$ are the $h=1/4$ bosonic twist fields, $S^\alpha$ are the $h=1/4$ fermionic spin fields, implementing the change of boundary conditions on $\p X^\mu$ and $\psi^\mu$, respectively.

Note that if we were to consider D$(-1)$ branes instead of $\overline{\text{D}(-1)}$ branes, the stretched string modes would give rise to states $w^{\dot{\alpha}}\Delta S_{\dot{\alpha}}$, $\bar{w}^{\dot{\alpha}}\bar{\Delta} S_{\dot{\alpha}}$ instead of $w_{{\alpha}}\Delta S^{{\alpha}}$, $\bar{w}_{{\alpha}}\bar{\Delta} S^{{\alpha}}$, where $\dot{\alpha}\in \{\dot{+},\dot{-}\}$ is the anti-chiral Weyl spinor index. See Appendix \ref{app:spin} for our conventions on 4d euclidean spinors and Appendix \ref{app:ope} for some OPE and correlators of spin and twist fields. Also, imposing reality condition on the string field, 
we obtain reality conditions
\begin{align}
(A_\mu)^\dagger  = A_\mu\,,\qquad (a_\mu)^\dagger  = a_\mu\,,\qquad
(\bar{w}_\alpha)^\dagger  = w^\alpha\,,
\label{eq:realN4}
\end{align}
on the polarizations of $\mathbb{V}_{1/2}$, where the last condition is equivalent to the reality condition (3.4) of \cite{Douglas:1996sw}.\footnote{For the anti-chiral stretched worldsheet fermions, the corresponding reality condition can be easily checked to read $(\bar{w}_{\dot{\alpha}})^\dagger=w^{\dot{\alpha}}$.}
We take the localising $R$-current to be
\begin{align}
    J=J_1+J_2 = \sum_{r=1}^2 :\!\psi_{r-}\psi_{r+}\!:\,= -i\sum_{r=1}^2\p h_r\,,
\end{align}
of the free field $\mathcal{N}=2$ worldsheet superconformal algebra with $c=6$ along the four Dirichlet-Neumann directions (together with the stress-energy tensor $T=T_1+T_2$ and charged supercurrents $G^\pm=G^\pm_1+G^\pm_2$)\footnote{Note that had we started with $\text{D}(-1)$ branes instead of $\overline{\text{D}(-1)}$ branes, we would have to take $J_1 - J_2$ as our localizing $R$-current (and correspondingly $G_1^\pm + G_2^\mp$ as the two charged supercurrents).},
where we have bosonized the two complex worldsheet fermions along the D3 worldvolume as
\begin{align}
    \psi^{r\pm} = e^{\pm i h_r}\,,
\end{align}
where
\begin{align}
    \psi^{r\pm}&=\frac{1}{\sqrt{2}}(\psi^{2r-1}\pm i\psi^{2r})\,.
\end{align}
We then have $\Psi_1 = \Psi_1^++\Psi_1^-$, $e=e_++ e_-$ where
\begin{subequations}
\begin{align}
    \Psi_1^\pm &= c\mathbb{V}_{\frac{1}{2}}^\pm e^{-\phi}= c\begin{pmatrix}
    A_{r\pm} \psi^{r\pm} & w_{\pm} \Delta S^{(\pm\frac{1}{2},\pm\frac{1}{2})} \\
   \bar{w}_{{\pm}} \bar{\Delta} S^{(\pm\frac{1}{2},\pm\frac{1}{2})}  & a_{r\pm} \psi^{r\pm}
    \end{pmatrix}e^{-\phi}\,,\\
    e^\pm &= c \tilde{\mathbb{V}}_{\frac{1}{2}}^\pm e^{-\phi}=c\begin{pmatrix}
    B_{r\pm} \psi^{r\pm} &  v_{\pm} \Delta S^{(\pm\frac{1}{2},\pm\frac{1}{2})}  \\
    \bar{v}_{{\pm}} \bar{\Delta} S^{(\pm\frac{1}{2},\pm\frac{1}{2})}   & b_{r\pm} \psi^{r\pm}
    \end{pmatrix} e^{-\phi}\,,
\end{align}
\end{subequations}
where we have explicitely indicated the $(J_1,J_2)$ charges of the stretched spin-fields and we have denoted
\begin{subequations}
\begin{align}
A_{r\pm} &= \frac{1}{\sqrt{2}}(A_{2r-1}\mp i A_{2r})\,,\\
B_{r\pm} &= \frac{1}{\sqrt{2}}(B_{2r-1}\mp i B_{2r})
\end{align}
\end{subequations}
together with
\begin{subequations}
\begin{align}
a_{r\pm} &= \frac{1}{\sqrt{2}}(a_{2r-1}\mp i a_{2r})\,,\\
b_{r\pm} &= \frac{1}{\sqrt{2}}(b_{2r-1}\mp i b_{2r})\,.
\end{align}
\end{subequations}
The reality conditions \eqref{eq:realN4} then give $(A_{r\pm})^\dagger = A_{r\mp}$, $(a_{r\pm})^\dagger = a_{r\mp}$ together with $(\bar{w}_+)^\dagger = w_-$ and $(\bar{w}_-)^\dagger = -w_+$, so that we can work only with $A_{r+}$, $a_{r+}$, $w_+$ and $\bar{w}_+$.

\subsubsection{Substituting into the localized obstruction}
%
Let us first evaluate the obstruction using the form \eqref{eq:obstr} which was a consequence of the $\mathcal{N}=2$ decomposition technique. Substituting into \eqref{eq:Hs} and using the OPE from Appendix~\ref{app:ope}, we obtain (displaying only those Chan-Paton sectors of the auxiliary fields, which are non-zero)
\begin{subequations}
\begin{align}
(\mathbb{H}_1^+)_{\text{D3},\text{D3}} &= 
\big([A_{1+},A_{2+}]-w_{{+}} \bar{w}_{{+}}\big):\!\psi^{1+}\psi^{2+}\!:\,,\\[+1mm]
(\mathbb{H}_1^-)_{\text{D3},\text{D3}} &=\big( 
[A_{1-},A_{2-}]-w_{{-}} \bar{w}_{{-}}\big) :\!\psi^{1-}\psi^{2-}\!:\,,\\[+1mm]
(\mathbb{H}_0)_{\text{D3},\text{D3}} &= 
[A_{r-},A_{r+}]+w_{{+}} \bar{w}_{{-}}+w_{{-}}\bar{w}_{{+}} \,,
\end{align}
\end{subequations}
and
\begin{subequations}
\begin{align}
(\mathbb{H}_1^+)_{\overline{\text{D}(-1)},\overline{\text{D}(-1)}} &= \big(
 [a_{1+},a_{2+}]+\bar{w}_{+} w_{+}\big)
:\!\psi^{1+}\psi^{2+}\!:\,,\\[+1mm]
(\mathbb{H}_1^-)_{\overline{\text{D}(-1)},\overline{\text{D}(-1)}} &= \big( [a_{1-},a_{2-}]+\bar{w}_{-} w_{-}\big)
:\!\psi^{1-}\psi^{2-}\!:\,,\\[+1mm]
(\mathbb{H}_0)_{\overline{\text{D}(-1)},\overline{\text{D}(-1)}} &= [a_{r-},a_{r+}]-\bar{w}_{+} w_{-}-\bar{w}_{-} w_{+}\,.
\end{align}
\end{subequations}
Note that the reality conditions \eqref{eq:realN4} on $A_{r\pm}$, $w_\alpha$, $\bar{w}_\alpha$ imply that $(\mathbb{H}_1^\pm)^\dagger=\mathbb{H}_1^\mp$ and $(\mathbb{H}_0)^\dagger=\mathbb{H}_0$.
Analogously, for the test state auxiliary fields, we obtain
\begingroup
\allowdisplaybreaks
\begin{subequations}
\begin{align}
(\tilde{\mathbb{H}}_1^+)_{\text{D}3,\text{D}3} &=
\big([A_{1+},B_{2+}]-w_{{+}} \bar{v}_{{+}}+[B_{1+},A_{2+}]-v_{{+}} \bar{w}_{{+}}\big):\!\psi^{1+}\psi^{2+}\!:\,,\\[+1mm]
(\tilde{\mathbb{H}}_1^-)_{\text{D}3,\text{D}3} &=
\big([A_{1-},B_{2-}]-w_{{-}} \bar{v}_{{-}}+[B_{1-},A_{2-}]-v_{{-}} \bar{w}_{{-}}\big):\!\psi^{1-}\psi^{2-}\!:\,,\\[+1mm]
(\tilde{\mathbb{H}}_0)_{\text{D}3,\text{D}3} &=
[A_{r-},B_{r+}]+w_{{+}} \bar{v}_{{-}}+w_{{-}}\bar{v}_{{+}}+[B_{r-},A_{r+}]+v_{{+}} \bar{w}_{{-}}+v_{{-}}\bar{w}_{{+}}\,,
\end{align}
\end{subequations}
and
\begin{subequations}
\begin{align}
(\tilde{\mathbb{H}}_1^+)_{\overline{\text{D}(-1)},\overline{\text{D}(-1)}} &=
\big([a_{1+},b_{2+}]+\bar{w}_{+} v_{{+}}+[b_{1+},a_{2+}]+\bar{v}_{{+}} w_{{+}}\big):\!\psi^{1+}\psi^{2+}\!:\,,\\[+1mm]
(\tilde{\mathbb{H}}_1^-)_{\overline{\text{D}(-1)},\overline{\text{D}(-1)}} &=
\big([a_{1-},b_{2-}]+\bar{w}_{{-}} v_{{-}}+[b_{1-},a_{2-}]+\bar{v}_{{-}} w_{{-}}\big):\!\psi^{1-}\psi^{2-}\!:\,,\\[+1mm]
(\tilde{\mathbb{H}}_0)_{\overline{\text{D}(-1)},\overline{\text{D}(-1)}} &=
[a_{r-},b_{r+}]\!-\!\bar{w}_{{+}} v_{{-}}\!-\!\bar{w}_{{-}} v_{{+}}\!+\![b_{r-},a_{r+}]\!-\!\bar{v}_{{+}} w_{{-}}\!-\!\bar{v}_{{-}} w_{{+}}\,.
\end{align}
\end{subequations}
\endgroup
Note that the charged the auxilliary fields $\mathbb{H}_1^\pm$ precisely encode the complex hyper-K\"{a}hler moment maps for the instanton moduli space, while the coefficient inside the neutral auxilliary field $\mathbb{H}_0$ encodes the real hyper-K\"{a}hler moment maps. Or, put in different words, we see that the $\langle\mathbb{H}_1^+|\mathbb{H}_1^-\rangle $ term in the quartic part of the classical effective action (see \eqref{eq:O2eff}) can be identified with the F-term of the corresponding 4d ${N}=2$ low-energy effective action, while the $\langle \mathbb{H}_0|\mathbb{H}_0\rangle$ gives the D-term. That is, the necessary conditions $\mathbb{H}_1^\pm = \mathbb{H}_0=0$ for the exact marginality of the deformation in fact read (cf.\ equations (8.2) and (8.3) of \cite{Douglas:1996sw})
\begin{subequations}
\label{eq:adhm1}
\begin{align}
\mu^{\mathbb{C}} \equiv [a_{1+},a_{2+}]+\bar{w}_+ w_+ &=0\,,\label{eq:C1}\\[+1mm]
\tilde{\mu}^{\mathbb{C}} \equiv [A_{1+},A_{2+}]-{w}_+ \bar{w}_+ &=0\label{eq:C2}\,,
\end{align}
\end{subequations}
and
\begin{subequations}
\label{eq:RealMu}
\begin{align}
{\mu}^{\mathbb{R}} \equiv [(a_{r+})^\dagger,a_{r+}]+(w_+)^\dagger w_+-\bar{w}_+(\bar{w}_+)^\dagger &=0\,,\\[+1mm]
\tilde{\mu}^{\mathbb{R}} \equiv [(A_{r+})^\dagger,A_{r+}] -w_+ (w_+)^\dagger + (\bar{w}_+)^\dagger \bar{w}_+ &=0\,.
\end{align}
\end{subequations}
Note that the relative sign in front of the second term of \eqref{eq:C1} and \eqref{eq:C2} can be eliminated by relabeling $A_{r+}\to iA_{r+}$ (remember that $A_{r+}$ can be any complex matrix) without changing the signs inside \eqref{eq:RealMu}. The moduli space of the brane configuration is then obtained by solving \eqref{eq:adhm1} and \eqref{eq:RealMu} modulo the zero-momentum gauge transformations. It was also noticed in 
\cite{Billo:2002hm,Maccaferri:2018vwo} that the auxiliary fields can be re-expressed as
\begin{subequations}
\begin{align}
\mathbb{H}_1^\pm &= \mp \frac{i}{4}\eta_\mp^{\mu\nu} T_{\mu\nu}:\!\psi^{1\pm}\psi^{2\pm}\!:\,,\\
\mathbb{H}_0 &= -\frac{i}{2}\eta_3^{\mu\nu}T_{\mu\nu}\,,
\end{align}
\end{subequations}
and
\begin{subequations}
\begin{align}
\tilde{\mathbb{H}}_1^\pm &= \mp \frac{i}{2}\eta_\mp^{\mu\nu} \tilde{T}_{\mu\nu}:\psi^{1\pm}\psi^{2\pm}:\,,\\[+3pt]
\tilde{\mathbb{H}}_0 &= -i\eta_3^{\mu\nu}\tilde{T}_{\mu\nu}\,,
\end{align}
\end{subequations}
where the ladder t'Hooft symbols are defined as $\eta_\pm^{\mu\nu} \equiv \eta_1^{\mu\nu}\pm i \eta_2^{\mu\nu}$
and we denote
\begin{align}
T_{\mu\nu} &= \begin{pmatrix}
[A_\mu,A_\nu]-\frac{1}{2}w_\alpha (\sigma_{\mu\nu})^{\alpha\beta}\bar{w}_{\beta} & 0\\
0 & [a_\mu,a_\nu]+\frac{1}{2}\bar{w}_\alpha (\sigma_{\mu\nu})^{\alpha\beta} {w}_{\beta} 
\end{pmatrix}\,,\\
\tilde{T}_{\mu\nu} &= \begin{pmatrix}
[A_\mu,B_\nu]\!-\!\frac{1}{4}w_\alpha (\sigma_{\mu\nu})^{\alpha\beta}\bar{v}_{\beta}\!-\!\frac{1}{4}v_\alpha (\sigma_{\mu\nu})^{\alpha\beta}\bar{w}_{\beta} & 0\\
0 & \!\!\!\!\!\!\!\!\!\!\!\!\!\!\!\!\!\!\!\!\!\!\!\![a_\mu,b_\nu]\!+\!\frac{1}{4}\bar{w}_\alpha (\sigma_{\mu\nu})^{\alpha\beta} {v}_{\beta} \!+\!\frac{1}{4}\bar{v}_\alpha (\sigma_{\mu\nu})^{\alpha\beta} {w}_{\beta} 
\end{pmatrix}\,.
\end{align}
We therefore obtain
\begin{align}
\mathcal{O} = +\frac{1}{4}\mathrm{tr}[\tilde{T}_{\mu\nu} \eta^{\mu\nu}_a\eta^{\rho\sigma a}T_{\rho\sigma}] = +\frac{1}{2}\mathrm{tr}[\tilde{T}_{\mu\nu}T^{\mu\nu}]+\frac{1}{4}\varepsilon^{\mu\nu\rho\sigma}\mathrm{tr}[\tilde{T}_{\mu\nu}T_{\rho\sigma}]\,. 
\label{eq:513}
\end{align}
Eq. \eqref{eq:obstr} says that the obstruction vanishes if and only if $\mathbb{H}_1^\pm=\mathbb{H}_0=0$, i.e. if $\eta_a^{\mu\nu} T_{\mu\nu}=0$, that is if and only if
\begin{subequations}
\label{eq:flat}
\begin{align}
\eta_a^{\mu\nu}\bigg([A_\mu,A_\nu]-\frac{1}{2}w_\alpha (\sigma_{\mu\nu})^{\alpha\beta}\bar{w}_{\beta}\bigg)&=0\,,\\
\eta_a^{\mu\nu}\bigg([a_\mu,a_\nu]+\frac{1}{2}\bar{w}_\alpha (\sigma_{\mu\nu})^{\alpha\beta} {w}_{\beta}\bigg)&=0\,.
\end{align}
\end{subequations}
These are precisely the flatness conditions (6.16) and (6.17) of \cite{Maccaferri:2019ogq} for the quartic effective potential. A particular solution
\begingroup\allowdisplaybreaks
\begin{subequations}
\label{eq:solN4inst}
\begin{align}
\tensor{w}{^{\alpha i}} &= \rho\begin{pmatrix}
1 & 0\\
0 & 1
\end{pmatrix}\,,\\
\tensor{{\bar{w}}}{_\alpha^{i}} &=  \rho\begin{pmatrix}
1 & 0\\
0 & 1
\end{pmatrix}\,,\\[+2mm]
A_\mu &= \frac{{\rho}}{\sqrt{2}}{\sigma}_\mu\,,
\end{align}
\end{subequations}
\endgroup
of \eqref{eq:flat} for the $SU(2)$ gauge group with $k=1$ was found
in \cite{Maccaferri:2019ogq}, which corresponds to a blown-up instanton with size $\rho$. Upon substituting the polarizations \eqref{eq:solN4inst} into $\Psi_1$, we therefore obtain that $\Psi(\lambda)=\lambda\Psi_1+\lambda^2 \Psi_2 +\lambda^3 \Psi_3 +\mathcal{O}(\lambda^4)$ (where $\Psi_2$ and $\Psi_3$ are determined in terms of $\Psi_1$ by \eqref{eq:solSecOrd} and \eqref{eq:solThirdOrd}, respectively) is a solution of the classical equations of motion of the $A_\infty$ OSFT which is consistent up to third order in $\lambda$. Thus, we can conclude that our findings represent evidence that finite-size instantons provide consistent open superstring backgrounds.

\subsubsection{Direct evaluation}

We will now show that identical results for $\mathcal{O}$ are obtained by using the formulae \eqref{eq:OprimeDir} and \eqref{eq:OpropDir}. As we shall see, the way the intermediate results recombine into \eqref{eq:513} turns out to be somewhat non-trivial. This should therefore serve as a convincing check of the validity of the $\mathcal{N}=2$ localization method. See Appendix \ref{app:ope} for the various twist and spin field correlators and OPEs which we are going to use.
Let us first focus on terms coming from the (33)(33)(33)(33) correlators (that is, the Chan-Paton sectors localized on the $\text{D}3$ brane stack). Note taht this is a calculation which has already been done by \cite{Berkovits:2003ny}. We have
\begingroup
\allowdisplaybreaks
\begin{subequations}
\begin{align}
\mathcal{O}^\text{prop}\supset &-2\int_0^\infty dt\,e^{-t}(a^2 e^{-2t}-{a^{-2}})(a^{-1}+e^{-t}a)^{-4}\mathrm{tr}[A_\mu B_{\nu}A^\mu A^\nu]+\nonumber\\
&-2\int_0^\infty dt\,e^{-t}(a^2 e^{-2t}-a^{-2})(a^{-1}-e^{-t}a)^{-4}\mathrm{tr}[B_\mu A_{\nu}A^\nu A^\mu]+\nonumber\\
&-2\int_0^\infty dt\,e^{-t}(a^2 e^{-2t}-a^{-2})(a^{-1}-e^{-t}a)^{-4}\mathrm{tr}[A_\mu B_{\nu}A^\nu A^\mu]+\nonumber\\
&-2\int_0^\infty dt\,e^{-t}(a^2 e^{-2t}-a^{-2})(a^{-1}+e^{-t}a)^{-4}\mathrm{tr}[B_\mu A_{\nu}A^\mu A^\nu]\\
=&+\frac{1}{4}\mathrm{tr}[A_\mu B_{\nu}A^\mu A^\nu]+\frac{1}{2}\mathrm{tr}[B_\mu A_{\nu}A^\nu A^\mu]+\nonumber\\
&\hspace{2cm}  +\frac{1}{2}\mathrm{tr}[A_\mu B_{\nu}A^\nu A^\mu]+\frac{1}{4}\mathrm{tr}[B_\mu A_{\nu}A^\mu A^\nu]\,,\label{eq:part1}
\end{align}
\end{subequations}
\endgroup
where we have used the integrals
\begin{subequations}
\label{eq:intBer}
\begin{align}
\int_0^\infty dt\, e^{-t}(a^2 e^{-2t}-a^{-2})(a^{-1}-ae^{-t})^{-4} &= -\frac{1}{4}\,,\\
\int_0^\infty dt\, e^{-t}(a^2 e^{-2t}-a^{-2})(a^{-1}+ae^{-t})^{-4} &= -\frac{1}{8}\,.
\end{align}
\end{subequations}
We can also show that
\begingroup
\allowdisplaybreaks
\begin{align}
\mathcal{O}' &\supset 
-\mathrm{tr}[  A_\mu B_\nu A^\nu A^\mu]-\mathrm{tr}[ B_\mu A_\nu A^\nu A^\mu]\nonumber\\
&\hspace{2cm} +\frac{1}{4} \mathrm{tr}[B_\mu A_\nu A^\mu A^\nu]+\frac{1}{4} \mathrm{tr}[A_\mu B_\nu A^\mu A^\nu]\,.\label{eq:part2}
\end{align}
\endgroup
Finally, putting \eqref{eq:part1} and \eqref{eq:part2} together, we find that
\begin{align}
\mathcal{O}^\text{prop}+\mathcal{O}'\supset&\,
+\frac{1}{2}\mathrm{tr}[[A_\mu ,B_{\nu}][A^\mu, A^\nu]]\,,
\end{align}
which is precisely the (33)(33)(33)(33) contribution to \eqref{eq:513}. By an identical computation, we obtain the (11)(11)(11)(11) contribution (that is, contribution from the Chan-Paton sectors localized on the $\text{D}(-1)$ brane stack) to $\mathcal{O}^\text{prop}+\mathcal{O}'$ 
\begin{align}
\mathcal{O}^\text{prop}+\mathcal{O}' \supset +\frac{1}{2}\mathrm{tr}[[a_\mu ,b_{\nu}][a^\mu, a^\nu]]\,,
\end{align}
which reproduces the corresponding term in \eqref{eq:513}. Next, let us consider the contributions of the type $(33)(31)(11)(13)$. Using that $(\varepsilon\sigma^\mu\bar{\sigma}^\nu)^{\alpha\beta}+(\varepsilon\sigma^\nu\bar{\sigma}^\mu)^{\beta\alpha}=2\delta^{\mu\nu}\varepsilon^{\alpha\beta}$ and various integrals of the type \eqref{eq:intBer}, we obtain the corresponding contributions to both $\mathcal{O}^\text{prop}$ and $\mathcal{O}'$ vanish.
As for the (33)(33)(31)(13) and (11)(11)(13)(31) terms, let us only focus on contributions proportional to $\text{tr}[A_\mu B_\nu w_\alpha\bar{w}_\beta]$ -- the remaining 7 contributions will follow using very similar calculation. Evaluating the corresponding correlators using formulae from appendix \ref{app:ope}, we first obtain
\begingroup
\allowdisplaybreaks
\begin{align}
\mathcal{O}'
\supset & 
+\sqrt{2}\bigg[\delta^{\mu\nu}\varepsilon^{\alpha\beta}-\frac{1}{4}(\varepsilon\sigma^\mu\bar{\sigma}^\nu)^{\alpha\beta}\bigg]\mathrm{tr}[A_\mu B_\nu w_\alpha\bar{w}_\beta]\,.\label{eq:ABwwbp}
\end{align}
\endgroup
Computing the corresponding contribution to $\mathcal{O}^\text{prop}$ again involves various integrals of the type \eqref{eq:intBer}. One eventually obtains
\begin{align}
   \mathcal{O}^\text{prop}\supset &- \mathrm{tr}[A_\mu B_\nu w_\alpha\bar{w}_\beta] \Bigg\{
    \left(\frac{2-\sqrt{2}}{8}\right)(\varepsilon\sigma^\mu\bar{\sigma}^
    \nu)^{\alpha\beta}+\left(\frac{\sqrt{2}-1}{4}\right)(\varepsilon\sigma^\nu
    \bar{\sigma}^\mu)^{\alpha\beta}+\nonumber\\
    &\hspace{+7cm}
    +\frac{\sqrt{2}}{8}(\varepsilon\sigma^\mu\bar{\sigma}^\nu)^{\alpha\beta}+\frac{\sqrt{2}}{2}\varepsilon^{\alpha\beta}\delta^{\mu\nu}
    \Bigg\}\,,
    \label{eq:ABwwb4}
\end{align}
where the four terms inside of the curly brackets precisely correspond to the four terms constituting \eqref{eq:OpropDir}. It is then straigtforward to show that \eqref{eq:ABwwb4} combines with \eqref{eq:ABwwbp} to give
\begingroup
\allowdisplaybreaks
\begin{align}
\mathcal{O}^\text{prop}+\mathcal{O}'
\supset & 
-\frac{1}{2}(\varepsilon\sigma^{\mu\nu})^{\alpha\beta}\mathrm{tr}[A_\mu B_\nu w_\alpha\bar{w}_\beta]\,,
\end{align}
\endgroup
which is indeed the correct contribution to \eqref{eq:513}.
Finally, we consider the (31)(13)(31)(13) and (13)(31)(13)(31) contributions to $\mathcal{O}^\text{prop}+\mathcal{O}'$. 
We obtain
\begingroup
\allowdisplaybreaks
\begin{align}
\mathcal{O}^\text{prop}+\mathcal{O}' 
 \supset&
 -\frac{1}{2}\mathrm{tr}[v_\alpha \bar{w}_\beta w_\gamma \bar{w}_\delta+w_\alpha \bar{v}_\beta w_\gamma \bar{w}_\delta](\varepsilon^{\alpha\beta}\varepsilon^{\gamma\delta}-\varepsilon^{\alpha\delta}\varepsilon^{\beta\gamma})
\nonumber\\
&\hspace{1cm}-\frac{1}{2}\mathrm{tr}[\bar{v}_\alpha w_\beta \bar{w}_\gamma w_\delta+\bar{w}_\alpha v_\beta \bar{w}_\gamma w_\delta](\varepsilon^{\alpha\beta}\varepsilon^{\gamma\delta}-\varepsilon^{\alpha\delta}\varepsilon^{\beta\gamma})\nonumber\\
&\hspace{1cm}-\frac{1}{2}\mathrm{tr}[\bar{v}_\alpha w_\beta \bar{w}_\gamma w_\delta+\bar{w}_\alpha v_\beta \bar{w}_\gamma w_\delta]\varepsilon^{\alpha\delta}\varepsilon^{\beta\gamma}\nonumber\\
&\hspace{1cm}-\frac{1}{2}\mathrm{tr}[v_\alpha\bar{w}_\beta w_\gamma \bar{w}_\delta+w_\alpha\bar{v}_\beta w_\gamma \bar{w}_\delta]\varepsilon^{\alpha\delta}\varepsilon^{\beta\gamma}\,.
\label{eq:524}
\end{align}
\endgroup
where the first two terms in \eqref{eq:524} were supplied by $\mathcal{O}'$ while the rest of the expression comes from $\mathcal{O}^\text{prop}$. Using the cyclic properties of the trace, we have
\begin{subequations}
\begin{align}
\frac{1}{2}\mathrm{tr}[v_\alpha\bar{w}_\beta w_\gamma \bar{w}_\delta+w_\alpha\bar{v}_\beta w_\gamma \bar{w}_\delta]\varepsilon^{\alpha\delta}\varepsilon^{\beta\gamma} &= -\frac{1}{2}\mathrm{tr}[\bar{w}_\alpha v_\beta\bar{w}_\gamma w_\delta +\bar{v}_\alpha w_\beta \bar{w}_\gamma w_\delta]\varepsilon^{\alpha\beta}\varepsilon^{\gamma\delta}\,,\\
\frac{1}{2}\mathrm{tr}[v_\alpha\bar{w}_\beta w_\gamma \bar{w}_\delta+w_\alpha\bar{v}_\beta w_\gamma \bar{w}_\delta]\varepsilon^{\alpha\beta}\varepsilon^{\gamma\delta} &= -\frac{1}{2}\mathrm{tr}[\bar{w}_\alpha v_\beta\bar{w}_\gamma w_\delta +\bar{v}_\alpha w_\beta \bar{w}_\gamma w_\delta]\varepsilon^{\alpha\delta}\varepsilon^{\beta\gamma}\,,
\end{align}
\end{subequations}
so that \eqref{eq:524} can be rewritten as
\begingroup
\allowdisplaybreaks
\begin{align}
\mathcal{O}^\text{prop}+\mathcal{O}' 
 \supset&
- \frac{1}{2}\mathrm{tr}[\bar{v}_\alpha w_\beta \bar{w}_\gamma w_\delta+\bar{w}_\alpha v_\beta \bar{w}_\gamma w_\delta](\varepsilon^{\alpha\beta}\varepsilon^{\gamma\delta}-\varepsilon^{\alpha\delta}\varepsilon^{\beta\gamma})\,.
\label{eq:3131fin}
\end{align}
\endgroup
But at the same time, we note that the (31)(13)(31)(13) and (13)(31)(13)(31) contributions to \eqref{eq:513} can be written as
\begin{align}
&+\frac{1}{16}\mathrm{tr}[w_\alpha \bar{v}_\beta w_\gamma\bar{w}_\delta +v_\alpha \bar{w}_\beta w_\gamma\bar{w}_\delta]\bigg[\tensor{(\sigma_{\mu\nu})}{^{\alpha\beta}}\tensor{(\sigma^{\mu\nu})}{^{\gamma\delta}}+\frac{1}{2}\varepsilon^{\mu\nu\rho\sigma}\tensor{(\sigma_{\mu\nu})}{^{\alpha\beta}}\tensor{(\sigma_{\rho\sigma})}{^{\gamma\delta}}\bigg]+\nonumber\\
&\hspace{+1.5cm}+\frac{1}{16}\mathrm{tr}[\bar{w}_\alpha {v}_\beta \bar{w}_\gamma {w}_\delta +\bar{v}_\alpha {w}_\beta \bar{w}_\gamma {w}_\delta]\bigg[\tensor{(\sigma_{\mu\nu})}{^{\alpha\beta}}\tensor{(\sigma^{\mu\nu})}{^{\gamma\delta}}+\frac{1}{2}\varepsilon^{\mu\nu\rho\sigma}\tensor{(\sigma_{\mu\nu})}{^{\alpha\beta}}\tensor{(\sigma_{\rho\sigma})}{^{\gamma\delta}}\bigg]\,.\label{eq:step1}
\end{align}
We can also show that
\begin{subequations}
\begin{align}
\tensor{(\sigma_{\mu\nu})}{^{\alpha\beta}}\tensor{(\sigma^{\mu\nu})}{^{\gamma\delta}} &= 4(\varepsilon^{\alpha\gamma}\varepsilon^{\beta\delta}+\varepsilon^{\alpha\delta}\varepsilon^{\beta\gamma})\,,\\
\varepsilon^{\mu\nu\rho\sigma}\tensor{(\sigma_{\mu\nu})}{^{\alpha\beta}}\tensor{(\sigma_{\rho\sigma})}{^{\gamma\delta}} &= 8(\varepsilon^{\alpha\gamma}\varepsilon^{\beta\delta}+\varepsilon^{\alpha\delta}\varepsilon^{\beta\gamma})\,,
\end{align}
\end{subequations}
so that, using cyclicity of the trace, one can show that \eqref{eq:step1} is indeed equal to \eqref{eq:3131fin}.
\begin{align}
+\frac{1}{2}\mathrm{tr}[\bar{w}_\alpha {v}_\beta \bar{w}_\gamma {w}_\delta +\bar{v}_\alpha {w}_\beta \bar{w}_\gamma {w}_\delta](\varepsilon^{\alpha\delta}\varepsilon^{\beta\gamma}-\varepsilon^{\alpha\beta}\varepsilon^{\gamma\delta})\,,
\end{align}
We have therefore shown that in the case of the D$(-1)$/D3 system, one obtains identical results for the obstruction whether one uses the localization method or evaluates \eqref{eq:O2propcon} directly.

\subsection{Instantons on unresolved ALE spaces}
\label{subsec:N4orb}

Here we will consider blowing up D-instantons inside D3 branes which are placed inside ALE spaces in their orbifold limit \cite{Douglas:1996sw,Johnson:1996py,kronheimer1990yang,Nakajima:1994nid}. For the sake of concreteness, we will focus on the A-series of the ADE classification of the orbifold singularities, but our results are straightforwardly extendable to D-type and E-type ALE spaces as well (see \cite{Johnson:1996py}). In their singular limit, the $A_{n-1}$-type ALE spaces coincide with the $\mathbb{C}^2 /\mathbb{Z}_n$ supersymmetric orbifold where $n=1,2,\ldots$ We will consider the $\mathbb{C}^2$ to extend along the directions $X^1,X^2,X^3,X^4$. 
Defining the complexified coordinates $X^{r\pm}=(X^{2r-1}\pm i X^{2r})/\sqrt{2}$, the $\mathbb{Z}_n$ acts as
\begin{subequations}
\begin{align}
gX^{1\pm}g^{-1} = \xi^\pm X^{1\pm}\,,\\
gX^{2\pm}g^{-1} = \xi^\mp X^{2\pm}\,,
\end{align}
\end{subequations}
where $\xi=e^{2\pi i/n}$ is the $n^\mathrm{th}$ root of unity.
We will consider placing $k$ $\overline{\text{D}(-1)}$ branes at the fixed point of the $\mathbb{Z}_n$ action together with $N$ euclidean D3 branes extending along the $\mathbb{C}^2$ directions: see Table~\ref{tab:ALE}.
\begin{table}[htpb!]
\centering
\caption{Instantons on ALE spaces.}
\label{tab:ALE}
\begin{tabular}{c||cc|cc|cc|cc||cc}
 & $X^1$ & $X^2$ & $X^3$ & $X^4$& $X^5 $& $X^6$& $X^7$& $X^8$& $X^9$& $X^0$\\
\hline
 $\overline{\text{D}(-1)}$ & $\bigcdot$ & $\bigcdot$ & $\bigcdot$ & $\bigcdot$ & $\bigcdot$ & $\bigcdot$ & $\bigcdot$ & $\bigcdot$ & $\bigcdot$& $\bigcdot$\\
 \hline
  D$3$ & $\times$ & $\times$ & $\times$ & $\times$ & $\bigcdot$ & $\bigcdot$ & $\bigcdot$ & $\bigcdot$ & $\bigcdot$& $\bigcdot$\\
   \hline
  $\mathbb{C}^2/\mathbb{Z}_n$ & $\times$ & $\times$ & $\times$ & $\times$ & $\bigcdot$ & $\bigcdot$ & $\bigcdot$ & $\bigcdot$ & $\bigcdot$& $\bigcdot$
\end{tabular}
\end{table}
Such background has the same spacetime supersymmetry as the $\text{D}(-1)$/D3 system on the flat space, so we should again expect to be able to decompose $\Psi_1 = \Psi_1^++\Psi_1^-$. We will denote by $k_I$ and $N_I$ for $I=1,\ldots ,n$ the number of $\overline{\text{D}(-1)}$ branes and D3 branes carrying the $n$ distinct twisted RR-charges (that is $k_1 +k_2+\ldots + k_n = k$ and $N_1+N_2+\ldots+N_n=N$). The matter part of the most general marginal deformation in the $\mathbb{C}^2$ directions is then written as
\begin{align}
\mathbb{V}^\pm_\frac{1}{2}=\begin{pmatrix}
A^{IA_I,JB_J}_{r\pm}\psi^{r\pm} & w_{{\pm}}^{IA_I,J b_J} \Delta S^{(\pm\frac{1}{2},\pm\frac{1}{2})} \\
 \bar{w}_{{\pm}}^{Ia_I,JB_J} \bar{\Delta} S^{(\pm\frac{1}{2},\pm\frac{1}{2})} & a^{Ia_I,Jb_J}_{r\pm}\psi^{r\pm} \,,
\end{pmatrix}
\end{align}
where $I,J=1,\ldots,n$ and $A_I = 1,\ldots,N_I$, $a_I=1,\ldots,k_I$ are the fundamental $U(N_I)$ and $U(k_I)$ indices. Invariance under the $\mathbb{Z}_n$ action 
\begingroup\allowdisplaybreaks
\begin{subequations}
\begin{align}
\gamma(g) A^{IA_I,JB_J}_{1\pm}\gamma(g)^{-1} &= \xi^{I-J\pm 1}A^{IA_I,JB_J}_{1\pm}\,,\\
\gamma(g) A^{IA_I,JB_J}_{2\pm}\gamma(g)^{-1} &= \xi^{I-J\mp 1}A^{IA_I,JB_J}_{2\pm}\,,\\
\gamma(g) a^{Ia_J,Jb_J}_{1\pm}\gamma(g)^{-1} &= \xi^{I-J\pm 1} a^{Ia_I,Jb_J}_{1\pm}\,,\\
\gamma(g) a^{Ia_J,Jb_J}_{2\pm}\gamma(g)^{-1} &= \xi^{I-J\mp 1} a^{Ia_I,Jb_J}_{2\pm}\,,\\
\gamma(g) w^{IA_I,Jb_J}_{\alpha}\gamma(g)^{-1} &= \xi^{I-J}w^{IA_I,Jb_J}_{\alpha}\,,\\
\gamma(g) \bar{w}^{Ia_I,JB_J}_{\alpha}\gamma(g)^{-1} &= \xi^{I-J} \bar{w}^{Ia_I,JB_J}_{\alpha}\,,
\end{align}
\end{subequations}
\endgroup
however, implies various selection rules on $I,J$. Keeping only the non-zero entries of $\mathbb{V}_{1/2}^\pm$ we actually have (suppressing the $U(N_I)$ and $U(k_I)$ indices)
\begin{align}
\mathbb{V}_\frac{1}{2}^\pm =\begin{pmatrix}
A^{I,I\pm 1}_{1\pm}\psi^{1\pm}+A^{I,I\mp 1}_{2\pm}\psi^{2\pm} &  w_{{\pm}}^{I,I} \Delta S^{(\pm\frac{1}{2},\pm\frac{1}{2})} \\
 \bar{w}_{{\pm}}^{I,I} \bar{\Delta} S^{(\pm\frac{1}{2},\pm\frac{1}{2})} & a^{I,I\pm 1}_{1\pm}\psi^{1\pm} +a^{I,I\mp 1}_{2\pm}\psi^{2\pm}
\end{pmatrix}\,,
\end{align}
that is, the $(33)$ and $(11)$ Chan-Paton sectors are block upper and lower diagonal for $\mathbb{V}^+_{1/2}$ and $\mathbb{V}^-_{1/2}$ along $X^1,X^2$, respectively, and vice versa along $X^3,X^4$. The $(31)$ and $(13)$ Chan-Paton sectors are block diagonal. Imposing reality condition on the string field (that is $(\mathbb{V}_{1/2}^\pm)^\dagger=\mathbb{V}_{1/2}^\mp$) imposes reality conditions
\begin{align}
(A_{r\pm}^{I,I\pm 1})^\dagger = A_{r\mp}^{I\pm 1, I}\,,\qquad (a_{r\pm}^{I,I\pm 1})^\dagger = a_{r\mp}^{I\pm 1, I}\,,\qquad (\bar{w}_\alpha^{I,I})^\dagger = (w^\alpha)^{I,I}\,.
\label{eq:realityALE}
\end{align}
It is then straightforward to evaluate the auxiliary fields using formulae \eqref{eq:Hs}, which yields (again, displaying the non-zero Chan-Paton sectors only)
\begin{subequations}
\begin{align}
(\mathbb{H}_1^+)_{\text{D3}_I ,\text{D3}_J} &= \delta_{IJ}\left(A^{1+}_{I,I+1}A^{2+}_{I+1,I}-A^{2+}_{I,I-1}A^{1+}_{I-1,I} -w_+^{I,I}\bar{w}_+^{I,I}\right):\!\psi^{1+}\psi^{2+}\!:\,,\\
(\mathbb{H}_1^-)_{\text{D3}_I, \text{D3}_J} &= \delta_{IJ}\left(A^{1-}_{I,I-1}A^{2-}_{I-1,I}-A^{2-}_{I,I+1}A^{1-}_{I+1,I} -w_-^{I,I}\bar{w}_-^{I,I}\right):\!\psi^{1-}\psi^{2-}\!:\,,\\
(\mathbb{H}_0)_{\text{D3}_I ,\text{D3}_J} &= \delta_{IJ}\left(A_{I,I-1}^{1-}A_{I-1,I}^{1+}-A_{I,I+1}^{1+}A_{I+1,I}^{1-}+\right.\nonumber\\
&\hspace{+2cm}\left.+A_{I,I+1}^{2-}A_{I+1,I}^{2+}-A_{I,I-1}^{2+}A_{I-1,I}^{2-} +w_-^{I,I}\bar{w}_+^{I,I}+w_+^{I,I}\bar{w}_-^{I,I}\right)\,,
\end{align}
\end{subequations}
and
\begin{subequations}
\begin{align}
(\mathbb{H}_1^+)_{\overline{\text{D}(-1)}_I, \overline{\text{D}(-1)}_J} &= \delta_{IJ}\left(a^{1+}_{I,I+1}a^{2+}_{I+1,I}-a^{2+}_{I,I-1}a^{1+}_{I-1,I}+\bar{w}_+^{I,I} w_+^{I,I}\right):\!\psi^{1+}\psi^{2+}\!:\,,\\
(\mathbb{H}_1^-)_{\overline{\text{D}(-1)}_I, \overline{\text{D}(-1)}_J} &= \delta_{IJ}\left( a^{1-}_{I,I-1}a^{2-}_{I-1,I}-a^{2-}_{I,I+1}a^{1-}_{I+1,I} +\bar{w}_-^{I,I} w_-^{I,I}\right):\!\psi^{1-}\psi^{2-}\!:\,,\\
(\mathbb{H}_0)_{\overline{\text{D}(-1)}_I, \overline{\text{D}(-1)}_J} &= \delta_{IJ}\left(a_{I,I-1}^{1-}a_{I-1,I}^{1+}-a_{I,I+1}^{1+}a_{I+1,I}^{1-}+\right.\nonumber\\
&\hspace{2cm}\left.+a_{I,I+1}^{2-}a_{I+1,I}^{2+}-a_{I,I-1}^{2+}a_{I-1,I}^{2-}-\bar{w}_-^{I,I}{w}_+^{I,I}-\bar{w}_+^{I,I}{w}_-^{I,I}\right)\,.
\end{align}
\end{subequations}
One can also check that the reality conditions \eqref{eq:realityALE} imply $(\mathbb{H}_1^\pm)^\dagger = \mathbb{H}_1^\mp$ and $(\mathbb{H}_0)^\dagger = \mathbb{H}_0$. Note that analgously to the ${N}=4$ SYM instanton case, the charged and neutral auxiliary fields $\mathbb{H}^\pm$ and $\mathbb{H}_0$ encode the well-known forms for the complex and real hyper-K\"{a}hler moment maps for the unresolved ALE background at hand, which were first written down by Kronheimer and Nakajima. Equivalently, $\mathbb{H}_1^\pm$ and $\mathbb{H}_0$ yield the D-term and the F-term, respectively, of the corresponding 4d ${N}=2$ low-energy effective action upon being inserted into the localized classical effective action of \cite{Maccaferri:2018vwo,Maccaferri:2019ogq} (see also \eqref{eq:O2eff}). It is again easy to read off the corresponding algebraic constraints on the moduli (cf.\ Introduction section of \cite{kronheimer1990yang}, as well as eq. (8.6) of \cite{Douglas:1996sw})
\begin{subequations}
\begin{align}
\mu^\mathbb{C}_I &\equiv a^{1+}_{I,I+1}a^{2+}_{I+1,I}-a^{2+}_{I,I-1}a^{1+}_{I-1,I}+\bar{w}_+^{I,I} w_+^{I,I} =0\,, \\
\tilde{\mu}^\mathbb{C}_I &\equiv A^{1+}_{I,I+1}A^{2+}_{I+1,I}-A^{2+}_{I,I-1}A^{1+}_{I-1,I} -w_+^{I,I}\bar{w}_+^{I,I} =0\,,
\end{align}
\end{subequations}
and
\begin{subequations}
\begin{align}
{\mu}^\mathbb{R}_I &\equiv (a_{I-1,I}^{1+})^\dagger a_{I-1,I}^{1+}-a_{I,I+1}^{1+}(a_{I,I+1}^{1+})^\dagger+\nonumber\\
&\hspace{1cm}+(a_{I+1,I}^{2+})^\dagger a_{I+1,I}^{2+}-a_{I,I-1}^{2+}(a_{I,I-1}^{2+})^\dagger-\bar{w}_+^{I,I}(\bar{w}_+^{I,I})^\dagger+(w_+^{I,I})^\dagger {w}_+^{I,I} =0\,,\\
\tilde{\mu}^\mathbb{R}_I &\equiv (A_{I-1,I}^{1+})^\dagger A_{I-1,I}^{1+}-A_{I,I+1}^{1+}(A_{I,I+1}^{1+})^\dagger+\nonumber\\
&\hspace{1cm}+(A_{I+1,I}^{2+})^\dagger A_{I+1,I}^{2+}-A_{I,I-1}^{2+}(A_{I,I-1}^{2+})^\dagger +(\bar{w}_+^{I,I})^\dagger \bar{w}_+^{I,I}-w_+^{I,I}({w}_+^{I,I})^\dagger =0\,.
\end{align}
\end{subequations}
We can again relabel $A^{r+}\to i A^{r+}$ so as to make the signs agree with \cite{Douglas:1996sw}.
We have therefore again reproduced previously known flatness conditions for the open string background at hand.

\subsection{Spiked instantons at zero $B$-field}
\label{subsec:spiked}

Here\footnote{I thank Ondra Hul\'{i}k for this suggestion.} we will consider a configuration of several stacks of (euclidean) D$(-1)$, D3 and D7 branes (see Table \ref{tab:spiked}) which wrap a direct product of four complex planes with coordinates $X^{r\pm}=(X^{2r-1}\pm i X^{2r})/\sqrt{2}$ for $r\in \underline{4}\equiv \{1,2,3,4\}$. We will denote by $\text{D}3_{a}$, where $a\in \underline{6}\equiv \{(12),(13),(14),(23),(24),(34)\}$ a D3 brane stack spanning the complex 2-plane $\mathbb{C}^2$ indexed by $a$. Also denote by $\bar{a}$ the conjugate of $a$ in $\underline{6}$, that is e.g.\ $\overline{(12)}=(34)$. Such a brane configuration (minus the D7 branes) will give rise to moduli spaces of \emph{spiked instantons} \cite{Nekrasov:2015wsu,Nekrasov:2016qym,Nekrasov:2016gud}. As we are only equipped to deal with the dynamics of massless modes, we will require the NSNS $B$-field to be turned off everywhere.
\begin{table}[htpb!]
\centering
\caption{Brane configuration corresponding to general spiked instantons.}
\label{tab:spiked}
\begin{tabular}{c||cc|cc|cc|cc||cc}
 & $X^1$ & $X^2$ & $X^3$ & $X^4$& $X^5 $& $X^6$& $X^7$& $X^8$& $X^9$& $X^0$\\
\hline
 D$(-1)$ & $\bigcdot$ & $\bigcdot$ & $\bigcdot$ & $\bigcdot$ & $\bigcdot$ & $\bigcdot$ & $\bigcdot$ & $\bigcdot$ & $\bigcdot$& $\bigcdot$\\
 \hline
  D$3_{(12)}$ & $\times$ & $\times$ & $\times$ & $\times$ & $\bigcdot$ & $\bigcdot$ & $\bigcdot$ & $\bigcdot$ & $\bigcdot$& $\bigcdot$\\
 \hline
   D$3_{(13)}$ & $\times$ & $\times$ & $\bigcdot$ & $\bigcdot$ & $\times$ & $\times$ & $\bigcdot$ & $\bigcdot$ & $\bigcdot$& $\bigcdot$\\
 \hline
   D$3_{(14)}$ & $\times$ & $\times$ & $\bigcdot$ & $\bigcdot$ & $\bigcdot$ & $\bigcdot$ & $\times$ & $\times$ & $\bigcdot$& $\bigcdot$\\
 \hline
   D$3_{(23)}$ & $\bigcdot$ & $\bigcdot$ & $\times$ & $\times$ & $\times$ & $\times$ & $\bigcdot$ & $\bigcdot$ & $\bigcdot$& $\bigcdot$\\
 \hline
   D$3_{(24)}$ & $\bigcdot$ & $\bigcdot$ & $\times$ & $\times$ & $\bigcdot$ & $\bigcdot$ & $\times$ & $\times$ & $\bigcdot$& $\bigcdot$\\
 \hline
   D$3_{(34)}$ & $\bigcdot$ & $\bigcdot$ & $\bigcdot$ & $\bigcdot$ & $\times$ & $\times$ & $\times$ & $\times$ & $\bigcdot$& $\bigcdot$\\
 \hline
 D$7$ & $\times$ & $\times$& $\times$& $\times$& $\times$& $\times$& $\times$& $\times$& $\bigcdot$& $\bigcdot$
\end{tabular}
\end{table}
In order to preserve some amount of spacetime supersymmetry, we will have to take some of the brane stacks to consist of antibranes. One can show that there are only two inequivalent\footnote{In the sense that they cannot be mapped to each other by T-dualities.} choices of distributing the RR charges, both of which have $N=(2,0)$ supersymmetry in the two non-compact dimensions $X^9,X^0$: 
\begin{subequations}
\begin{align}
\text{C}_1:&\quad \overline{\text{D}(-1)}\,,\!\quad \text{D}3_{(12)}\,,\!\quad \text{D}3_{(13)}\,,\!\quad \text{D}3_{(14)}\,,\!\quad \text{D}3_{(23)}\,,\!\quad \text{D}3_{(24)}\,,\!\quad \text{D}3_{(34)}\,,\!\quad \overline{\text{D}7}\,,\\
\text{C}_2:&\quad{\text{D}(-1)}\,,\!\quad {\text{D}3_{(12)}}\,,\!\quad {\text{D}3_{(13)}}\,,\!\quad {\text{D}3_{(14)}}\,,\!\quad \overline{\text{D}3}_{(23)}\,,\!\quad \overline{\text{D}3}_{(24)}\,,\!\quad \overline{\text{D}3}_{(34)}\,,\!\quad \overline{\text{D}7}\,.
\end{align}
\end{subequations}
Therefore, recalling our discussion from subsection \ref{subsubsec:prel}, we expect to be able to decompose all marginal NS boundary fields into eigenstates carrying charges $\pm 1$ under a localising $U(1)$ $R$-current of a global $\mathcal{N}=2$ worldsheet superconformal algebra. 
Indeed, noting that the fermionic NS twist fields surviving the GSO projection for strings stretched between the $\overline{\text{D}(-1)}$ and $\text{D}3_{(rs)}$ branes, as well as between $\overline{\text{D}7}$ and $\text{D}3_{(rs)}$ branes 
and also between $\text{D}3_{(rs)}$ and $\text{D}3_{(rt)}$ branes (for $s\neq t$)
are always \emph{chiral} (that is, they can be bosonized with $U(1)$ charges $\pm (1/2,1/2)$ under the $U(1)$ currents $(J_r,J_s)=(:\!\psi_{r-}\psi_{r+}\!:\,,\,:\!\psi_{s-}\psi_{s+}\!:)$ where $\psi_{r\pm}$ and $\psi_{s\pm}$ span the four Neumann-Dirichlet directions at hand), it is easy to see that for $\text{C}_1$, all marginal NS boundary fields carry charges $\pm 1$ under the localizing $R$-current 
\begin{align}
J_1 +J_2+J_3+J_4 \,,
\end{align}
of the free field $\mathcal{N}=2$ worldsheet superconformal algebra with $c=12$ along the directions $X^1,\ldots,X^8$ (remember that $J_r = -i\p h_r\,=\,:\!\psi_{r-}\psi_{r+}\!:$, where $h_r$ is such that $\psi_{r\pm}=e^{\pm i h_r}$). Analogously, for $\text{C}_2$, all NS boundary fields carry charges $\pm 1$ with respect to the $R$-current
\begin{align}
J_1-J_2-J_3-J_4 \,.
\end{align}
While it is straightforward to evaluate the generalized ADHM equations $\mathbb{H}_1^\pm = \mathbb{H}_0$ for both $\text{C}_1$ and $\text{C}_2$, below we will only do so explicitly for simpler configurations which are termed by \cite{Nekrasov:2016qym,Nekrasov:2016gud} as crossed and folded instantons. These two configurations will each conserve four spacetime supercharges giving rise to $N=(4,0)$ and $N=(2,2)$ supersymmetry, respectively, in the two non-compact dimensions $X^9, X^0$. 

\subsubsection{Crossed instanton scenario}

Here we will keep only the ${\mathrm{D(-1)}}$, D$3_{(12)}$ and D$3_{(34)}$ brane stacks. There are three T-duality inequivalent possibilities of distributing the RR charges:
\begingroup\allowdisplaybreaks
\begin{subequations}
\begin{align}
\text{CC}_1:\qquad \overline{\text{D}(-1)}\,,\quad \mathrm{D3}_{(12)}\,,\quad \mathrm{D3}_{(34)}\,,\\
\text{CC}_2:\qquad {\text{D}(-1)}\,,\quad \mathrm{D3}_{(12)}\,,\quad \overline{\mathrm{D3}}_{(34)}\,,\\
\text{CC}_3:\qquad {\text{D}(-1)}\,,\quad \mathrm{D3}_{(12)}\,,\quad \mathrm{D3}_{(34)}\,.
\end{align}
\end{subequations}
\endgroup
All three configurations preserve $N=(4,0)$ supersymmetry in the two dimensions spanned by $X^9,X^0$. First, note that there are no massless NS modes for the strings stretched between the $\text{D3}_{(12)}$ and $\text{D3}_{(34)}$ branes. In the case of $\text{CC}_1$, all massless NS stretched fermions are chiral. For $\text{CC}_2$, the massless NS fermions stretched between the $\text{D}(-1)$ and $\text{D3}_{(12)}$ branes are anti-chiral while those stretched between the $\text{D}(-1)$ and $\overline{\text{D3}}_{(34)}$ branes are chiral. Finally, for $\text{CC}_3$ all massless NS stretched fermions are anti-chiral. In the three respective cases, all boundary fields therefore carry charges $\pm 1$ with respect to the $U(1)$ currents
\begingroup\allowdisplaybreaks
\begin{subequations}
\begin{align}
J_{\text{C},1}=+J_1+J_2+J_3+J_4\,,\\
J_{\text{C},2}=-J_1+J_2+J_3+J_4\,,\\
J_{\text{C},3}=-J_1+J_2-J_3+J_4\,.
\end{align}
\end{subequations}
\endgroup
However, analyzing the generalized ADHM equations $\mathbb{H}_1^\pm =\mathbb{H}_0=0$ for each of the three crossed-instanton configurations, one eventually finds that all of them give rise to the same constraints on the moduli space. We will now therefore work these out for $\text{CC}_1$, only making brief comments along the way on how one should proceed had we started with $\text{CC}_2$ or $\text{CC}_3$. Let us introduce the following notation for the matter part of the marginal boundary fields which we will work with:
\begin{subequations}
\begin{align}
(\mathbb{V}_\frac{1}{2}^\pm)_{{\text{D}3_{(rs)}},{\text{D}3_{(rs)}}} &= \sum_{u\in\underline{4}}{A}^{(rs)}_{u\pm}\psi^{u\pm}\,,\\[+0.5mm] 
(\mathbb{V}_\frac{1}{2}^\pm)_{\overline{\text{D}(-1)},\overline{\text{D}(-1)}} &= \sum_{r\in\underline{4}}{a}_{r\pm}\psi^{r\pm}\,,
\end{align}
\end{subequations} 
together with 
\begin{subequations}
\begin{align}
(\mathbb{V}_\frac{1}{2}^\pm)_{{\text{D}3_{(rs)}},\overline{\text{D}(-1)}} &= w_{\pm}^{(rs)}\psi^{(rs)\pm} \,,\\
(\mathbb{V}_\frac{1}{2}^\pm)_{\overline{\text{D}(-1)},{\text{D}3_{(rs)}}} &= \bar{w}_{\pm}^{(rs)}\psi^{(\tilde{r}\tilde{s})\pm} \,,
\end{align}
\end{subequations}
where, as usual, the notation $\mathbb{V}_{1/2}^\pm$ is taken to mean that the respective states carry charge $\pm 1$ under $J_{\text{C},1}$. We have also introduced the following notation for the boundary condition changing operators
\begin{subequations}
\begin{align}
\psi^{(rs)\alpha} = \Sigma^r\Sigma^s S^{(rs)\alpha}\,,\\
\psi^{(\tilde{r}\tilde{s})\alpha} = \bar{\Sigma}^r\bar{\Sigma}^s S^{(rs)\alpha}\,,
\end{align}
\end{subequations}
with bosonic twist fields $\Sigma^r = \sigma^{2r-1}\sigma^{2r}$, $\overline{\Sigma}^r = \overline{\sigma}^{2r-1}\overline{\sigma}^{2r}$ and $S^{(rs)\alpha}$ the chiral components of the 4d euclidean spin fields in the complex 2-plane $(rs)$. Observe that in the case of $\text{CC}_2$, we would have $a_{\pm}\psi^{1\mp}$, $A^{(12)}_{\pm}\psi^{1\mp}$, $A^{(34)}_{\pm}\psi^{1\mp}$ (where we have arbitrarily relabelled $a_\pm\to a_\mp$ etc.) and $w^{(12)}_{\dot{\pm}}\psi^{(12)\dot{\pm}}$, $\bar{w}^{(12)}_{\dot{\pm}}\psi^{(\tilde{1}\tilde{2})\dot{\pm}}$ contributing into $\mathbb{V}_{1/2}^\pm$ instead. Similar changes would have to be in place for $\text{CC}_3$.
We have the reality conditions
\begin{align}
({A}^{(12)}_{r\pm})^\dagger = {A}^{(12)}_{r\mp}\,,\qquad ({A}^{(34)}_{r\pm})^\dagger = {A}^{(34)}_{r\mp}\,,\qquad (a_{r\pm})^\dagger = a_{r\mp}\,,
\end{align}
together with 
\begin{align}
(\bar{{w}}^{(12)}_\alpha)^\dagger={w}^{(12)\alpha}\,,\qquad (\bar{{w}}^{(34)}_\alpha)^\dagger={w}^{(34)\alpha}\,.
\end{align}
For $\text{CC}_2$ and $\text{CC}_3$, we would have also $(\bar{{w}}^{(12)}_{\dot{\alpha}})^\dagger={w}^{(12)\dot{\alpha}}$ and $(\bar{{w}}^{(34)}_{\dot{\alpha}})^\dagger={w}^{(34)\dot{\alpha}}$, respectively.
The charged auxiliary fields then give constraints
\begingroup\allowdisplaybreaks
\begin{subequations}
\begin{align}
[a_{1+},a_{2+}]+\bar{w}_{+}^{(12)}w_{+}^{(12)} &=0\,,\\
[a_{1+},a_{3+}] &=0\,,\\
[a_{1+},a_{4+}] &=0\,,\\
[a_{2+},a_{3+}] &=0\,,\\
[a_{2+},a_{4+}] &=0\,,\\
[a_{3+},a_{4+}]+\bar{w}_{+}^{(34)}w_{+}^{(34)} &=0\,,
\end{align}
\end{subequations}
together with
\begin{subequations}
\begin{align}
[A^{(12)}_{1+},A^{(12)}_{2+}]-{w}_{+}^{(12)}\bar{w}_{+}^{(12)}&=0\,,\\
[A^{(12)}_{3+},A^{(12)}_{4+}]&=0\,,\\
[A^{(34)}_{3+},A^{(34)}_{4+}]-{w}_{+}^{(34)}\bar{w}_{+}^{(34)}&=0\,,\\
[A^{(34)}_{1+},A^{(34)}_{2+}]&=0\,,
\end{align}
\end{subequations}
and
\begin{subequations}
\begin{align}
[A^{(12)}_{1+},A^{(12)}_{3+}]&=0\,,\\
[A^{(12)}_{1+},A^{(12)}_{4+}]&=0\,,\\
[A^{(12)}_{2+},A^{(12)}_{3+}]&=0\,,\\
[A^{(12)}_{2+},A^{(12)}_{4+}]&=0\,,\\
[A^{(34)}_{1+},A^{(34)}_{3+}]&=0\,,\\
[A^{(34)}_{2+},A^{(34)}_{3+}]&=0\,,\\
[A^{(34)}_{1+},A^{(34)}_{4+}]&=0\,,\\
[A^{(34)}_{2+},A^{(34)}_{4+}]&=0\,.
\end{align}
\end{subequations}
\endgroup
The diagonal part of the neutral auxiliary field then gives
\begin{subequations}
\begin{align}
\sum_{r\in \underline{4}}[(a_{r+})^\dagger,a_{r+}]-\bar{w}_+^{(12)}(\bar{w}_+^{(12)})^\dagger+({w}_+^{(12)})^\dagger{w}_+^{(12)}+\hspace{2cm}&\nonumber\\[-2mm]
-\bar{w}_+^{(34)}(\bar{w}_+^{(34)})^\dagger+({w}_+^{(34)})^\dagger{w}_+^{(34)}
&=0\,,\label{eq:neuta}\\[+4mm]
\sum_{r\in \underline{4}}[(A^{(12)}_{r+})^\dagger,A^{(12)}_{r+}]+(\bar{{w}}_+^{(12)})^\dagger\bar{{w}}_+^{(12)}-{w}_+^{(12)}({w}_+^{(12)})^\dagger&=0\,,\\
\sum_{r\in \underline{4}}[(A^{(34)}_{r+})^\dagger,A^{(34)}_{r+}]+(\bar{{w}}_+^{(34)})^\dagger\bar{{w}}_+^{(34)}-{w}_+^{(34)}({w}_+^{(34)})^\dagger&=0\,.
\end{align}
\end{subequations}
Finally, from the off-diagonal part of the charged auxiliary fields, we get
\begin{align}
w_+^{(12)}\bar{w}_+^{(34)}=w_+^{(34)}\bar{w}_+^{(12)}&=0\,,
\label{eq:1234}
\end{align}
together with
\begingroup
\allowdisplaybreaks
\begin{subequations}
\begin{align}
a_{3+}\bar{w}_+^{{(12)}}-\bar{w}_+^{{(12)}}A^{{(12)}}_{3+}&=0\,,\\
a_{4+}\bar{w}_+^{{(12)}}-\bar{w}_+^{{(12)}}A^{{(12)}}_{4+}&=0\,,\\
a_{1+}\bar{w}_+^{{(34)}}-\bar{w}_+^{{(34)}}A^{{(34)}}_{1+}&=0\,,\\
a_{2+}\bar{w}_+^{{(34)}}-\bar{w}_+^{{(34)}}A^{{(34)}}_{2+}&=0\,,
\end{align}
\end{subequations}
and
\begin{subequations}
\begin{align}
A^{(12)}_{3+}w_+^{(12)}-w_+^{(12)}a_{3+} &=0\,,\\
A^{(12)}_{4+}w_+^{(12)}-w_+^{(12)}a_{4+} &=0\,,\\
A^{(34)}_{1+}w_+^{(34)}-w_+^{(34)}a_{1+} &=0\,,\\
A^{(34)}_{2+}w_+^{(34)}-w_+^{(34)}a_{2+} &=0\,.
\end{align}
\end{subequations}
\endgroup
Identical equations would be obtained for $\text{CC}_2$ except for the replacements
\begin{align}
 w_+^{(12)}\to w_{\dot{+}}^{(12)}\quad\text{and}\quad   \bar{w}_+^{(12)}\to \bar{w}_{\dot{+}}^{(12)}\,.
\end{align}
Similarly for $\text{CC}_3$. Note that as observed by \cite{Nekrasov:2016qym}, it is impossible to find a single-instanton solution of these constraints which would have non-zero vevs in both stretched sectors ((12) and (34)) simultaneously: equation \eqref{eq:1234} implies that any such solution would have to have either $w_+^{(12)}=w_+^{(34)}=0$ or $\bar{w}_+^{(12)}=\bar{w}_+^{(34)}=0$. But then, equation \eqref{eq:neuta} fixes the rest of the stretched moduli to be zero as well. In other words, it is only possible to dissolve a single D-instanton into only one D3-brane stack at a time. Analysis of the above constraints for a general number of D-instantons, as interesting as it may be, lies beyond the scope of this paper.

\subsubsection{Folded instanton scenario}

Here we will keep only the ${\mathrm{D(-1)}}$, D$3_{(12)}$ and D$3_{(13)}$ brane stacks. There are two T-duality inequivalent possibilities of distributing the RR charges:
\begin{subequations}
\begin{align}
\text{FC}_1:\qquad \overline{\text{D}(-1)}\,,\quad \mathrm{D3}_{(12)}\,,\quad \mathrm{D3}_{(13)}\,,\\
\text{FC}_2:\qquad {\text{D}(-1)}\,,\quad \mathrm{D3}_{(12)}\,,\quad {\mathrm{D3}}_{(13)}\,,
\end{align}
\end{subequations}
both of which preserve $N=1$ supersymmetry in the four dimensions spanned by $X^7$, $X^8$, $X^9$, $X^0$, which gives rise to $N=(2,2)$ supersymmetry in the two dimensions spanned by $X^9,X^0$. In the case of $\text{FC}_1$, all massless NS stretched fermions are chiral. For $\text{FC}_2$, the massless NS fermions stretched between the $\text{D}(-1)$ and $\text{D3}_{(12)}$, $\text{D}(-1)$ and $\text{D3}_{(13)}$ branes are anti-chiral, while the massless the NS fermions stretched between the $\text{D3}_{(12)}$ and $\text{D3}_{(13)}$ branes are chiral. That is, in the two respective cases, all boundary fields carry charges $\pm 1$ with respect to the $U(1)$ currents
\begin{subequations}
\begin{align}
J_{\text{F},1}=+J_1+J_2+J_3\,,\\
J_{\text{F},2}=-J_1+J_2+J_3\,.
\end{align}
\end{subequations}
However, similarly to the crossed-instanton scenario, upon evaluating the auxilliary fields $\mathbb{H}^\pm_1$, $\mathbb{H}_0$, we obtain structurally identical constraints for both $\text{FC}_1$ and $\text{FC}_2$. Let us therefore focus on $\text{FC}_1$ only. On top of the moduli introduced in the crossed instanton case, we denote
\begin{subequations}
\begin{align}
(\mathbb{V}_\frac{1}{2}^\pm)_{{\text{D}3_{(12)}},\text{D}3_{(13)}} &= W_{\pm}^{(23)1}\psi^{(2\tilde{3})\pm} \,,\\ 
(\mathbb{V}_\frac{1}{2}^\pm)_{\text{D}3_{(13)},{\text{D}3_{(12)}}} &= \bar{W}_{\pm}^{(23)1}\psi^{(\tilde{2}3)\pm}\,,
\end{align}
\end{subequations}
where we introduce BCCOs 
\begin{align}
    \psi^{({2}\tilde{3})\alpha} = {\Sigma}^2\bar{\Sigma}^3 S^{(23)\alpha}\,,\\
\psi^{(\tilde{2}{3})\alpha} = \bar{\Sigma}^2{\Sigma}^3 S^{(23)\alpha}\,.
\end{align} 
We have reality conditions
\begin{align}
({A}^{(12)}_{r\pm})^\dagger = {A}^{(12)}_{r\mp}\,,\qquad ({A}^{(13)}_{r\pm})^\dagger = {A}^{(13)}_{r\mp}\,,\qquad (a_{r\pm})^\dagger = a_{r\mp}\,,
\end{align}
together with 
\begin{align}
(\bar{{w}}^{(12)}_\alpha)^\dagger={w}^{(12)\alpha}\,,\qquad (\bar{{w}}^{(13)}_\alpha)^\dagger={w}^{(13)\alpha}\,,\qquad (\bar{W}^{(23)1}_\alpha)^\dagger=W^{(23)1\alpha}\,.
\end{align}
The constraints on moduli coming from the diagonal part of the charged auxilliary fields then read
\begin{subequations}
\label{eq:e92}
\begin{align}
[a_{1+},a_{2+}]+\bar{w}_{+}^{(12)}w_{+}^{(12)} &=0\,,\\
[a_{1+},a_{3+}]+\bar{w}_{+}^{(13)}w_{+}^{(13)} &=0\,,\\
[a_{1+},a_{4+}] &=0\,,\\
[a_{2+},a_{3+}] &=0\,,\\
[a_{2+},a_{4+}] &=0\,,\\
[a_{3+},a_{4+}] &=0\,,
\end{align}
\end{subequations}
together with
\begin{subequations}
\label{eq:e93}
\begin{align}
[A^{(12)}_{1+},A^{(12)}_{2+}]-{w}_{+}^{(12)}\bar{w}_{+}^{(12)}&=0\,,\\
[A^{(12)}_{3+},A^{(12)}_{4+}]&=0\,,\\
[A^{(13)}_{1+},A^{(13)}_{3+}]-{w}_{+}^{(13)}\bar{w}_{+}^{(13)}&=0\,,\\
[A^{(13)}_{2+},A^{(13)}_{4+}]&=0\,,
\end{align}
\end{subequations}
and
\begingroup\allowdisplaybreaks
\begin{subequations}
\label{eq:e94}
\begin{align}
[A^{(12)}_{1+},A^{(12)}_{3+}]&=0\,,\\
[A^{(12)}_{1+},A^{(12)}_{4+}]&=0\,,\\
[A^{(12)}_{2+},A^{(12)}_{3+}]-W_{+}^{(23)1}\bar{W}_{+}^{(23)1}&=0\,,\\
[A^{(12)}_{2+},A^{(12)}_{4+}]&=0\,,\\
[A^{(13)}_{1+},A^{(13)}_{2+}]&=0\,,\\
[A^{(13)}_{1+},A^{(13)}_{4+}]&=0\,,\\
[A^{(13)}_{2+},A^{(13)}_{3+}]+\bar{W}_{+}^{(23)1}{W}_{+}^{(23)1}&=0\,,\\
[A^{(13)}_{3+},A^{(13)}_{4+}]&=0\,.
\end{align}
\end{subequations}
\endgroup
The diagonal part of the neutral auxiliary field then gives
\begin{subequations}
\label{eq:e96}
\begin{align}
\sum_{r\in \underline{4}}[(a_{r+})^\dagger,a_{r+}]-\bar{w}_+^{(12)}(\bar{w}_+^{(12)})^\dagger+({w}_+^{(12)})^\dagger{w}_+^{(12)}+\hspace{2cm}&\nonumber\\[-3mm]
-\bar{w}_+^{(13)}(\bar{w}_+^{(13)})^\dagger+({w}_+^{(13)})^\dagger{w}_+^{(13)}&=0\,,\\
\sum_{r\in \underline{4}}[(A^{(12)}_{r+})^\dagger,A^{(12)}_{r+}]+(\bar{{w}}_+^{(12)})^\dagger\bar{{w}}_+^{(12)}-{w}_+^{(12)}({w}_+^{(12)})^\dagger +\hspace{2cm}&\nonumber\\[-3mm]
+({\bar{W}}_+^{(23)1})^\dagger{\bar{W}}_+^{(23)1}-{W}_+^{(23)1}({W}_+^{(23)1})^\dagger  &=0\,,\\
\sum_{r\in \underline{4}}[(A^{(13)}_{r+})^\dagger,A^{(13)}_{r+}]+(\bar{{w}}_+^{(13)})^\dagger\bar{{w}}_+^{(13)}-{w}_+^{(13)}({w}_+^{(13)})^\dagger +\hspace{2cm}&\nonumber\\[-3mm]
+({{W}}_+^{(23)1})^\dagger{{W}}_+^{(23)1}-\bar{W}_+^{(23)1}(\bar{W}_+^{(23)1})^\dagger  &=0\,.
\end{align}
\end{subequations}
Before we write down the constraints coming from the non-diagonal part of the charged auxilliary fields, we note that one first needs to fix possible phases (cocycles) $c_r^\pm$, $\tilde{c}_r^\pm$ for $r=1,2,3$ arising in the OPE
\begingroup\allowdisplaybreaks
\begin{subequations}
\label{eq:opeAmbig}
\begin{align}
\psi^{(2\tilde{3})\pm}(z)\psi^{(13)\pm}(0)&\sim c_3^\pm\psi^{(12)\pm}\psi^{3\pm}(0)\,,\\
\psi^{(\tilde{1}\tilde{3})\pm}(z)\psi^{(\tilde{2}3)\pm}(0)&\sim \tilde{c}_3^\pm\psi^{(\tilde{1}\tilde{2})\pm}\psi^{3\pm}(0)\,,\\
\psi^{(12)\pm}(z)\psi^{(\tilde{1}\tilde{3})\pm}(0)&\sim {c}_1^\pm\psi^{(2\tilde{3})\pm}\psi^{1\pm}(0)\,,\\
\psi^{(13)\pm}(z)\psi^{(\tilde{1}\tilde{2})\pm}(0)&\sim \tilde{c}_1^\pm\psi^{(\tilde{2}3)\pm}\psi^{1\pm}(0)\,,\\
\psi^{(\tilde{1}\tilde{2})\pm}(z)\psi^{(2\tilde{3})\pm}(0)&\sim c_2^\pm\psi^{(\tilde{1}\tilde{3})\pm}\psi^{2\pm}(0)\,,\\
\psi^{(\tilde{2}{3})\pm}(z)\psi^{({1}{2})\pm}(0)&\sim \tilde{c}_2^\pm\psi^{({1}{3})\pm}\psi^{2\pm}(0)\,.
\end{align}
\end{subequations}
\endgroup
Associativity of the OPE requires that
\begin{align}
  \tilde{c}_2^\pm c_3^\pm =c_1^\pm c_3^\pm=\tilde{c}_1^\pm\tilde{c}_2^\pm =-\tilde{c}_1^\pm\tilde{c}_3^\pm = -c_2^\pm\tilde{c}_3^\pm=-c_1^\pm c_2^\pm &=+1\,.
  \label{eq:relPhase}
\end{align}
We will now see that the relations \eqref{eq:relPhase} can be used to eliminate all potential phase ambiguities from the algebraic constraints on the moduli. Using the OPE \eqref{eq:opeAmbig}, from the non-diagonal part of the charged auxiliary fields one obtains constraints
\begingroup
\allowdisplaybreaks
\begin{subequations}
\begin{align}
a_{3+}\bar{w}_+^{{(12)}}-\bar{w}_+^{{(12)}}A^{{(12)}}_{3+}+\tilde{c}_3^+\bar{w}_+^{(13)} \bar{W}_+^{(23)1}&=0\,,\\
a_{4+}\bar{w}_+^{{(12)}}-\bar{w}_+^{{(12)}}A^{{(12)}}_{4+}&=0\,,\\
a_{2+}\bar{w}_+^{{(13)}}-\bar{w}_+^{{(13)}}A^{{(13)}}_{2+}+c_2^+\bar{w}_+^{(12)} W_+^{(23)1}&=0\,,\\
a_{4+}\bar{w}_+^{{(13)}}-\bar{w}_+^{{(13)}}A^{{(13)}}_{4+}&=0\,,\\
A^{(12)}_{3+}w_+^{(12)}-w_+^{(12)}a_{3+}+c_3^+ W_+^{(23)1}w_+^{(13)} &=0\,,\\
A^{(12)}_{4+}w_+^{(12)}-w_+^{(12)}a_{4+} &=0\,,\\
A^{(13)}_{2+}w_+^{(13)}-w_+^{(13)}a_{2+}+\tilde{c}_2^+\bar{W}_+^{(23)1}w_+^{(12)} &=0\,,\\
A^{(13)}_{4+}w_+^{(13)}-w_+^{(13)}a_{4+} &=0\,,
\end{align}
\end{subequations}
together with
\begin{subequations}
\begin{align}
A^{(12)}_{1+}W_+^{(23)1}-W_+^{(23)1} A^{(13)}_{1+}+c_1^+ w_+^{(12)}\bar{w}_+^{(13)}&=0\,,\\
A^{(12)}_{4+}W_+^{(23)1}-W_+^{(23)1} A^{(13)}_{4+}&=0\,,\\
A^{(13)}_{1+}\bar{W}_+^{(23)1}-\bar{W}_+^{(23)1} A^{(12)}_{1+}+\tilde{c}_1^+ w_+^{(13)}\bar{w}_+^{(12)}&=0\,,\\
A^{(13)}_{4+}\bar{W}_+^{(23)1}-\bar{W}_+^{(23)1} A^{(12)}_{4+}&=0\,.
\end{align}
\end{subequations}
\endgroup
Using the consistency relations \eqref{eq:relPhase} and replacing
\begin{subequations}
\begin{align}
(\tilde{c}_3^+)^{-1}W_+^{(23)1}\to W_+^{(23)1}\,,\\
\tilde{c}_3^+ \bar{W}_+^{(23)1}\to \bar{W}_+^{(23)1}\,,
\end{align}
\end{subequations}
(note that this rescaling does not have any effect on the previously derived constraints \eqref{eq:e92}, \eqref{eq:e93}, \eqref{eq:e94} and \eqref{eq:e96}) we can rewrite these constraints as
\begingroup
\allowdisplaybreaks
\begin{subequations}
\begin{align}
a_{3+}\bar{w}_+^{{(12)}}-\bar{w}_+^{{(12)}}A^{{(12)}}_{3+}+\bar{w}_+^{(13)} \bar{W}_+^{(23)1}&=0\,,\\
a_{4+}\bar{w}_+^{{(12)}}-\bar{w}_+^{{(12)}}A^{{(12)}}_{4+}&=0\,,\\
a_{2+}\bar{w}_+^{{(13)}}-\bar{w}_+^{{(13)}}A^{{(13)}}_{2+}-\bar{w}_+^{(12)} W_+^{(23)1}&=0\,,\\
a_{4+}\bar{w}_+^{{(13)}}-\bar{w}_+^{{(13)}}A^{{(13)}}_{4+}&=0\,,\\
A^{(12)}_{3+}w_+^{(12)}-w_+^{(12)}a_{3+}- W_+^{(23)1}w_+^{(13)} &=0\,,\\
A^{(12)}_{4+}w_+^{(12)}-w_+^{(12)}a_{4+} &=0\,,\\
A^{(13)}_{2+}w_+^{(13)}-w_+^{(13)}a_{2+}-\bar{W}_+^{(23)1}w_+^{(12)} &=0\,,\\
A^{(13)}_{4+}w_+^{(13)}-w_+^{(13)}a_{4+} &=0\,,
\end{align}
\end{subequations}
together with
\begin{subequations}
\begin{align}
A^{(12)}_{1+}W_+^{(23)1}-W_+^{(23)1} A^{(13)}_{1+}+ w_+^{(12)}\bar{w}_+^{(13)}&=0\,,\\
A^{(12)}_{4+}W_+^{(23)1}-W_+^{(23)1} A^{(13)}_{4+}&=0\,,\\
A^{(13)}_{1+}\bar{W}_+^{(23)1}-\bar{W}_+^{(23)1} A^{(12)}_{1+}- w_+^{(13)}\bar{w}_+^{(12)}&=0\,,\\
A^{(13)}_{4+}\bar{W}_+^{(23)1}-\bar{W}_+^{(23)1} A^{(12)}_{4+}&=0\,,
\end{align}
\end{subequations}
\endgroup
which are free of the cocycle factors. Note that setting $W_+^{(23)1}=\bar{W}_+^{(23)1}=0$ and keeping only a single D-instaton, we can use analogous arguments as we did for the crossed scenario to show that it is impossible to find a solution with both $w_+^{(12)}\neq 0$ and $w_+^{(13)}\neq 0$ simultaneously. This reproduces the result of \cite{Nekrasov:2016qym}. That is, we can again only dissolve a single D-instanton into one of the two D3 brane stacks at a time. However, it would be very interesting to investigate whether a non-trivial solution can be found if we allow for non-zero $W_+^{(23)1}$, $\bar{W}_+^{(23)1}$, that is, if it is possible to dissolve all three brane stacks into each other.

\section{Discussion and outlook}
\label{sec:disc}

The aim of this paper was to take up the line of development initiated by \cite{Maccaferri:2018vwo}. In the context of the Berkovits formulation of open superstring field theory, the authors of \cite{Maccaferri:2018vwo} (motivated by \cite{Sen:2015uoa}) introduce the $\mathcal{N}=2$ $R$-charge decomposition technique and derive the simple expression \eqref{eq:O2eff} for algebraic couplings of the quartic classical effective action, noting that it localizes on the boundary of the worldsheet moduli space. In particular, they apply these results in the case of the D$(-1)$/D3 system and show that their result reproduces the ADHM equations as flatness conditions for the quartic effective potential. Ref.\ \cite{Mattiello:2019gxc} then goes on to investigate (within the framework of the $A_\infty$ formulation of open superstring field theory) obstructions to exact marginality of deformations of the D$(-1)$/D3 system by massless open string modes. They also compute open string gauge field profiles for the finite-size $SU(2)$ instanton and compare their result with the findings of \cite{Billo:2002hm}. Finally, ref.\ \cite{Maccaferri:2019ogq} shows that it is possible to repeat the analysis of \cite{Maccaferri:2018vwo} using $A_\infty$ OSFT so that one manifestly stays in the small Hilbert space throughout the derivation. They also derive relation \eqref{eq:ObstActRel} between the third-order obstruction and the quartic part of the classical effective action and note that it implies that all marginal deformations which are unobstructed at third order must correspond to flat directions of the quartic effective action. The main contributions of the present paper can then be listed as follows:
\begin{enumerate}
    \item We have shown that assuming that the background at hand supports a global $\mathcal{N}=2$ worldsheet superconformal symmetry such that all NS marginal fields have $R$-charge $\pm 1$ (this we argued to include backgrounds preserving at least $N=(2,0)$ in two non-compact dimensions),  then the flatness conditions $\mathbb{H}_1^\pm = \mathbb{H}_0 = 0$ of \cite{Maccaferri:2018vwo,Maccaferri:2019ogq} (generalized ADHM equations) are in fact equivalent to the conditions which are neccessary and sufficient for the vanishing of the obstruction to exact marginality at third order in the deformation parameter $\lambda$. In particular, applying these results in the case of the D$(-1)$/D3 system, we have confirmed that the solution of classical equations of motion of open superstring field theory describing a finite-size instanton is consistent up to third order in~$\lambda$.
    \item Apart from discussing the situation in the context of both the usual $A_\infty$ and Berkovits formulation of open superstring field theory, we have investigated the changes which arise in the context of $A_\infty$ OSFT deformed by adding stubs to the Witten star product. Our findings show that even though this necessitates an addition of higher fundamental products into the theory which involve integration over bosonic moduli, the final result remains unchanged and localized on the boundary of the moduli space. As per the discussion of \cite{Maccaferri:2019ogq}, this should open the way to generalizing these results to a closed superstring setting.
    \item We have discussed computation of the third-order obstruction (and therefore the quartic part of the effective action) in more general cases where the above-mentioned global $\mathcal{N}=2$ worldsheet superconformal symmetry is not present. This was based on the method of \cite{Berkovits:2003ny} using a Schwinger parametrization of the Siegel gauge propagator.
    \item We have further used the generalized ADHM equations $\mathbb{H}_1^\pm=\mathbb{H}_0=0$ to compute algebraic constraints on the moduli in the case of two more complicated backgrounds: the D$(-1)$/D3 system on the background of the $\mathbb{C}^2/\mathbb{Z}_n$ orbifold and also the D-brane confgurations giving rise to the crossed and folded instantons at zero $B$-field. Doing so we have demonstrated that the generalized ADHM equations can be used for a quick derivation of algebraic constraints on the moduli in a wide variety of backgrounds. We have also noted that in two of our three examples, where the background had $N=2$ supersymmetry in 4d, the auxiliary fields $\mathbb{H}_1^\pm$ were related to the F-term of the low-energy effective action while the auxiliary field $\mathbb{H}_0$ was related to the D-term.
\end{enumerate}

Finally, let us conclude by briefly discussing possible future directions. It would be of great interest to investigate if there are additional contraints on the marginal couplings appearing at higher orders in the deformation parameter. In particular, it is a priori not clear whether consistency at higher orders produces corrections to the generalized ADHM equations or whether, under certain assumptions, the vanishing of the third-order obstruction already implies exact marginality at all orders (for the D$-1$/D3 system, this was already argue to be the case in \cite{Seiberg:1999vs} from a different point of view). One might also try to obtain analogous constraints on the fermionic moduli (or, equivalently, fermionic sector of the quartic effective action) using a formulation of open superstring field theory in the Ramond sector \cite{Kunitomo:2015usa,Erler:2016ybs} (see also \cite{Asada:2017ykq} which generalizes the computation of \cite{Berkovits:2003ny} to the Ramond sector). Together with the bosonic constraints discussed in this paper, these findings would provide information about the whole supermoduli space. These considerations would also allow for a discussion of how spacetime supersymmetry manifests itself at the level of complete effective actions \cite{Erler:2016rxg}. It would be also interesting to see whether our results generalize to the case of NS marginal deformations in heterotic string and NSNS and RR marginal deformations in closed type II superstring (see \cite{Sen:2019jpm} for a related recent discussion of marginal deformations in closed (super)string field theory). 
Ultimately one should be interested in looking at moduli spaces of D-brane systems on marginally deformed classical closed string backgrounds. This can be done by including the effects of non-dynamical background on-shell closed string field (satisfying closed SFT equations of motion) into a classical open superstring field theory action through open-closed disk vertices (see \cite{Moosavian:2019ydz} for some related recent progress). 
We hope to report on our progress in this direction soon \cite{Maccaferri:2019xx}. 
Unobstructed open string marginal deformations yield new consistent open string backgrounds. As such, these need to be described by superconformal boundary states. It would be interesting to recover these boundary states by computing suitable gauge invariant observables in open superstring field theory \cite{Hashimoto:2001sm,Ellwood:2008jh,Kudrna:2012re}. In the case of finite-size instantons, one should expect that the corresponding boundary states will be highly non-trivial as they cannot satisfy the usual linear gluing conditions on the free-field oscillators $\alpha_n^\mu$, $\psi_r^\mu$ (all such cases are already exhausted by the conventional D$p$-branes). OSFT methods may therefore yield valuable insights into the structure of these boundary states complementary to other techniques (see e.g.\ \cite{Schnabl:2019oom}). We hope to report on our progress soon \cite{Maccaferri:2019xx}. 

\acknowledgments 

I would like to thank Carlo Maccaferri for motivating me to work on this topic, for numerous discussions, as well as for suggesting many improvements of the manuscript. Many thanks as well to Ted Erler, Mat\v{e}j Kudrna, Renann Lipinski-Jusinskas, Ondra Hul\'{i}k, Martin Schnabl, Ashoke Sen and especially Luca Mattiello and Ivo Sachs for useful comments and discussions. I also thank the Galileo Galilei Institute for Theoretical Physics and INFN for hospitality and partial support during the workshop ``String Theory from a worldsheet perspective'' where this work has been initiated. Many thanks as well to the Arnold Sommerfeld Center for Theoretical Physics for their hospitality during the later stages of this work. This research has been supported by the Czech Science Foundation (GA\v{C}R) grant 17-22899S.

\appendix

\section{Spinors in 4d}

\label{app:spin}

We will consider the euclidean Clifford algebra in 4d $\{\gamma^\mu,\gamma^\nu\}=2\delta^{\mu\nu}$. In terms of the Pauli matrices
\begin{align}
	\tau^1 = \begin{pmatrix}
	0 & 1\\
	1 & 0
	\end{pmatrix}\,,\quad
	\tau^2 = \begin{pmatrix}
	0 & -i\\
	i & 0
	\end{pmatrix}\,,\quad
	\tau^3 = \begin{pmatrix}
	1 & 0\\
	0 & -1
	\end{pmatrix}\,,
\end{align}
we define 
\begin{align}
(\sigma^\mu)_{\alpha\dot{\beta}} &= (+i\tau^a, \bm{1})_{\alpha\dot{\beta}}\,,\nonumber\\
(\bar{\sigma}^\mu)^{\dot{\alpha}\beta} &= (-i\tau^a, \bm{1})^{\dot{\alpha}\beta}\,,\nonumber
\end{align}
so that $\sigma^\mu\bar{\sigma}^\nu+\sigma^\nu\bar{\sigma}^\mu=2\delta^{\mu\nu}$ and
\begin{align}
\gamma^\mu = \begin{pmatrix}
0 & \sigma^\mu\\
\bar{\sigma}^\mu & 0
\end{pmatrix}\,.
\end{align}
Here $\alpha,\beta,\ldots$ are the chiral 2d Weyl spinor indices with $\alpha,\beta,\ldots\in\{+,-\}$, while $\dot{\alpha},\dot{\beta},\ldots$ are the anti-chiral spinor indices with $\dot{\alpha},\dot{\beta},\ldots\in\{\dot{+},\dot{-}\}$. Defining the charge-conjugation matrix so that $\varepsilon^{+-}=\varepsilon^{\dot{-}\dot{+}}=\varepsilon_{+-}=\varepsilon_{\dot{-}\dot{+}}=+1$, we have
\begin{align}
\psi^\alpha = \varepsilon^{\alpha\beta}\psi_\beta\,,\qquad \psi_\alpha = \psi^\beta\varepsilon_{\beta\alpha}\,,\qquad \psi_{\dot{\alpha}} = \varepsilon_{\dot{\alpha}\dot{\beta}}\psi^{\dot{\beta}}\,,\qquad \psi^{\dot{\alpha}} = \psi_{\dot{\beta}}\varepsilon^{\dot{\beta}\dot{\alpha}}\,,
\end{align}
together with
\begin{align}
\varepsilon^{\alpha\beta}\varepsilon_{\beta\gamma} = -\delta^{\alpha}_\gamma\,,\qquad \varepsilon^{\dot{\alpha}\dot{\beta}}\varepsilon_{\dot{\beta}\dot{\gamma}} = -\delta^{\dot{\alpha}}_{\dot{\gamma}}\,.
\end{align}
We further define
\begin{align}
\tensor{(\sigma^{\mu\nu})}{_\alpha^\beta} &= \frac{1}{2}(\sigma^\mu\bar{\sigma}^\nu-\sigma^\nu\bar{\sigma}^\mu)\,,\nonumber\\
\tensor{(\bar{\sigma}^{\mu\nu})}{^{\dot{\alpha}}_{\dot{\beta}}} &= \frac{1}{2}(\bar{\sigma}^\mu\sigma^\nu-\bar{\sigma}^\nu\sigma^\mu)\,,
\end{align}
which satisfy the (anti-)selfduality relations
\begin{align}
\frac{1}{2}\varepsilon_{\mu\nu\rho\sigma} \sigma^{\rho\sigma} &= + \sigma_{\mu\nu}\,,\\
\frac{1}{2}\varepsilon_{\mu\nu\rho\sigma} \bar{\sigma}^{\rho\sigma} &= - \bar{\sigma}_{\mu\nu}\,.
\end{align}
Finally, defining the (anti-)selfdual t'Hooft symbols
\begin{align}
\eta_{a\mu\nu} &= \varepsilon_{a\mu\nu 4}+\delta_{a\mu}\delta_{\nu 4} - \delta_{a\nu}\delta_{\mu 4}\,,\\
\bar{\eta}_{a\mu\nu} &= \varepsilon_{a\mu\nu 4}-\delta_{a\mu}\delta_{\nu 4} +\delta_{a\nu}\delta_{\mu 4}\,,
\end{align}
we have
\begin{align}
\sigma_{\mu\nu} = i\tau_c \eta_{c\mu\nu} \,,\qquad \bar{\sigma}_{\mu\nu} = i \tau_c\bar{\eta}_{c\mu\nu} \,,
\end{align}
together with
\begin{align}
\eta_{c\mu\nu}\eta_{c\rho\sigma} &= \delta_{\mu\rho}\delta_{\nu\sigma} - \delta_{\nu\rho}\delta_{\mu\sigma}+\varepsilon_{\mu\nu\rho\sigma}\,,\\
\bar{\eta}_{c\mu\nu}\bar{\eta}_{c\rho\sigma} &= \delta_{\mu\rho}\delta_{\nu\sigma} - \delta_{\nu\rho}\delta_{\mu\sigma}-\varepsilon_{\mu\nu\rho\sigma}\,.
\end{align}

\section{Conventions for the $A_\infty$ OSFT}
\label{app:conv}
We denote by $\xi$ and $X$ the $\xi$-ghost and PCO charges
\begin{align}
\xi = \oint_{|z|=1}\frac{dz}{2\pi i}\frac{1}{z}\xi(z)\,,\qquad  X = \oint_{|z|=1}\frac{dz}{2\pi i}\frac{1}{z}X(z)\,.
\end{align}
Note that $X=[Q,\xi]$.
We will work in the suspended Hilbert space. Let $d(A)=|A|+1$ 
denote the degree of state $A$. We define
\begin{align}
    m_2(A,B) \equiv (-1)^{d(A)}A\ast B
\label{eq:prop1}
\end{align}
and denote 
\begin{align}
    \langle A,B\rangle \equiv \mathrm{Tr}(A\ast B) \equiv\mathrm{Tr}(A B)
\end{align}
the BPZ inner product. Also,
\begin{align}
    [A,B]\equiv AB - (-1)^{|A||B|}BA
\end{align}
will be used to denote the graded $\ast$-commutator of string fields $A,B$. The symplectic form is then defined as 
\begin{align}
\omega(A,B) \equiv (-1)^{d(A)}\langle A,B\rangle\,.
\label{eq:prop2}
\end{align}
Then we have
\begin{subequations}
\begin{align}
    0&= Q^2 A\,,\\
    0&=Qm_2(A,B)+m_2(QA,B)+(-1)^{d(A)}m_2(A,QB)\,,
\end{align}
\end{subequations}
together with
\begin{subequations}
\begin{align}
    0&= Q^2 A\,,\\
    0&=QM_2(A,B)+M_2(QA,B)+(-1)^{d(A)}M_2(A,QB)\,,\\
    0&=M_2(M_2(A,B),C)+(-1)^{d(A)}M_2(A,M_2(B,C))+QM_3(A,B,C)+\\
    &\hspace{1cm}+M_3(QA,B,C)+(-1)^{d(A)}M_3(A,QB,C)+(-1)^{d(A)+d(B)}M_3(A,B,QC)\,,
\end{align}
\end{subequations}
where
\begin{align}
    M_2(A,B)&=\frac{1}{3}(Xm_2(A,B)+m_2(XA,B)+m_2(A,XB))\,.
\end{align}
In the large Hilbert space, $M_2$ is exact, i.e. 
\begin{align}
    M_2(A,B)&=Q\overline{M}_2(A,B)-\overline{M}_2(QA,B)-(-1)^{d(A)}\overline{M}_2(A,QB)\,,
\end{align}
(here $A,B$ need to be in the small Hilbert space) where we define
    \begin{align}
    \overline{M}_2(A,B)&\equiv \frac{1}{3}(\xi m_2(A,B)-m_2(\xi A,B)-(-1)^{d(A)}m_2(A,\xi B))\,.
\end{align}
The BPZ product on the small Hilbert space can be expressed in terms of the BPZ product on the large Hilbert space as
\begin{align}
\langle A,B\rangle_S = \langle \xi A,B\rangle_L= (-1)^{|A|}\langle  A,\xi B\rangle_L\,.
\end{align}
The corresponding relation for the symplectic form reads
\begin{align}
\omega_S (A,B) = -\omega_L(\xi A, B)=-(-1)^{d(A)} \omega_L( A, \xi B)\,.
\end{align}
Further, we have
\begin{subequations}
\begin{align}
    \langle A,B\rangle &= (-1)^{|A||B|}\langle B,A\rangle\nonumber\\
    \langle A,B\ast C\rangle &= \langle A\ast B,C\rangle
\end{align}
\end{subequations}
so that the cubic vertex $m_2$ enjoys the following cyclic property
\begin{subequations}
\begin{align}
\langle A, m_2(B, C)\rangle &=  (-1)^{d(A)(d(B)+d(C))}\langle B,m_2(C,A)\rangle\\
&=(-1)^{d(C)(d(A)+d(B))}\langle C,m_2(A,B)\rangle\,.
\end{align}
\end{subequations}
We also note that we have
\begin{align}
    \langle A,QB\rangle &= -(-1)^{|A|}\langle QA,B\rangle\,,
    \end{align}
    together with
    \begin{subequations}
\begin{align}
    \langle A,\xi B\rangle_L &= (-1)^{|A|}\langle \xi A,B\rangle_L\,,\\[+0.5mm]
    \langle A,XB\rangle_L &= \langle XA,B\rangle_L\,.
\end{align}
\end{subequations}
The corresponding relations for the symplectic form read
\begin{subequations}
\begin{align}
\omega(A,B) &= (-1)^{d(A)d(B)+1}\omega(B,A)\\
\omega(A,B\ast C) &= (-1)^{d(B)+1}\omega (A\ast B, C)\\
\omega(A,m_2(B, C)) &= (-1)^{d(A)+1}\omega (m_2(A, B), C)
\end{align}
\end{subequations}
together with
\begin{subequations}
\begin{align}
\omega( A, m_2(B, C)) &=  (-1)^{d(A)+d(B)}(-1)^{d(A)(d(B)+d(C))}\omega( B,m_2(C,A))\\
&=(-1)^{d(A)+d(C)}(-1)^{d(C)(d(A)+d(B))}\omega( C,m_2(A,B))
\end{align}
\label{eq:cyc}
\end{subequations}
and
\begin{align}
\omega( A,QB) &= -(-1)^{d(A)}\omega( QA,B)\,,
\end{align}
together with
\begin{subequations}
\begin{align}
    \omega_L( A,\xi B) &=  (-1)^{d(A)}\omega_L( \xi A,B)\,,\\[+0.5mm]
    \omega_L( A,XB) &=\omega_L( XA,B)\,.
\end{align}
\label{eq:BPZpar}
\end{subequations}

\section{Some useful OPE and correlators}
\label{app:ope}

We will work with the symmetric conventions where
\begingroup\allowdisplaybreaks
\begin{subequations}
\label{eq:ghostOPE}
\begin{align}
c(z)c(-z) &= -2z\, c\p c(0) +\mathcal{O}(z^3)\,,\\
\xi (z)\xi(-z) &= -2z\, \xi \p \xi(0) +\mathcal{O}(z^3)\,,\\
e^{-\phi}(z)e^{-\phi}(-z)&= (2z)^{-1}e^{-2\phi}(0)+\mathcal{O}(z)\,,\\
e^{+\phi}(z)e^{-\phi}(-z)&= 2z+(2z)^2\p \phi+\mathcal{O}(z^3)\,.
\end{align}
\end{subequations}
\endgroup
Define $V=c\mathbb{V}_\frac{1}{2}e^{-\phi}$, $W=c\mathbb{W}_\frac{1}{2}e^{-\phi}$, where the $h=1/2$ matter fields $\mathbb{V}_\frac{1}{2}$, $\mathbb{W}_\frac{1}{2}$ satisfy
\begin{align}
\mathbb{V}_\frac{1}{2}(+z)\mathbb{W}_\frac{1}{2}(-z) = (2z)^{-1}\{\mathbb{V}_\frac{1}{2}\mathbb{W}_\frac{1}{2}\}_1(0) + \{\mathbb{V}_\frac{1}{2}\mathbb{W}_\frac{1}{2}\}_0(0)+\ldots\,,
\end{align}
where $\{\mathbb{V}_\frac{1}{2}\mathbb{W}_\frac{1}{2}\}_n$ denotes the coefficient of the pole of order $n$ in the OPE of  (that is, $\{\mathbb{V}_\frac{1}{2}\mathbb{W}_\frac{1}{2}\}_1$ is proportional to the identity). 
We also denote $\mathbb{V}_1 = G_{-1/2}\mathbb{V}_{1/2}$, $\mathbb{W}_1 = G_{-1/2}\mathbb{W}_{1/2}$.
Using the formula (3.9) of \cite{Schnabl:2002gg} and the OPE \eqref{eq:ghostOPE}, one can first show that
\begin{align}
    P_0 m_2(V,W) = c\p c  \{\mathbb{V}_\frac{1}{2}\mathbb{W}_\frac{1}{2}\}_0 e^{-2\phi}\,,\label{eq:simpleStar}
\end{align}
where, in particular, we note the absence of $\{\mathbb{V}_\frac{1}{2}\mathbb{W}_\frac{1}{2}\}_1$ on the r.h.s.\ of \eqref{eq:simpleStar}. Further, we have
\begin{subequations}
\label{eq:PVW}
\begin{align}
P_0\xi m_2( V, W) &=+\xi c\p c\, \{\mathbb{V}_\frac{1}{2}\mathbb{W}_\frac{1}{2}\}_0 e^{-2\phi}\,,\\
P_0 m_2( \xi V, W) &= -\xi c\p c\, \{\mathbb{V}_\frac{1}{2}\mathbb{W}_\frac{1}{2}\}_0 e^{-2\phi}-(1/2)\p\xi c\p c\,\{\mathbb{V}_\frac{1}{2}\mathbb{W}_\frac{1}{2}\}_1 e^{-2\phi}\,,\\
P_0m_2(V, \xi W) &= -\xi c\p c\, \{\mathbb{V}_\frac{1}{2}\mathbb{W}_\frac{1}{2}\}_0 e^{-2\phi}+(1/2)\p\xi c\p c\,\{\mathbb{V}_\frac{1}{2}\mathbb{W}_\frac{1}{2}\}_1 e^{-2\phi}\,,
\end{align}
\end{subequations}
which give
\begin{align}
P_0 \overline{M}_2(V,W)\equiv \frac{1}{3}[P_0\xi m_2(V,W)-P_0m_2(\xi V,W)-P_0 m_2(V,\xi W)] = P_0\xi m_2(V,W)\,,
\label{eq:xiCon}
\end{align}
where we have noted that $d(V)=-d(\xi)=+1$. Finally, we also have
\begin{align}
    P_0 m_2 (\xi V,\xi W) =- c\p c \xi \p \xi \{\mathbb{V}_\frac{1}{2}\mathbb{W}_\frac{1}{2}\}_1 e^{-2\phi}\,.
    \label{eq:xixiVW}
\end{align}
so that $P_0 \overline{M}_2 (V,W) = \xi P_0m_2 (V, W)$. Let us further denote
\begin{align}
G(z) \mathbb{V}_\frac{1}{2}(-z) &= (2z)^{-1}\mathbb{V}_1(0)+\ldots\,,
\end{align}
so that using
\begin{align}
Q &= \oint \frac{dz}{2\pi i}\left[
c(T_\text{m}+T_{\xi\eta}+T_\phi)+b c\p c +\eta e^{\phi}G-\eta\p\eta e^{2\phi}b
\right]\,,
\end{align}
we have
\begin{subequations}
\begin{align}
Q(c{\mathbb{V}}_{\frac{1}{2}}e^{-\phi}) &= 0\,,\\
Q(c\p c\p \xi e^{-2\phi}) &= -2c\eta \,,\\[+0.75mm]
X(c{\mathbb{V}}_{\frac{1}{2}}e^{-\phi}) &= c{\mathbb{V}}_1- e^\phi\eta{\mathbb{V}}_{\frac{1}{2}}\,,\\
X(c\p c\p \xi e^{-2\phi}) &= 2c\p \phi - \p c\,.
\end{align}
\end{subequations}
It will be also useful to note that
\begin{subequations}
\label{eq:ghostCorr}
\begin{align}
\big\langle c\p c\xi e^{-2\phi}(z)c(w)\big\rangle_L &= -(z-w)^2\,,\\
\big\langle c\p c\xi\p \xi e^{-2\phi}(z)c\eta (w)\big\rangle_L &= -1\,,\\
\big\langle\eta(z_1)\xi(z_2)\xi(z_3)\big\rangle_L &= {z_{23}}({z_{12} z_{13}})^{-1}\,.
\end{align}
\end{subequations}
Setting $\alpha'=2$, we also have
\begin{subequations}
\begin{align}
i\p X^\mu(z) i\p X^\nu(w) &= +\delta^{\mu\nu}(z-w)^{-2} +\ldots\,,\\
\psi^\mu(z)\psi^\nu(w)&= +\delta^{\mu\nu}(z-w)^{-1}+\ldots\,,
\end{align}
\end{subequations}
together with
\begin{subequations}
\begin{align}
S^\alpha (z)S^\beta(w) &= +\varepsilon^{\alpha\beta}(z-w)^{-\frac{1}{2}}+(1/4)(z-w)^{+\frac{1}{2}}(\varepsilon\sigma^{\mu\nu})^{\alpha\beta}:\psi_\mu\psi_\nu:(w)+\ldots\\[+1.5pt]
S^{\dot{\alpha}} (z)S^{\dot{\beta}}(w) &=- \varepsilon^{\dot{\alpha}\dot{\beta}}(z-w)^{-\frac{1}{2}}-(1/4)(z-w)^{+\frac{1}{2}}(\bar{\sigma}^{\mu\nu}\varepsilon)^{\dot{\alpha}\dot{\beta}}:\psi_\mu\psi_\nu:(w)+\ldots\\[+1.5pt]
S_\alpha(z)S_{\dot{\beta}}(w) &= +({1}/{\sqrt{2}}){\sigma}^\mu_{\alpha\dot{\beta}}\psi_\mu(w)+\ldots\,,\\[-3pt]
\psi^\mu (z)S^\alpha(w) &= +({1}/{\sqrt{2}}){(z-w)^{-\frac{1}{2}}}(\sigma^\mu)^{\alpha\dot{\beta}}S_{\dot{\beta}}(w)+\ldots\,,
\end{align}
\end{subequations}
and
\begin{subequations}
\begin{align}
\bar{\Delta}(z)\Delta(w)=-{\Delta}(z)\bar{\Delta}(w)&= +{(z-w)^{-\frac{1}{2}}}+\ldots\,,\\
i\p X^\mu (z)\Delta(w) &= \tau^\mu(w)(z-w)^{-\frac{1}{2}}+\ldots\,,\\
i\p X^\mu (z)\tau_\nu(w)&=(1/2)\delta^\mu_\nu \Delta(w)(z-w)^{-\frac{3}{2}}+\ldots\,,\\
\bar{\tau}^\mu(z) \tau^\nu(w) &= (1/2)(z-w)^{-\frac{3}{2}}
\end{align}
\end{subequations}
It will also come in useful to note the following RNS and spin field correlators (see \cite{Hartl:2011tza} and \cite{Mattiello:2018kue})
\begin{subequations}
\begin{align}
\big\langle \psi^\mu(z_1)\psi^\nu(z_2)\psi^\rho(z_3)\psi^\sigma(z_4)\big\rangle &=\frac{\delta^{\mu\nu}\delta^{\rho\sigma}}{z_{12}z_{34}} - \frac{\delta^{\mu\rho}\delta^{\nu\sigma}}{z_{13}z_{24}}+\frac{\delta^{\mu\sigma}\delta^{\nu\rho}}{z_{14}z_{23}}\,,\\
\big\langle \psi^\mu(z_1)\psi^\nu(z_2)i\p X^\rho(z_3)i\p X^\sigma(z_4)\big\rangle &= \frac{\delta^{\mu\nu}\delta^{\rho\sigma}}{z_{12}z_{34}^2}\,,
\end{align}
\end{subequations}
together with
\begin{subequations}
\begin{align}
\big\langle\psi^\mu(z_1)\psi^\nu(z_2) S^{\alpha}(z_3) S^{\beta}(z_4) \big\rangle\! &=\!+ \frac{z_{34}}{(z_{13}z_{14}z_{23}z_{24}z_{34})^{1/2}}\!\bigg[\delta^{\mu\nu}\varepsilon^{\alpha\beta}\frac{z_{13}z_{24}}{z_{12}z_{34}}\!-\!\frac{1}{2}(\varepsilon\sigma^\mu\bar{\sigma}^{\nu})^{\alpha\beta}\bigg]\,,\\
\big\langle\psi^\mu(z_1)\psi^\nu(z_2) S^{\dot{\alpha}}(z_3) S^{\dot{\beta}}(z_4) \big\rangle\! &=\!- \frac{z_{34}}{(z_{13}z_{14}z_{23}z_{24}z_{34})^{1/2}}\!\bigg[\delta^{\mu\nu}\varepsilon^{\dot{\alpha}\dot{\beta}}\frac{z_{13}z_{24}}{z_{12}z_{34}}\!-\!\frac{1}{2}(\bar{\sigma}^\mu{\sigma}^{\nu}\varepsilon)^{\dot{\alpha}\dot{\beta}}\bigg]\,,
\end{align}
and
\begin{align}
\big\langle S^\alpha(z_1)S^\beta(z_2)S^\gamma(z_3)S^\delta(z_4)\big\rangle &=\bigg(\frac{z_{12}z_{13}z_{23}z_{24}}{z_{13}z_{24}}\bigg)^\frac{1}{2}\bigg(\frac{\varepsilon_{\alpha\beta}\varepsilon_{\gamma\delta}}{z_{12}z_{34}}-\frac{\varepsilon_{\alpha\delta}\varepsilon_{\beta\gamma}}{z_{14}z_{23}}\bigg)\,,\\
\big\langle S^\alpha(z_1)S^\beta(z_2)S^{\dot{\gamma}}(z_3)S^{\dot{\delta}}(z_4)\big\rangle &= -\varepsilon^{\alpha\beta}\varepsilon^{\dot{\gamma}\dot{\delta}}(z_{12}z_{34})^{-\frac{1}{2}}\,,
\end{align}
\end{subequations}
and
\begin{align}
\big\langle\bar{\Delta}(z_1) i \p X^\mu (z_2)i \p X^\nu (z_3) \Delta  (z_4)\big\rangle  \!&=\! - \frac{1}{2}\frac{\delta^{\mu\nu}}{z_{14}^\frac{1}{2}z_{23}^2}\bigg(\sqrt{\frac{z_{13}z_{24}}{z_{12}z_{34}}}+\sqrt{\frac{z_{12}z_{34}}{z_{13}z_{24}}}\bigg)\,,
\end{align}
and
\begin{subequations}
\begin{align}
\big\langle \psi^\mu (z_1) S^\alpha(z_2)S^{\dot{\beta}}(z_3)\big\rangle&=+({1}/{\sqrt{2}})(\sigma^\mu)^{\alpha\dot{\beta}}z_{12}^{-\frac{1}{2}}z_{13}^{-\frac{1}{2}}\,,\\
\big\langle i\p X^\mu (z_1) \bar{\Delta}(z_2){\tau}_\nu(z_3)\big\rangle &=+(1/2)\delta^\mu_\nu z_{12}^{-\frac{1}{2}}z_{13}^{-\frac{3}{2}}\,.
\end{align}
\end{subequations}
Finally, we have
\begin{subequations}
\begin{align}
\big\langle \bar{\Delta}S^\alpha(z_1) \Delta S^\beta(z_2)\bar{\Delta}S^\gamma(z_3) \Delta S^\delta(z_4)\big\rangle &= \frac{\varepsilon^{\alpha\beta}\varepsilon^{\gamma\delta}}{z_{12}z_{34}}-\frac{\varepsilon^{\alpha\delta}\varepsilon^{\beta\gamma}}{z_{14}z_{23}}\,,\\
\big\langle \bar{\tau}_\mu S^{\dot{\alpha}}(z_1){\tau}_\nu S^{\dot{\beta}}(z_2)\bar{\Delta} S^{\gamma} (z_3){\Delta}S^\delta (z_4) \big\rangle &= -\frac{\delta_{\mu\nu}}{2}\frac{\varepsilon^{\dot{\alpha}\dot{\beta}}\varepsilon^{\gamma\delta} }{ z_{12}z_{34}^2}\,,\\[+1.5mm]
\big\langle\bar{\Delta} S^\alpha(z_1)\tau_{\mu}S^{\dot{\beta}}(z_2)\bar{\Delta}S^\gamma(z_3)\tau_\nu S^{\dot{\delta}}(z_4)\big\rangle &= 0\,.
\end{align}
\end{subequations}

\bibliography{worldsheet}{}
\bibliographystyle{JHEP}

\end{document}